\RequirePackage{fix-cm}
\documentclass[twocolumn,epjc3]{svjour3}

\smartqed  
\RequirePackage{graphicx}
\usepackage{color,amsmath,amssymb,epsfig,graphicx}
\usepackage{mathptmx, courier, pifont}
\usepackage{adjustbox}
\usepackage[scaled=0.92]{helvet}
\usepackage[T1]{fontenc}
\usepackage{textcomp}
\usepackage{bm}
\usepackage{mathptmx, courier, pifont}
\usepackage[scaled=0.92]{helvet}
\usepackage[T1]{fontenc}
\usepackage{textcomp}
\usepackage[colorlinks=true,linkcolor=blue,filecolor=blue,urlcolor=blue,citecolor=blue]{hyperref}

\usepackage{multirow}
\usepackage{longtable}
\def\bra{\,<\!} \def\ket{\!>\,} \def\ack{\,|\,}

\journalname{Eur. Phys. J. A}
\begin{document}
\sloppy

\title{ Fingerprints of the triaxial deformation from energies  and  $B(E2)$  transition probabilities of $\gamma$-bands in transitional and deformed nuclei }


\author{S.~P. ~Rouoof\thanksref{addr1} \and Nazira ~Nazir\thanksref{addr2} \and S.~Jehangir\thanksref{e1,addr1} \and G.H. Bhat\thanksref{addr3} \and J.A. Sheikh\thanksref{addr1,addr2} \and N. Rather\thanksref{addr1} \and and S. Frauendorf\thanksref{addr4}}

\thankstext{e1}{e-mail: sheikhahmad.phy@gmail.com}


\institute{Department of Physics, Islamic University of Science and Technology, Awantipora-192122, India\label{addr1}
\and
  Department of Physics, University of Kashmir, Srinagar,190 006, India \label{addr2}  
  \and
  Department of Physics, SP College  Srinagar, Jammu and Kashmir, 190 001, India\label{addr3}
  \and
Physics Department, University of Notre Dame, Notre Dame, Indiana 46556, USA\label{addr4}
}

\date{Received: date / Accepted: date}



\maketitle

\begin{abstract}
  The energies and $B(E2)$ transitions involving the states of the ground- and $\gamma$-bands  in thirty transitional
  and deformed nuclei are 
  calculated using the triaxial projected shell model (TPSM) approach. Systematic good agreement with the
  existing data substantiates 
  the reliability of the model predictions. The Gamma-rotor version of the collective Bohr Hamiltonian is discussed
  in order to quantify the 
  classification with respect to the triaxial shape degree of freedom. The pertaining criteria are applied
  to the TPSM results and  
  the staggering of the energies of the $\gamma$-bands is analyzed in detail. An analog staggering  of the
  intra-$\gamma$ $B(E2, I \rightarrow I-2)$
  is introduced for the first time. The emergence of the staggering phenomena in the transitions is explained in the terms of
  interactions between the bands.

\end{abstract}




\PACS  21.60.Cs, 21.10.Hw, 21.10.Ky, 27.50.+e


\maketitle

\section{Introduction}
A recurring theme in nuclear structure research is how to classify the collective excitation modes and 
how to discern the nature of the collective motion from the measured properties \cite{BM75}. 
The appearance and the consequences of triaxial quadrupole deformation are currently of considerable interest. 
The topic is mostly addressed in terms of some version of the  Bohr Hamiltonian \cite{BM75}, which describes the collective 
motion in terms of the deformation parameters of the nuclear shape. These approaches 
may be purely phenomenological or based on fitted parameters to a microscopic theory. 
A recent review of these approaches can be found in Ref. \cite{Frauendorf15}.
From the phenomenological perspective emerge the concepts of static and dynamic triaxiality that
are used to interpret the experimental results 
for energies and transition probabilities between the near-yrast states.

The spherical shell model (SSM) is an alternative approach that has been demonstrated of being capable of
accounting for the data on the collective excitations in lighter nuclei \cite{SSM,Brown22,Poves12}.
Recent progress in the shell model techniques has made it possible 
to carry out calculations for heavy deformed nuclei  \cite{Otsuka2019,Tsunoda2021}. Besides reproducing
the energies and transition probabilities,
the authors devote substantial effort to connect the SSM results with the established concepts of a triaxial shape and its fluctuations.

In the present work, we have employed the triaxial projected shell model (TPSM) \cite{JS16} which
incorporates a major part of the SSM correlations by generating the shell model basis space from 
quasiparticle configurations in a deformed potential.
The details of the TPSM approach can be found in our previous publications \cite{JS16,JS99,GH14,GH08,bh15,JG11,GH12,GS12,SJ18}.
The pairing plus quadrupole Hamiltonian is diagonalized in a basis consisting of angular-momentum projected \cite{RS80} quasiparticle configurations 
that are generated with a fixed triaxial deformation value. The basis space in the TPSM approach consists of
zero-, two- and four- quasiparticle configurations that allows one 
to describe the interplay between the collective and the quasiparticle excitations. The computational effort in TPSM approach
is negligible compared to the SSM approach, and it has been shown
that TPSM provides an accurate description of the experimental data of the states in the near-yrast region.
Nevertheless, like for the SSM approach, 
one would like to relate the results of the TPSM diagonalization to the concepts developed in the framework of  the
collective models, which is one of the major aims of the present study.

The present work is a continuation of our previous investigation of the collective
$\gamma$-degree of freedom for a large set of
deformed and transitional nuclei \cite{GH14,SJ21}. This
analysis was primarily focused on the excitation energies of twenty-three nuclei, where energies were known for
both even- and odd-spin branches of $\gamma$-bands so that it was possible to evaluate the energy staggering \cite{NV91}. 
It was shown that TPSM approach  provides an excellent description of the  systematics of 
the staggering parameter, defined as,
\begin{eqnarray}\label{eq:staggering1}
S(I)= \frac{[E(I)-E(I-1)]-[E(I-1)-E(I-2)]}{E(2^{+}_1)}~~~~.
\end{eqnarray}
In the framework of collective models, the staggering parameter is known to be strongly correlated with the rigidity of the triaxial shape \cite{NV91,McCutchan07,AS60,KM70,CB78,CB80}.
The odd-$I$-down pattern indicates the concentration of the collective 
wave function around a finite $\gamma$-value (static triaxiality) like for the Davydov model  \cite{Davydov} of $\gamma$-rigid motion.
The even-$I$-down pattern points to a spread of the wavefunction over the whole range of $\gamma$ (dynamic triaxiality) like for the Wilets-Jean model \cite{Wilets}.
This correlation is reviewed in Refs. \cite{Frauendorf15,McCutchan07,MC11,Stefanescu07,ab52,km66,jm87}, where the relevant literature is cited. It needs to be pointed out that the
first application of TPSM approach to $\gamma$-bands was performed in Refs. \cite{Sun2000,Boutachkov2002} .

It was shown \cite{GH14,SJ21} that restricting the TPSM basis to
 the angular-momentum projection from the vacuum configuration, always
generates the odd-$I$-down pattern of $\gamma$-rigidity.
However, it was observed \cite{GH14,SJ21} that the quasiparticle admixtures into the vacuum configuration changed the staggering
phase from the $\gamma$-rigid to that of
$\gamma$-soft pattern for all nuclei from a selection of twenty-three,  except for the four nuclei of $^{76}$Ge, $^{112}$Ru, $^{170}$Er and $^{232}$Th. In Ref. \cite{SJ21} transition probabilities were only studied  for the 
Mo- and Ru-isotopes using a phenomenological Bohr Hamiltonian \cite{MC11,GG69,PO80,AF65}, the parameters of which were fitted to  the TPSM 
energies $\biggl({\left[\frac{E(2^+_2)}{E(2^+_1)}\right]}$	, ${\left[\frac{E(2^+_2)}{E(4^+_1)}\right]}$ and $S(I)\biggl)$. It was
shown that this approach gives $B(E2)$ transition probabilities between the low-lying states, that are similar to the TPSM probabilities. 
In our more recent work \cite{Na23} we demonstrated that the TPSM approach provides an excellent description of the large set of $E2$ matrix elements from
COULEX experiments available for $^{104}$Ru \cite{Sr06}. 
Applying  the shape invariant analysis to the $E2$ 
matrix elements, we showed that the inclusion of the quasiparticle excitations transforms
 $^{104}$Ru  from $\gamma$-rigid to $\gamma$-soft.

In the present study, we complement the results of our previous work \cite{GH14,SJ21} by performing
an exhaustive analysis of the $B(E2)$ transition probabilities, both in-band and inter-band, of the yrast- and the $\gamma$-bands for
thirty nuclei, which encompasses both transitional and well-deformed nuclei. The focus
is whether the  $B(E2)$ transitions can delineate $\gamma$-soft versus $\gamma$-rigid characteristics.
We also discuss the TPSM energies of 
Os- and Pt-isotopes which were not considered in Ref. \cite{SJ21}.

In our previous study, the delineation between $\gamma$-soft and $\gamma$-rigid has been qualitative and was mainly based on the sign of the
staggering phase. Here we introduce a more quantitative 
classification  in  the framework 
of the Gamma-rotor model \cite{caprio11}, which is a simplified version of the algebraic collective model (ACM)
of Bohr Hamiltonian \cite{BM75,RW10}.
   In section \ref{sec:ACM}, we discuss several typical potentials and relations between energies and $B(E2)$  transition rates that characterize them.

 The TPSM calculations are presented in section \ref{sec:TPSM} and compared with the available experimental information.
 In section \ref{sec:discussion}, the results are discussed with the underlying question whether the
 spectroscopic signatures
 for $\gamma$-softness or $\gamma$-rigidity, as they  emerge from the phenomenological  Bohr Hamiltonian in section \ref{sec:ACM},  appear in the
 TPSM calculations and how do they
 correlate with the data. It is demonstrated that the TPSM accounts for the available experimental data
 on energies and the $B(E2)$ reduced transition probabilities
 remarkably well, without adjustment of any additional parameters. We discuss the $(N-Z)$ dependence of
 $\gamma$-softness versus $\gamma$-rigidity 
 in terms of the quasiparticle composition near the Fermi level. 
  The details of the Gamma-rotor model are provided in the Appendix.

\begin{figure*}[t]
 \centerline{\includegraphics[trim=0cm 0cm 0cm
0cm,width=1.1\textwidth,clip]{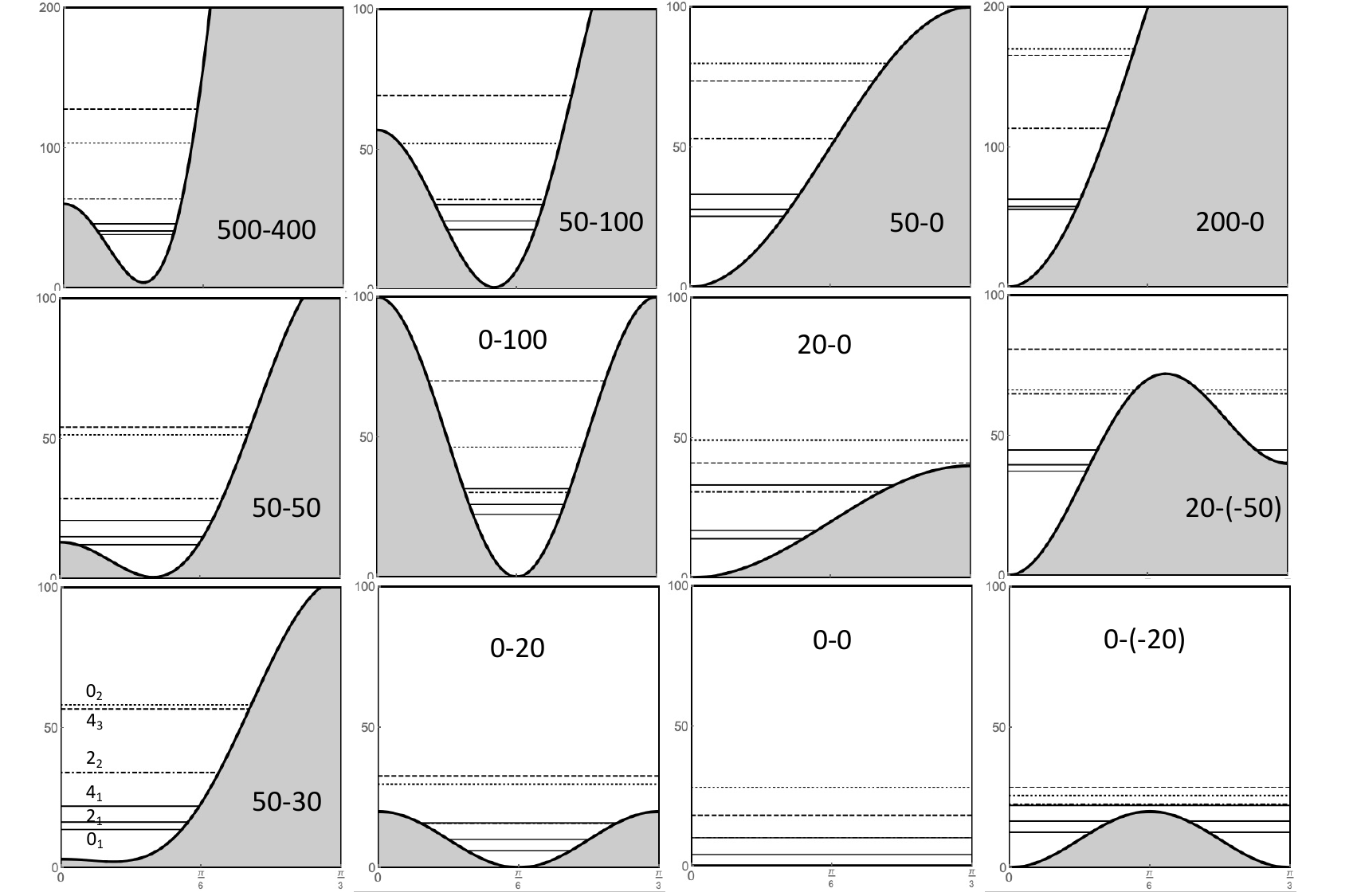}} \caption{ \label{fig:grot}
Energies of the low-lying states of the phenomenological Gamma-rotor Hamiltonian (\ref{eq:ACM})  \cite{caprio11} for characteristic potentials, which are labeled by their parameters $\chi-\kappa$. The abscissas represent the triaxiality parameter $0\leq \gamma  \leq \pi/3$. The ordinates represent the energies of the displayed quantities in units of  $\hbar^2/B\beta^2$, where the range is [0,100], except for  the potentials  500-400 and 200-0, where the range is [0,200].
The horizontal lines show the energies of the heads of the low-lying bands, which are labeled by their $I^+_n$ in the spectrum for the potential 50-30. Rescaling the energies by the factor $E(2^+_1)_{exp}/E(2^+_1)_{GR}$ adapts the potentials to the experimental energy scale, where $E(2^+_1)_{GR}$ is given in Table \ref{tab:gav}.
}  
\end{figure*}
\begin{figure*}[t]
 \centerline{\includegraphics[trim=0cm 0cm 0cm 0cm,width=1.06\textwidth,clip]{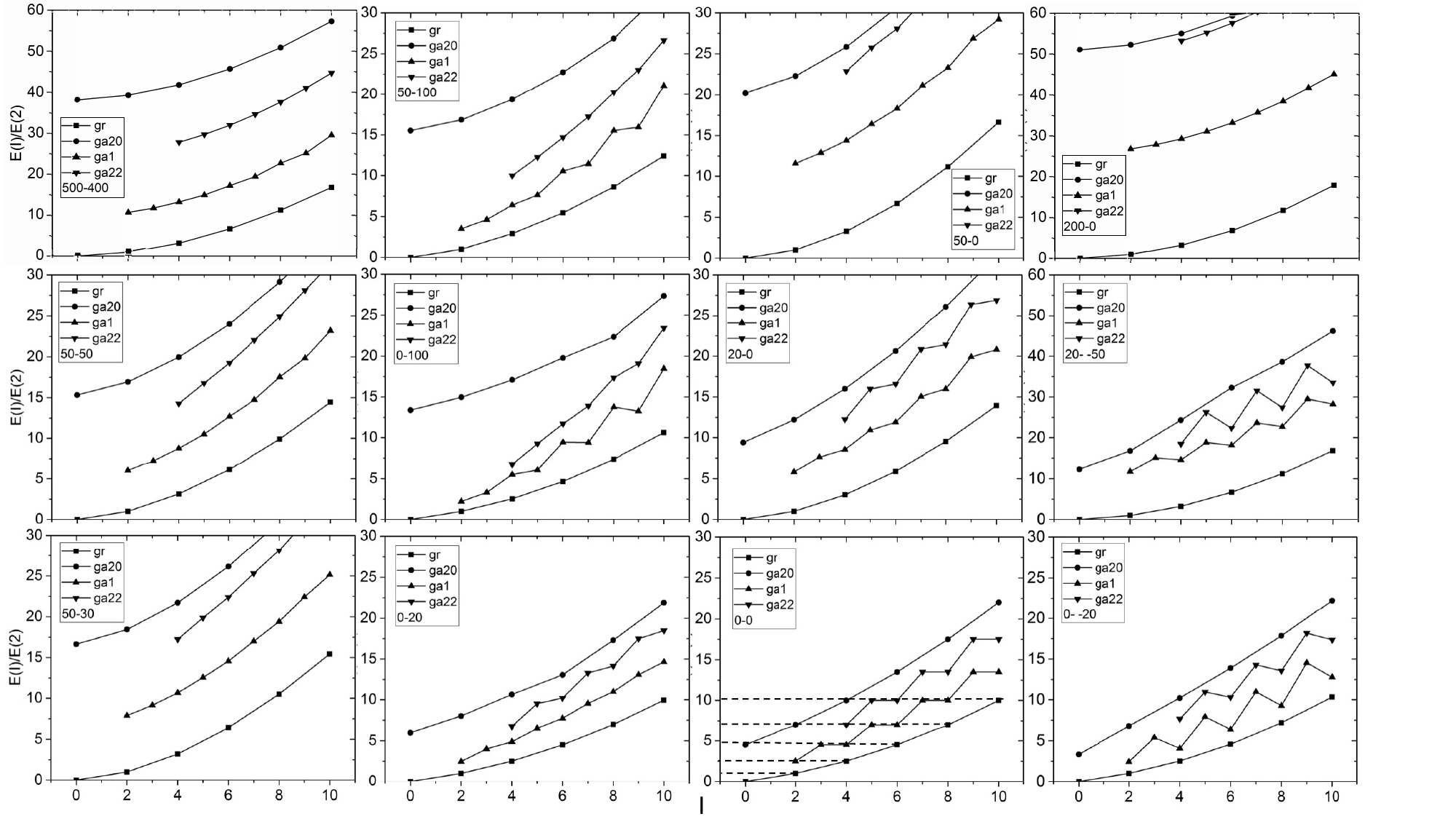}} 
\caption{ \label{fig:gband} Rotational sequences on the low-lying states of the potentials shown in Fig. \ref{fig:grot}. 
(Note the different scales of 500-400, 200-0 and  20-(-50)).  The dashed lines for the $\gamma$-independent potential 0-0 connect the states of 
the seniority $v$, which is equal to $I/2$ of the yrast states.  The labelling of the bands in the legend is as follows, gr : ground/yrast-band, ga1 : $\gamma$-band, ga22 : $\gamma\gamma$4-band and ga20 : $\gamma\gamma$0-band. }
\end{figure*} 

\begin{table}[htp!]
\LTcapwidth=0.4\textwidth
\caption{Energy characteristics of the potentials plotted in Fig. \ref{fig:grot} .
The staggering parameter $\bar S(I)$ is defined by Eq. (\ref{eq:staggering2}). The excitation energy $E(2^+_1)_{GR}$  is
 the difference $E(2^+_1)-E(0^+_1)$ of the eigenvalues of the Gamma-rotor Hamiltonian (\ref{eq:ACM}).
}
\resizebox{0.49\textwidth}{!}
  {
\begin{tabular}{|c|c|c|c|c|c|c|}
  \hline
 $\chi-\kappa$  &$E(2^+_1)_{GR}$ &$\gamma_m$	 &$\Delta \gamma$	 &${\left[\frac{E(2^+_2)}{E(2^+_1)}\right]}$	&${\left[\frac{E(2^+_2)}{E(4^+_1)}\right]}$		&$\bar S(6)$ \\
  \hline										
200-0        &2.15       & 0       &14               &26.8     &8.07     &-0.03  \\    %
100-0        &2.24       & 0       &17               &18.0    &3.31    &-0.14 \\    %
50-0          &2.39      &0          &20                &11.6   &3.55     &-0.51   \\ %
20-0         &2.87        &  0   & 24                &5.82      &1.93      &-1.75  \\  %
10-0         &3.42        &  0   & 27                 &3.61     &1.32       &-2.49  \\  %
0-0           &4.00         &30     &60              &2.50     &1.00         &-2.75  \\  %
0-200        &3.05       & 30       &16               &2.11     &0.81     &3.87  \\   %
0-100       &3.57      &30        &19               &2.21     &0.86      &3.12   \\ %
0-50       &3.80     &30        &21                  &2.34     &0.92     &1.61\\ %
0-20          &3.95      & 30    &22                  &2.45    &0.98       &0.50 \\   %
50-100     &3.10       &25          &19	 	   &3.50   &1.20     &1.85   \\%
500-400	&2.34	&17	        &15		&10.8   &3.17          &0.32 \\%
50-50        &2.73        &20      &26                &6.01   &1.92       &0.30 \\ %
50-30        & 2.58     &11     &26                  &7.78    &2.44       & -0.35  \\%
0- -20       &3.39      &30     &35                 &2.42     &0.96        &-5.79\\%
20- -50      &2.35     &34     &17                   &11.74     &3.62    &-5.55    \\ %
\hline								
\end{tabular}
}
\label{tab:gav}
\end{table}

\begin{table*}[htp!]
\LTcapwidth=0.4\textwidth
\caption{Transition probabilities that characterize the potentials in Fig. \ref{fig:grot} in
units of $B(E2, 2^+_1\rightarrow 0^+_1 )$. The $B(E2,4^+_3\rightarrow 2^+_1)$ values are smaller than 0.003.
}
\resizebox{\textwidth}{!}
  {
\begin{tabular}{|c|c|c|c|c|c|c|c|c|c|c|}
  \hline
 $\chi-\kappa$  &$Q(2^+_1)$	 &$Q(2^+_2)$	 &$B(E2, 2^+_2\rightarrow 0^+_1 )$	&$B(E2,4^+_3\rightarrow 2^+_2)$	&$B(E2,2^+_2\rightarrow 2^+_1)$
 &$B(E2,0^+_2\rightarrow 2^+_2)$	&$B(E2,0^+_2\rightarrow 2^+_1)$  &$B(E2,2^+_3\rightarrow 0^+_1)$	&$B(E2,3^+_1\rightarrow 2^+_1)$ &$B(E2,3^+_1\rightarrow 2^+_2)$	\\
 	
  \hline										
200-0         &-0.888      &0.873     &0.033     &0.094    &0.056     &0.195          &0.001    &  0.002     &0.059     &1.715\\%
100-0         &-0.878      &0.859     &0.047     &0.130    &0.086    &0.304          &0.004    &  0.003     &0.084     &1.675\\%
50-0          &-0.861       &0.840     &0.064    &0.164     &0.143   &0.527           &0.009      &0.008     &0.114     &1.595\\%
20-0          &-0.797      &0.789     &0.079      &0.106    &0.347   &1.193          &0.030       &0.004      &0.136     &1.362\\%
10-0          &-0.656      &0.655     &0.060      &0.030    &0.719   &1.550         &0.040       &0.003      &0.097    &1.235\\
0-0           &0.000        &0.000      &0.000     &0.000     &1.428   &1.666          &0.000       &0.000     &0.000     &1.190\\%
0-200       &0.000        &0.000     &0.000     &0.000    &1.422    &0.102          &0.000        &  0.000   &0.000     &1.763\\%
0-100        &0.000       &0.000      &0.000     &0.000    &1.416  &0.208         &0.000        &0.000    & 0.000     &1.723\\%
0-50          &0.000       &0.000      &0.000     &0.000    &1.413    &0.445          &0.000        &0.000    & 0.000    &1.663\\
0-20           & 0.000    &0.000      &0.000      &0.000      &1.421  &0.922          &0.000       &0.000     &0.000    &1.458\\
50-100      &-0.693     &0.684       &0.084      &0.180   &0.629    &  0.247         &0.011      &0.001      &0.148     &1.704\\%
500-400     &-0.861     &0.851	  &0.066	  &0.130    &0.143      &0.142	    &0.004       &0.003       &0.118    &1.736 \\%
50-50         &-0.807      &0.789    &0.091    &0.186      &0.316     &0.425          &0.016       &0.002     &0.160    &1.634\\%
50-30          &-0.835    &0.815      &0.082     &0.181     &0.227   &0.492          &0.014       &0.002     &0.144   &1.609 \\%
0- -20            &0.000      &0.000    &0.000     &0.000     &1.411   &2.810         &0.000       &0.000    &0.000     &0.785\\%
20- -50         &-0.868      &0.874    &0.032     &0.016     &0.089    &2.995          &0.009      &0.022    &0.0921   & 0.750\\%
\hline								
\end{tabular}
}\label{tab:gtrans}
\end{table*}

\begin{figure}[t]
 \centerline{\includegraphics[trim=0cm 0cm 0cm 0cm,width=1.1\linewidth,clip]{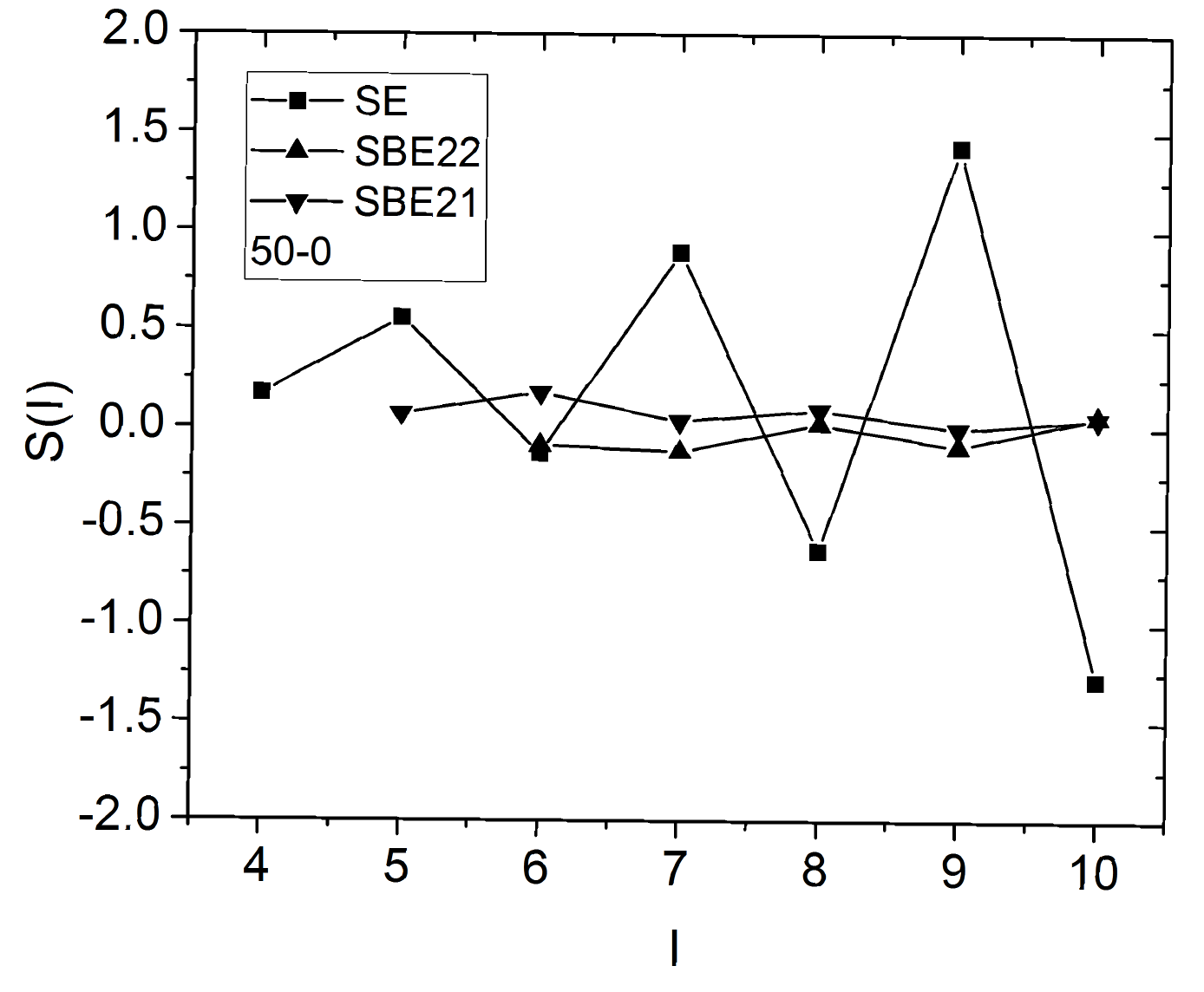}} 
  \centerline{\includegraphics[trim=0cm 0cm 0cm 0cm,width=1.1\linewidth,clip]{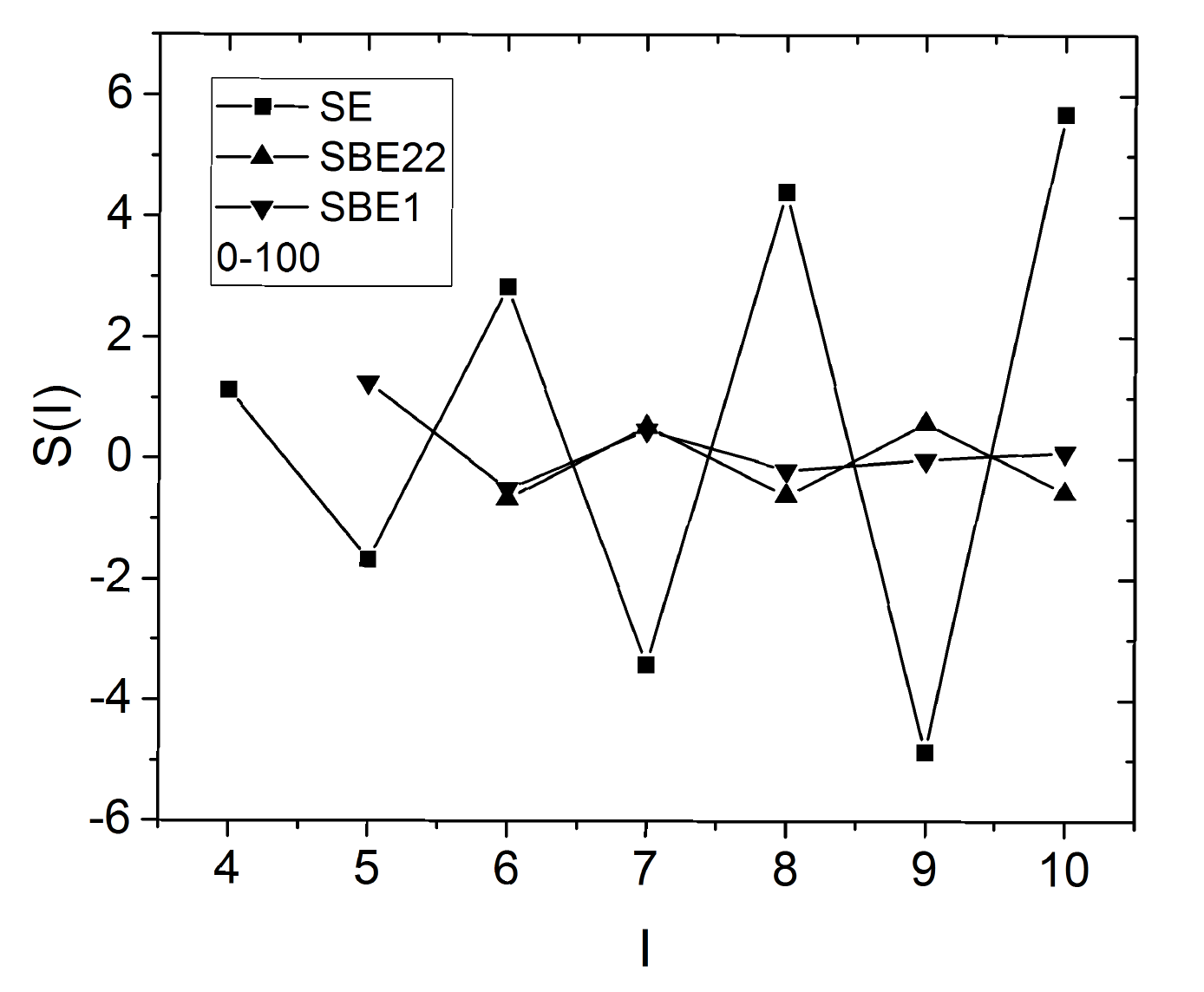}} 
\caption{ \label{fig:SBE2} Staggering parameter $S(I)$ calculated by Eq. (\ref{eq:staggering1}) from the energies ($SE$) and by the analog equations (\ref{eq:SBE22}) 
from the $B(E2, I\rightarrow I-2)$ values ($SBE22$) and (\ref{eq:SBE21})  from the $B(E2, I\rightarrow I-1)$ values ($SBE21$). The upper panel shows the soft 
prolate potential 50-0 and the lower panel triaxial potential 0-100.}
\end{figure} 
\begin{figure}[t]
 \centerline{\includegraphics[trim=0cm 0cm 0cm
0cm,width=0.52\textwidth,clip]{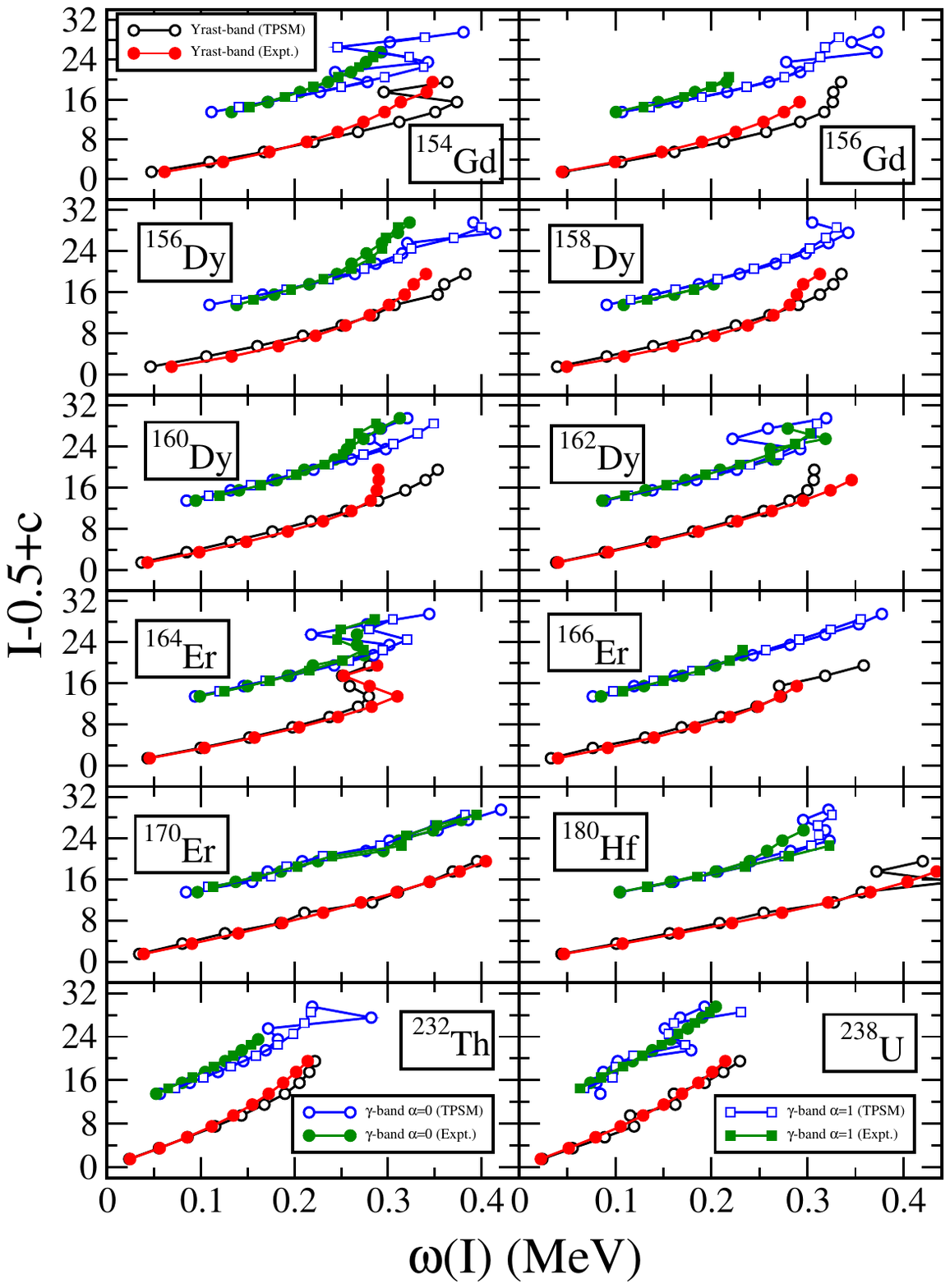}} \caption{(Color
online) Angular momentum as function of the angular frequency for  the yrast and the $\gamma$-bands after 
configuration mixing compared  with experimental data for the $^{154,156}$Gd, $^{156,158,160,162}$Dy, $^{164,166,170}$Er, $^{180}$Hf, $^{232}$Th and $^{238}$U isotopes. 
A shift $c=10$ is added for the $\gamma$-bands.
The even-I states $(\alpha=0)$ are shown as circles and the odd-I $(\alpha=1)$ states as squares. (Data taken from \cite{NNDC,Lee1980,Emling1981,Majola2015}).}
\label{fig:energy1}
\end{figure}
\begin{figure}[t]
 \centerline{\includegraphics[trim=0cm 0cm 0cm
0cm,width=0.52\textwidth,clip]{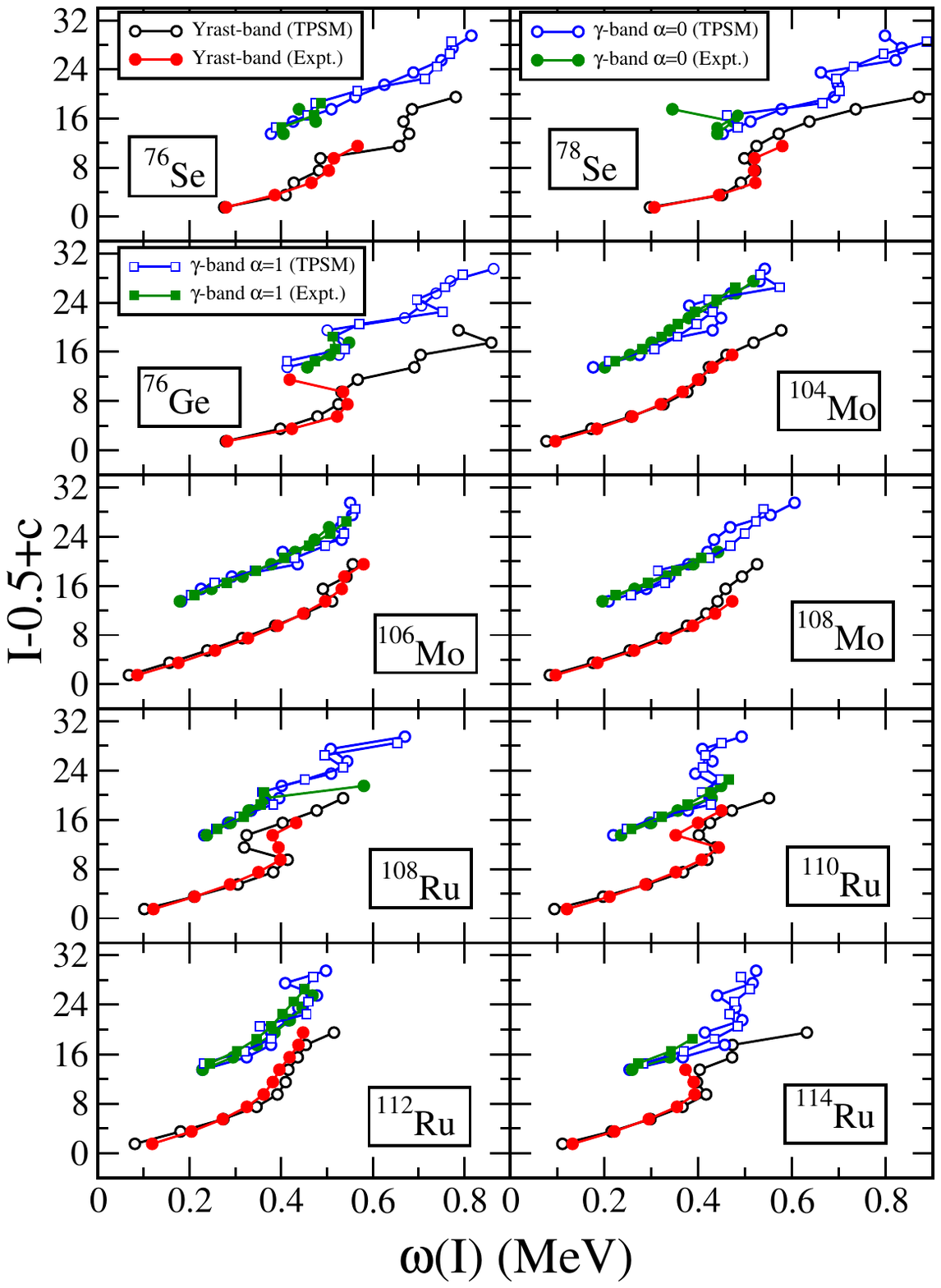}} \caption{(Color
online) Angular momentum as function of the angular frequency for  the yrast and the $\gamma$-bands after 
configuration mixing compared  with experimental data for the $^{76,78}$Se, $^{76}$Ge, $^{104,106,108}$Mo and $^{108,110,112,114}$Ru isotopes. 
A shift $c=10$ is added for the $\gamma$-bands. 
The even-I states $(\alpha=0)$ are shown as circles and the odd-I $(\alpha=1)$ states as squares. (Data taken from \cite{NNDC,Musangu2021}).
  }
\label{fig:energy2}
\end{figure}

\begin{figure}[t]
 \centerline{\includegraphics[trim=0cm 0cm 0cm
0cm,width=0.52\textwidth,clip]{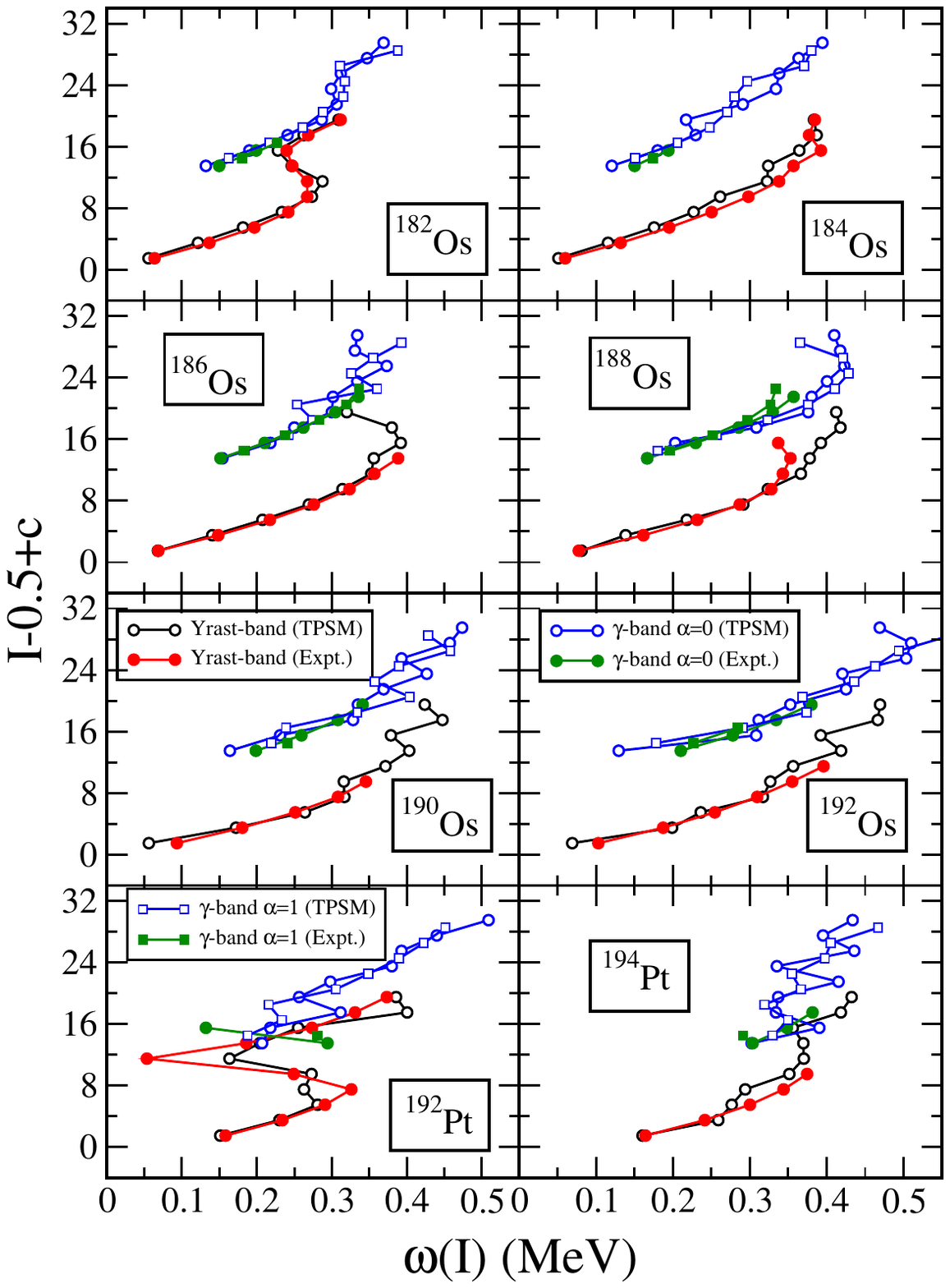}} \caption{(Color
online) Angular momentum as function of the angular frequency for  the yrast and the $\gamma$-bands after 
configuration mixing compared  with experimental data for the $^{182-192}$Os and $^{192,194}$Pt isotopes. A shift $c=10$ is added for the $\gamma$-bands. (Data taken from \cite{NNDC}).
  }
\label{fig:energy3}
\end{figure}

 \begin{figure}[htb]
  \centerline{\includegraphics[trim=0cm 0cm 0cm
 0cm,width=0.5\textwidth,clip]{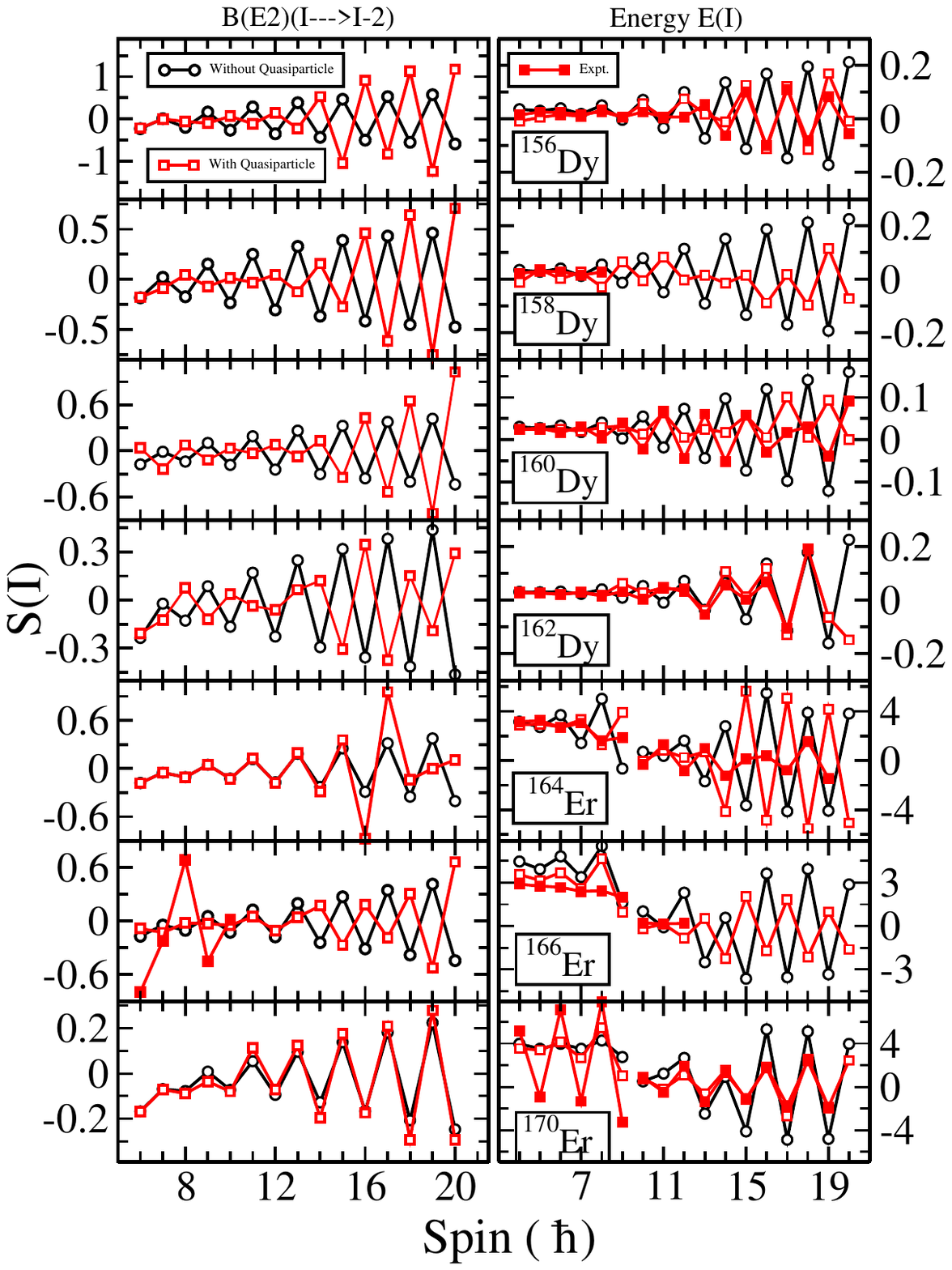}}  \caption{(Color
 online) Staggering parameters $S(I)$ calculated by means of Eq. (\ref{eq:staggering1}) from the energies of the $\gamma$-band (right panels) and
 by the analogue equation (\ref{eq:SBE22}) from the intra $\gamma$-band $B(E2, I\rightarrow I-2)$ values (left panels). For the Er isotopes, $10\times S(I)$ is displayed for $I\leq 9$ in order to resolve the details of the small staggering.  
   }
 \label{fig:gg2E1}
 \end{figure}

 \begin{figure}[htb]
  \centerline{\includegraphics[trim=0cm 0cm 0cm
 0cm,width=0.5\textwidth,clip]{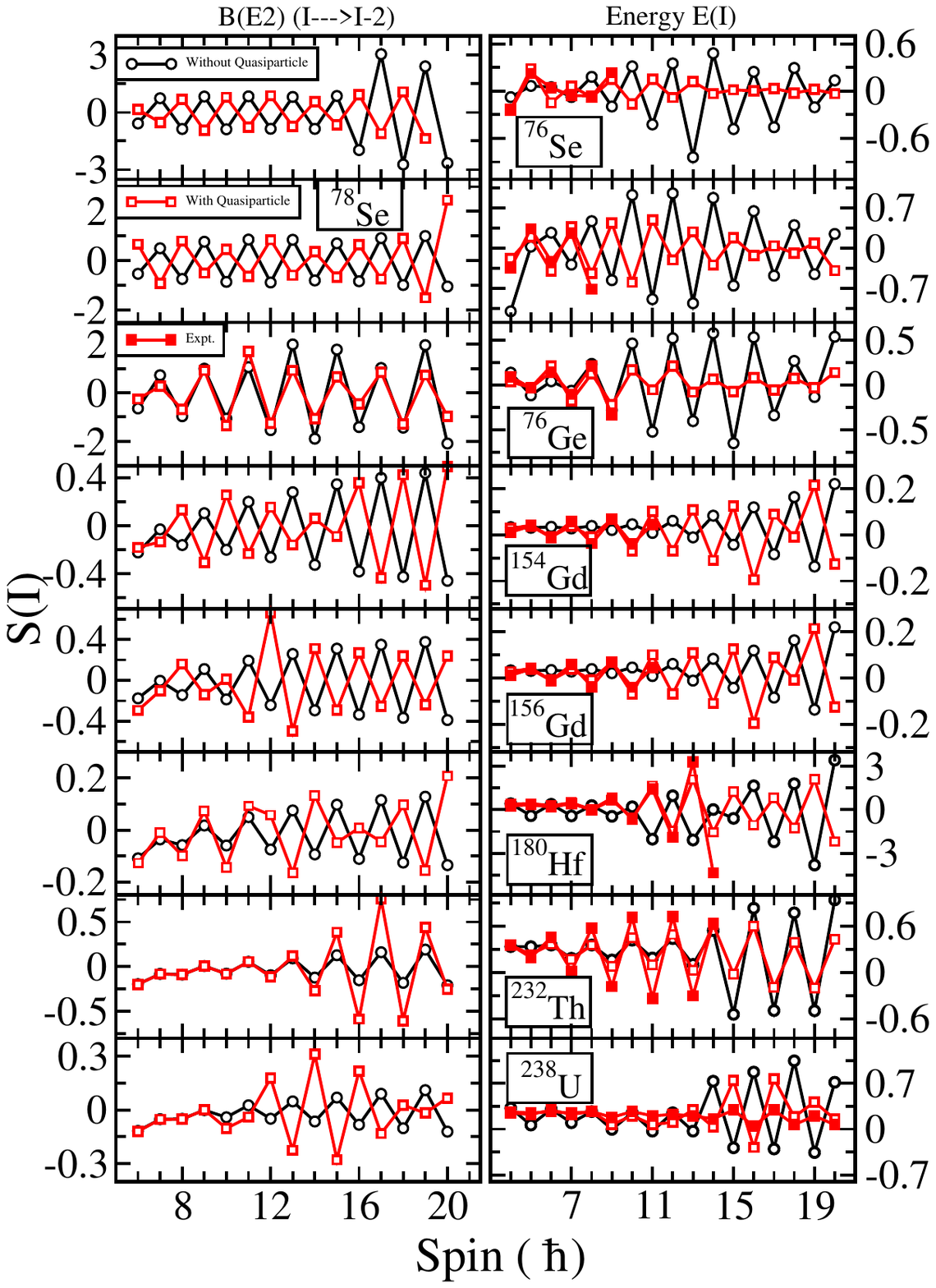}} \caption{(Color
 online) Staggering parameters $S(I)$ calculated by means of Eq. (\ref{eq:staggering1}) from the energies of the $\gamma$-band (right panels) and
 by the analogue equation (\ref{eq:SBE22}) from the intra $\gamma$-band $B(E2, I\rightarrow I-2)$ values (left panels).  
   }
 \label{fig:gg2E2}
 \end{figure}
 \begin{figure}[htb]
  \centerline{\includegraphics[trim=0cm 0cm 0cm
 0cm,width=0.5\textwidth,clip]{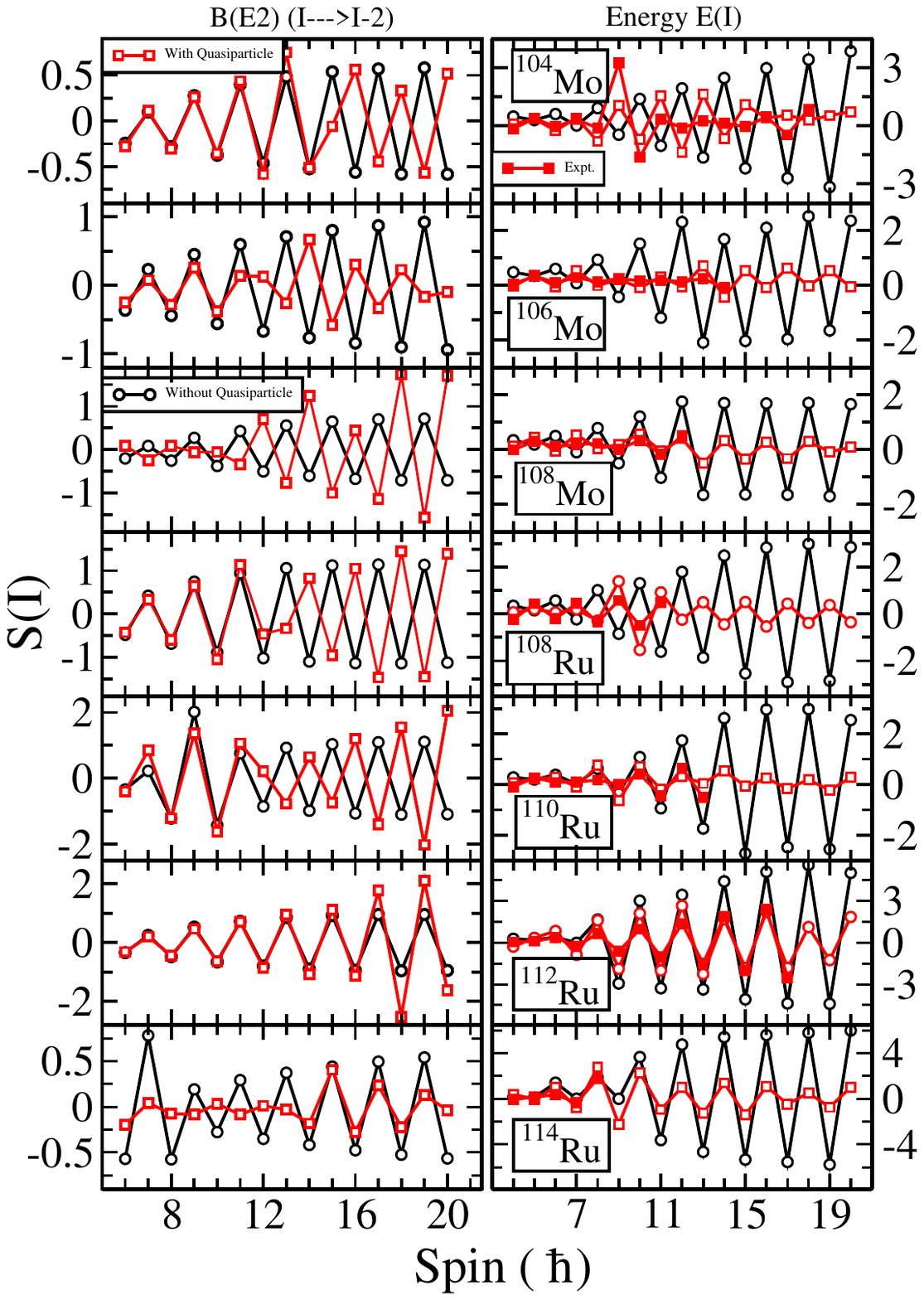}} \caption{(Color
 online) Staggering parameters $S(I)$ calculated by means of Eq. (\ref{eq:staggering1}) from the energies of the $\gamma$-band (right panels) and
 by the analogue equation (\ref{eq:SBE22}) from the intra $\gamma$-band $B(E2, I\rightarrow I-2)$ values (left panels).   
   }
 \label{fig:gg2E3}
 \end{figure}
 \begin{figure}[htb]
  \centerline{\includegraphics[trim=0cm 0cm 0cm
 0cm,width=0.5\textwidth,clip]{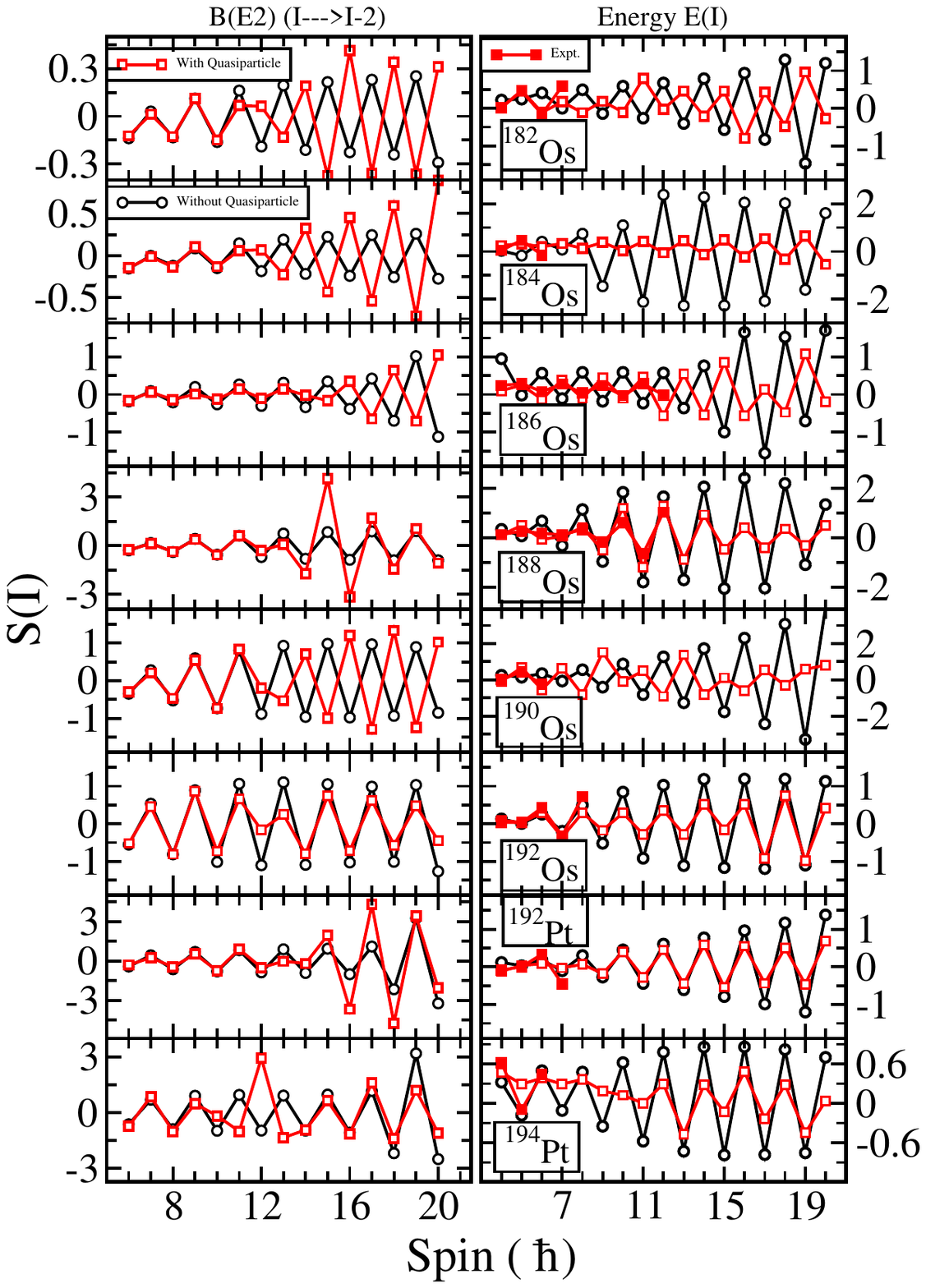}}  \caption{(Color
 online) Staggering parameters $S(I)$ calculated by means of Eq. (\ref{eq:staggering1}) from the energies of the $\gamma$-band (right panels) and
 by the analogue equation  (\ref{eq:SBE22}) from the intra $\gamma$-band $B(E2, I\rightarrow I-2)$ values (left panels).   
   }
 \label{fig:gg2E4}
 \end{figure}

\section{Triaxiality and $\gamma$-softness}\label{sec:ACM}

The concepts of triaxiality and  $\gamma$-softness have been developed in the context 
of the phenomenological Bohr Hamiltonian, see Refs. \cite{BM75}, p. 677 ff. and \cite{RW10}, p. 97 ff. In order to quantify them in terms of the low-energy
observables, we assume that the deformation parameter $\beta$ is fixed, i.e.,  only the triaxiality parameter $\gamma$ and
the orientation angles are considered as dynamical variables. The model is presented as
"Gamma-rotor" in Ref. \cite{caprio11}. The nature of the triaxiality
can be characterized by the $\gamma$-dependence of the potential.

The shape of the nuclear surface is defined by the quadrupole deformation parameters 
\begin{equation}\label{eqn-bohr-q}
\alpha_\mu = \beta \left[ \cos\gamma\, {\cal D}^{(2)}_{0\mu}(\Omega) +
\frac{1}{\sqrt{2}} \sin\gamma \left[ {\cal D}^{(2)}_{2\mu}(\Omega) +
{\cal D}^{(2)}_{-2\mu}(\Omega) \right] \right],
\end{equation}
where $\beta$ is the deformation parameter, $\gamma$ the triaxiality parameter and $\Omega$ are the Euler angles, specifying the orientation of the shape. The Hamiltonian in units of $\hbar^2/B\beta^2$ is given by
 
\begin{eqnarray}\label{eq:ACM}
H_{GR}&=&\hat\Lambda^2+V(\gamma)_{GR}\nonumber\\
&=&\hat\Lambda^2+\chi(1-\cos3\gamma)+\kappa(\cos^2 3\gamma-1),
\end{eqnarray}
where $B$ sets the scale  of the kinetic energy, and

\begin{equation}
\label{eqn-Lambdasqr}
\hat\Lambda^2=-\biggl(
\frac{1}{\sin 3\gamma} 
\frac{\partial}{\partial \gamma} \sin 3\gamma \frac{\partial}{\partial \gamma}
- \frac{1}{4}
\sum_{i=1,2,3} \frac{\hat{L}_i^{\prime2}}{\sin^2(\gamma -\frac{2}{3} \pi i)}.
\biggr).
\end{equation}
with ``$\hat{L}_i^{\prime}$'' being the angular momentum  components with respect to the body-fixed axes. The kinetic energy operator, $\hat\Lambda^2$, is the
Laplacian in five dimensions. It is
the Casimir operator for the five-dimensional rotation group SO(5), 
which contains the rotations in physical space, acting on
the Euler angle coordinates as an SO(3) subgroup. 
The potential energy
$V(\beta,\gamma)$ must be periodic in $\gamma$ with period
of $120^\circ$, and it must be symmetric about $\gamma=0^\circ$ and
$\gamma=60^\circ$.  

The matrix elements of the charge quadrupole moments, which generate the $B(E2)$ $\gamma$-transitions and the static electric quadrupole 
matrix elements, are modeled by a homogeneous charged droplet 
\begin{equation}\label{eq:E2}
Q_\mu=\frac{3ZR_0^2}{4\pi}\left(e\alpha_\mu^*-\frac{10}{\sqrt{70\pi}}e^2[\alpha\times\alpha]^{(2)*}_\mu\right),
\end{equation}
where ``$e$'' fixes the scale.

Fig. \ref{fig:grot} illustrates a selection of potential types, and Fig. \ref{fig:gband} shows
the low-lying bands that belong to these potentials. Tables \ref{tab:gav} and  \ref{tab:gtrans}  list 
quantities that characterize the collective mode. The results for several potentials, not shown in Fig. \ref{fig:grot}, 
are added to better display their parameter dependence.
In Table  \ref{tab:gav}, the position of the minimum (maximum) of the potentials is listed as $\gamma_m$.
The softness of the potentials $\Delta \gamma$ is quantified  as the distance between zero and the turning
point or the two turning points of 
the classical motion for the  Gamma-rotor Hamiltonian (\ref{eq:ACM}) with the angular momentum $L=0$  and energy $E$ equal to 
the quantal energy of the $0^+_1$ state, which is the length of the corresponding bars in Fig.~\ref{fig:grot}. It is a measure  
of the ground-state fluctuation in $\gamma$, which coincides with the oscillator length in the case of the harmonic limit. The next columns
list important energy criteria characterizing the nature of triaxiality, namely the ratios ${\left[\frac{E(2^+_2)}{E(2^+_1)}\right]}$, ${\left[\frac{E(2^+_2)}{E(4^+_1)}\right]}$		 
and  the modified staggering parameter $\bar S(6)$ , which is defined as
\begin{eqnarray}\label{eq:staggering2}
\bar S(I)=\frac{S(I)-S(I+1)}{2}.
\end{eqnarray}
The latter is more appropriate in tables that cite only the value for one $I$, because it oscillates around zero, while $S(I)$ oscillates around some positive value which
represents the curvature of the rotational energy.   

The  lowest rotational sequences in Fig. \ref{fig:gband} have the following general characteristics. Consider the potential $\chi-\kappa=200-0$, which  represents a rigid prolate nucleus with harmonic $\gamma$-vibrational excitations. 
The $\Delta I=2$ ground-band on the $0^+_1$ state represents the zero-phonon sequence.  The one-phonon 
 single $\gamma$-band ($\gamma$) is the $\Delta I=1$ rotational sequence on the $2^+_2$ state.
It represents a traveling wave, which classically corresponds to a triaxial distortion that rotates around the symmetry axis.
The two-phonon double $\gamma$-band ($\gamma\gamma 4$) is the $\Delta I=1$ rotational sequence on the $4^+_3$ state.
It represents a traveling wave generated by adding a second $K=2$ on top of the first with the same angular momentum along the symmetry axis.
The second two-phonon double $\gamma$-band ($\gamma\gamma 0$) is the $\Delta I=2$ rotational sequence on the $0^+_2$ state.  
It is generated by putting the second phonon with opposite angular momentum projection on the symmetry axis 
 on top of the first.  It  represents  a pulsating wave, which   classically corresponds  to an oscillation of the triaxial deformation
 between the prolate and oblate turning points. (For a discussion of the classical correspondence see also Ref. \cite{BM75} p. 656.)
 
 The lowest bands keep their character when the potential deviate from the harmonic limit. Making the axial potential more shallow by
 reducing the strength of the $\chi(1-\cos3\gamma)$ term (see 50-0 and 20-0))
 brings down the $\gamma$-excitations and generates couplings between them, which shift the energies. As discussed in detail in the Appendix, 
 the repulsion between the even-$I$ states of the $\gamma\gamma 0$ and the $\gamma$-bands prevail, which results in the even-$I$-down staggering  pattern
 that signifies $\gamma$-softness.  Adding the $\xi(\cos^2 3\gamma-1)$ term shifts the $\gamma$-band towards the ground-band (see the potentials 50-50 and 50-100 
 in Table \ref{tab:gav}). The repulsion between the even-$I$ states of the ground-band and the $\gamma$-band generates the even-$I$-up 
 staggering pattern that characterizes $\gamma$-rigidity. The competition between the two effects is reflected by the transition from a soft axial to a soft triaxial 
 potential.  As seen in Fig. \ref{fig:gband} and Table \ref{tab:gav},  the down-shift by the $\gamma\gamma 0$-band prevails for the 50-0 potential,
 the up-shift by the ground-band prevails for the  0-50 potential, and both shifts largely compensate each other for the potential  50-50.

 In the Appendix we  discuss in detail how characteristic relations between energies and $E2$ transition rates 
 between the states in the yrast region emerge for the various the potentials, and
  how they can be understood in terms of the couplings between the lowest bands. 
 Such an interpretation is useful because it applies in an analogous way to the TPSM results discussed later. Further details
 can be found in Ref. \cite{caprio11}.

Fig. \ref{fig:SBE2} shows the staggering parameters $S(I)$ calculated by means of Eq. (\ref{eq:staggering1})  from the energies of the $\gamma$-bands of the soft prolate potential 
50-0 and the triaxial potential 0-100. Added are plots of the staggering of  the intra-band $B(E2, I\rightarrow I-2)$ values ($SBE22$) and of intraband
 the $B(E2, I\rightarrow I-1)$ values ($SBE21$)
 \begin{eqnarray}
 SBE22(I)=[B(E2,I_\gamma\rightarrow I_\gamma-2) \nonumber  \\
 -2(B(E2,I_\gamma-1\rightarrow I_\gamma-3) \nonumber  \\
 +(B(E2,I_\gamma-2\rightarrow I_\gamma-4) ]/
B(E2,2^+_1\rightarrow0^+_1),\label{eq:SBE22}\\
 SBE21(I)=[B(E2,I_\gamma\rightarrow I_\gamma-1) \nonumber  \\
 -2(B(E2,I_\gamma-1\rightarrow I_\gamma-2)\nonumber \\
 +(B(E2,I_\gamma-2\rightarrow I_\gamma-3)]/B(E2,2^+_1\rightarrow0^+_1)\label{eq:SBE21}.
 \end{eqnarray}
  As seen the transition probabitlies show a weak staggering with the opposite phase as that of the energies.
 As discussed above, the staggering of the energies can be explained by the mixing of the even-$I$-states of $\gamma$-band with the states of the
 ground- and $\gamma\gamma 0$-bands. The staggering of the $B(E2)$  values can be explained in the same way, in particular its opposite phase. 
 The explanation is somewhat complex as it involves the phases of the mixed states and is given in the Appendix.

The SO(5) Laplacian (\ref{eqn-Lambdasqr}) is invariant under the transformation $\gamma\rightarrow60^\circ-\gamma$ and an exchange of the three principal axes, which leads
to a symmetry quantum number called $\gamma$-parity \cite{RW10}. The potential term ``$\cos 3\gamma$'' is odd under this transformation and it has no diagonal matrix elements in
the absence of a potential. As a consequence, changing from prolate to oblate shape by $\chi\rightarrow-\chi$, will give the same energies. Moreover, the $E2$ transition operator  
transforms in a simple way (see \cite{RW10} p. 222 ff. ) such that all $B(E2,I\rightarrow I')$ values are the same as well, while the static quadrupole moment change their signs.  
For this reason only the potentials with prolate preference are discussed. The  cases with oblate preference are given by this symmetry. 
The sign change of the quadrupole operator implies that for the potentials that are symmetric about $\gamma=30^\circ$ the static   quadrupole moments  
$Q(2^+_2)=-Q(2^+_1)=0$ as well as $B(E2, 4^+_2\rightarrow 2^+_2)=B(E2, 2^+_2\rightarrow 0^+_1)=0$. 

The following classification scheme seems appropriate. The potential is called "prolate" if $\gamma_m=0^\circ$, "triaxial"  if $\gamma_m>0^\circ$ and
"transitional" if its curvature at $\gamma=0^\circ$ is close to zero (50-30 in Fig. \ref{fig:grot}). For $\Delta\gamma\leq 16^\circ$, we
further specify. the potential as "rigid", 
 for $16^\circ \leq \Delta\gamma\leq 25^\circ$ as "soft", and for  $\Delta\gamma\geq 25^\circ$ as "shallow".  
 Of course, the boundaries are not sharp and to some extent arbitrarily chosen.
 The potentials determine the density distributions. Thus it appears appropriate  to classify, loosely 
speaking, the "nuclear shapes" in the same way.

 The signatures of triaxility and $\gamma$-softness from the perspective of a collective model can be summarized as follows:\\
\\
 1) The deviation from axial shape is indicated by the energy of the $2^+_2$ state ($\gamma$-band head) relative to the $4^+_1$ level of the ground-band.
  This relative differences changes from a larger value for narrow axial potential to $E(2^+_2)=E(4^+_1)$ for the $\gamma$-independent
  potential to  $E(2^+_2)<E(4^+_1)$ for a stiff triaxial potential.\\ 
  2) The staggering parameter $S(I)$ is a measure for the $\gamma$-softness.  A large even-$I$-down amplitude indicates a shallow  potential.
  A small even-$I$-down amplitude and $E(2^+_2)>>E(4^+_1)$ indicate a narrow axial potential.  A large odd-$I$-down amplitude
   indicates a narrow triaxial potential.  
  A small amplitude of $S(I)$ and $E(2^+_2)$ near $E(4^+_1)$ indicate a  soft potential transitional  between axial and triaxial.\\
 3) The symmetry of the potential with respect to $\gamma=30^\circ$ is reflected by the static quadrupole moments $Q$ of the $2^+$ states.
 For a potential preferring $\gamma <30^\circ$ one has $Q(2^+_1$)<0 and $Q(2^+_2)=-Q(2^+_1)$. For a potential preferring $\gamma >30^\circ$ one has $Q(2^+_1$>0 
 and $Q(2^+_2)=-Q(2^+_1)$.  For a symmetric potential $Q(2^+_2)=-Q(2^+_1)$=0 and $B(E2, 2^+_1\rightarrow0^+_1)$=0,
  and $B(E2, 2^+_2\rightarrow 2^+_1)$ is large. 
 Deviations from symmetry quickly remove the quenching and reduce the $B(E2, 2^+_2\rightarrow 2^+_1)$  value.\\
 4) For all potentials,  the band built on the $0^+_2$ represents  an oscillation of $\gamma$ around the potential minimum at $\gamma_m$ 
 with two-phonon nature $\gamma\gamma0$.
 The transition probability $B(E2, 0^+_2\rightarrow2^+_2)$ to the one-phonon state is large when the pattern of 
 staggering parameter $S(I)$ is strongly even-$I$-down.

\section{Triaxial projected shell model calculations} \label{sec:TPSM}


The TPSM approach employs the methodology similar to that of the standard spherical shell model, except that deformed angular momentum projected basis
are used to diagonalize the shell model Hamiltonian \cite{JS16,Hara1995}. The Hamiltonian in the TPSM approach contains the standard pairing plus
quadrupole-quadrupole interaction terms, and is given by
\begin{equation}
\hat H = \hat H_0 - {\chi \over 2}  \sum_\mu \hat Q^\dagger_\mu
\hat Q^{}_\mu - G_M \hat P^\dagger \hat P - G_Q \sum_\mu \hat
P^\dagger_\mu\hat P^{}_\mu .
\label{hamham}
\end{equation}
The terms in the above equation (\ref{hamham}) represent the modified harmonic oscillator single-particle Hamiltonian \cite{Ni69}, monopole pairing, quadrupole pairing, and quadrupole-quadrupole interaction, respectively. The value of $\chi$, the $QQ$-force strength, is fixed such that the quadrupole mean-field is
obtained through Hartree-Fock-Bogoliubov self-consistency condition \cite{Hara1995} :
\begin{eqnarray}\label{chi}
  \chi_{\tau\tau^\prime} = {{2\over3}\epsilon \hbar\omega_\tau\hbar\omega_{\tau^\prime}\over\hbar\omega_n\langle \hat{Q}_0\rangle_n+\hbar\omega_p\langle \hat{Q}_0\rangle_p}
\end{eqnarray}
where $\omega_\tau=\omega_0 a_\tau$, with $\hbar\omega_0=41.4678A^{-{1\over3}}$MeV, and
\begin{eqnarray}\label{iso}
 a_\tau=\left[1\pm{{N-Z}\over A}\right]^{{1\over 3}}
\end{eqnarray}
controls  isospin-dependence  with $+ (-)$ for neutrons (protons).

For the monopole pairing strength, the coupling constant $G_M$ is of the standard form
\begin{eqnarray}\label{pairing}
G_M = \left(G_1 \mp G_2{{N-Z}\over A}\right){1\over A}
\end{eqnarray}
with $- (+)$ referring to neutrons (protons). The values of $G_1$ and $G_2$ for a particular mass region are adjusted in such a way that the calculated 
gap parameters reproduce the experimental odd-even mass differences. The coupling constant $G_Q$ of quadrupole pairing is  $0.18 G_M$. 
The pairing parameters are taken from our previous work \cite{SJ21} and are listed in Table 1 of this reference.

The deformed shell model basis is constructed from BCS quasiparticle configurations generated from the eigenstates  of the triaxial 
Nilsson potential with the axial deformation parameter
$\varepsilon$ and the non-axial parameter $\varepsilon '$ and the BCS gaps determined by the selfconsistency conditions
$\Delta_M=G_M\langle P\rangle$ and $\Delta_\mu=G_Q\langle P_\mu\rangle$.
The quasiparticle configurations 
 are projected onto good angular momentum states
using the three-dimensional projection operator \cite{RS80}. The diagonalization of the shell model Hamiltonian (\ref{hamham}) 
is then performed using the angular momentum projected 
multi-quasiparticle basis states. The multi-quasiparticle basis space used in the present work are given by $:$
\begin{eqnarray}\label{basis}
&&\hat P^I_{MK}\ack\Phi\rangle;\nonumber\\
&&\hat P^I_{MK}~a^\dagger_{p_1} a^\dagger_{p_2} \ack\Phi\rangle;\nonumber\\
&&\hat P^I_{MK}~a^\dagger_{n_1} a^\dagger_{n_2} \ack\Phi\rangle;\\
&&\hat P^I_{MK}~a^\dagger_{p_1} a^\dagger_{p_2}a^\dagger_{n_1} a^\dagger_{n_2} \ack\Phi\rangle,\nonumber
\end{eqnarray}
where the subscript $n$ denotes the quasineutron and $p$ the quasiproton states, $\ack\Phi\rangle$ represents the triaxially-deformed quasiparticle  vacuum state, and $P^I_{MK}$ is the three-dimensional angular momentum projection operator in its standard form  \cite{RS80,HS79,HS80}
\begin{equation}
\hat P^I_{MK} = {2I+1 \over 8\pi^2} \int~d\Omega\,
D^{I}_{MK}(\Omega)\, \hat R(\Omega).
\label{PD}
\end{equation}
Here
\begin{eqnarray}
\hat R(\Omega) = e^{-i \alpha \hat J_z} e^{-i \beta \hat J_y}e^{-i \gamma \hat J_z}~~~.\label{rotop}
\end{eqnarray}
is the rotation operator, $D^{I}_{MK}(\Omega)$ is the Wigner $D$-function \cite{BM75}, $\Omega$ represents the set of Euler angles 
($\alpha, \gamma = [0,2\pi],\, \beta= [0, \pi]$), and
$\hat{J}_{x,y,z}$ are the angular momentum operators.

The Hamiltonian in Eq. (\ref{hamham}) is diagonalized using the non-orthogonal  
basis of Eq. (\ref{basis}), which leads to the generalized eigenvalue problem
\begin{eqnarray}
\sum_{K',\kappa'}( \bra \Phi_{\kappa}|&\hat H& \hat P^{I}_{KK'}|\Phi_{\kappa'}\ket
\\\nonumber&-&E_{l\sigma} \bra \Phi_{\kappa}|\hat P^{I}_{KK'}|\Phi_{\kappa^{\prime}}\ket)f^{\sigma I}_{K'\kappa'} = 0
\label{HW}
\end{eqnarray}
with the projected wavefunction 
\begin{equation}
\ack\psi^{\sigma I}\rangle = \sum_{K,\kappa}~f^{\sigma I}_{K\kappa}~\hat P^{I}_{MK}| 
~ \Phi_{\kappa} \ket.
\label{17}
\end{equation}
The symbol $\kappa$ represents quasiparticle configurations of the basis states  Eq.~(\ref{basis}). 
A new set of components is introduced \cite{WANG2020}
 \begin{equation}\label{eq:GKk}
  g^{\sigma I}_{K\kappa}=\sum_{K',\kappa'}\bra \Phi_{\kappa}|\hat P^{I}_{KK'}|\Phi_{\kappa'}\ket^{1/2}f^{\sigma I}_{K'\kappa'},
 \end{equation}
 which are orthonormal, and $\vert  g^I_{K,\kappa}\vert^2$ is the probability of a projected quasiparticle configuration in
 the wave function (\ref{17}).
 
 The electromagnetic transition probabilities are obtained using the expression $:$
\begin{eqnarray}\label{BE2}
  B(E2, I_i\rightarrow I_f)={e^2 \over {2I_i+1}}\ack \langle \psi^{\sigma_f I_f}\ack\ack \hat  Q_2\ack\ack\psi^{\sigma_i I_i}\rangle \ack^2
\end{eqnarray}
from an initial state $\psi^{\sigma_i I_i}$ to a final state $\psi^{\sigma_f I_f}$. Effective charges of 1.5e (0.5e) for protons (neutrons) similar to our previous publications \cite{GH14,bh15,SJ18} are used in our calculations.

For an irreducible spherical tensor, $\hat  Q$, of rank $L$, the reduced matrix element can be expressed as
\begin{eqnarray}\label{tensor}
  &&\langle \psi^{\sigma_f I_f}\ack\ack \hat  Q_L\ack\ack\psi^{\sigma_i I_i}\rangle\nonumber\\
&& ~~~~~~~~=\sum_{\kappa_i,\kappa_f,K_i,K_f}f_{\kappa_iK_i}^{\sigma_i I_i}f_{\kappa_fK_f}^{\sigma_f I_f}\sum_{M_i,M_f,M}(-)^{I_f-M_f}\nonumber\\
&&~~~~~~~~~~~~~~~~~~~~~~~~\times\begin{pmatrix}I_f&L&I_i\\-M_f&M&M_i\end{pmatrix}\nonumber\\
&&~~~~~~~~~~~~~~~~\times\langle \Phi_{\kappa_f}\ack \hat P_{K_fM_f}^{I_f}\hat  Q_{LM}\hat P_{K_iM_i}^{I_i}\ack\Phi_{\kappa_i}\rangle\\
  &&~~~~~~~~ =2\sum_{\kappa_i,\kappa_f,K_i,K_f}f_{\kappa_iK_i}^{\sigma_i I_i}f_{\kappa_fK_f}^{\sigma_f I_f}\sum_{M^\prime,M^{\prime\prime}}(-)^{I_f-K_f}(2I_f+1)^{-1}\nonumber\\
&&~~~~~~~~~~~~~~~~ \times\begin{pmatrix}I_f&L&I_i\\-K_f&M^\prime&M^{\prime\prime}\end{pmatrix}\int d\Omega D_{M^{\prime\prime}K_i}^{I_i}(\Omega)\nonumber\\
&&~~~~~~~~~~~~~~~~~~~~~~~~\times\langle \Phi_{\kappa_f}\ack \hat  Q_{LM^\prime}\hat R(\Omega)\ack\Phi_{\kappa_i}\rangle.\nonumber
\end{eqnarray}
The symbol (   ) in the above expression represents $3j$-coefficient.

It is important to point out that the triaxiality parameter $\gamma_N=\arctan(\varepsilon '/\varepsilon )$ merely controls the Nilsson Hamiltonian, which generates  the deformed 
single particle orbits. The "geometric" triaxiality parameter $\gamma$ of the nuclear charge and mass distributions calculated from the TPSM wavefunctions 
differs from $\gamma_N$. The authors of Ref. \cite{Shimizu08}
gave an approximate estimate based on the volume conserving deformed oscillator, which is the basis of the Nilsson Hamiltonian, 
\begin{align}
\omega_i=\omega_0\left[1-\frac{2}{3}\varepsilon \cos\left(\gamma_N+i\frac{2\pi}{3}\right)\right], \\
\tan(\gamma)=\frac{\sqrt{3}(\omega_2^{-2}-\omega_1^{-2})}{2\omega_3^{-2}-\omega_1^{-2}-\omega_2^{-2}},~~~\gamma\approx\left(1-\frac{3}{2}\varepsilon\right)\gamma_N, 
\label{eq;geo}
\end{align}
where the approximation hold for small $\gamma$. The triaxiality parameter $\gamma$, which concerns the collective Bohr Hamiltonian, is substantially smaller than $\gamma_N$.
For example, the TPSM parameters for $^{166}$Er are $\varepsilon$=0.325, $\varepsilon '$=0.120, that is  $\gamma_N=21^\circ$, which  corresponds to $\gamma=13^\circ$.

\begin{figure}[t]
 \centerline{\includegraphics[trim=0cm 0cm 0cm
0cm,width=0.5\textwidth,clip]{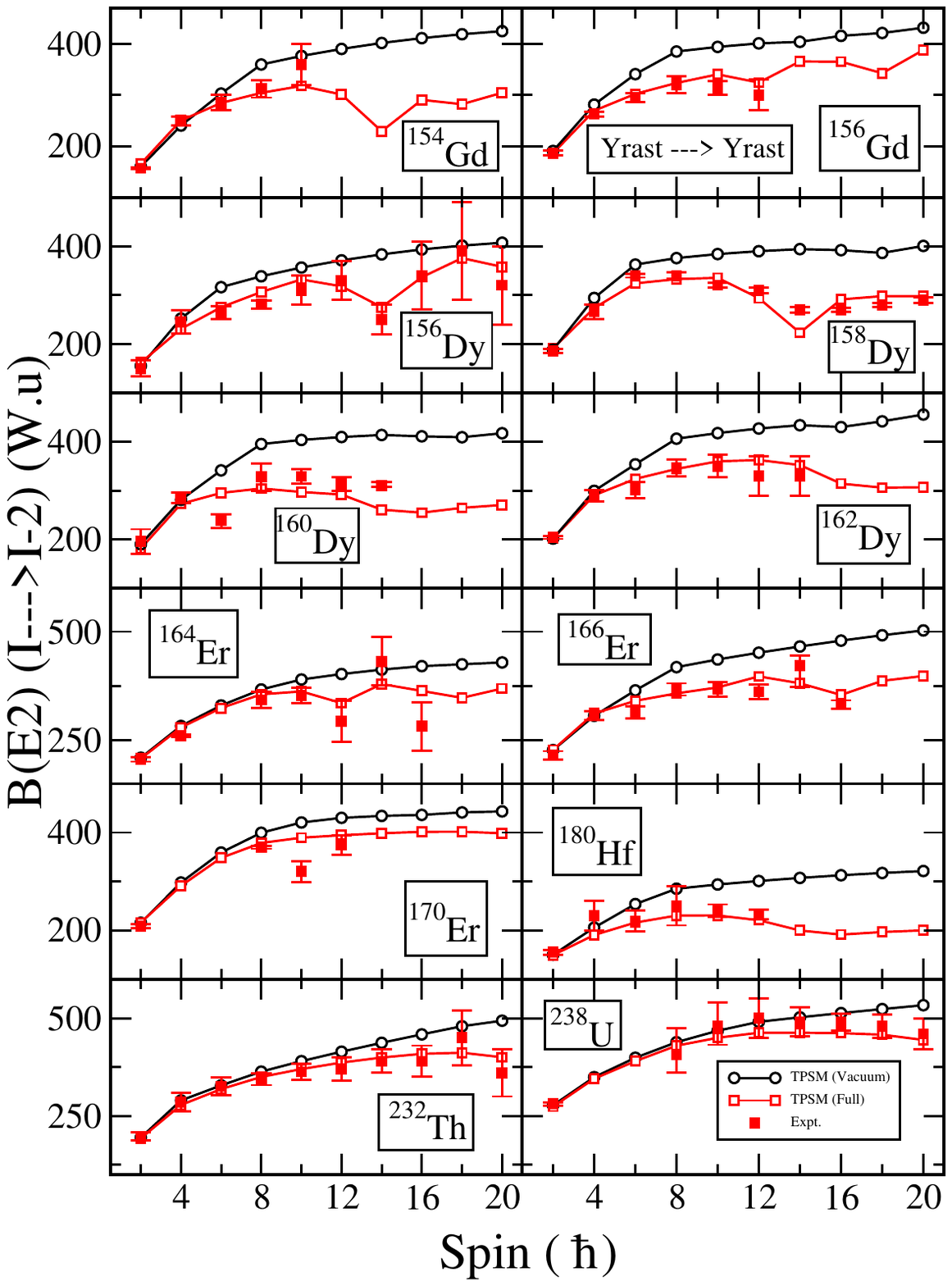}} \caption{(Color
online) $B(E2)$ transition probabilities (W.u) from yrast-band to the yrast-band for $^{154,156}$Gd, $^{156,158,160,162}$Dy, $^{164,166,170}$Er, $^{180}$Hf, $^{232}$Th and
$^{238}$U isotopes. The black  curves show the TPSM values for the vacuum configuration denoted by TPSM (Vacuum).
The red  curves show the TPSM values including the quasiparticle configuration denoted by TPSM (Full).
 The bold red squares show the experimental values (Data taken from \cite{Reich2009,Reich2012,Nica2017,Nica2021,Reich2007,Singh2018,Baglin2008,Baglin2018,Wu2003,Browne2006,Martin2002}).
  }
\label{fig:yyA}
\end{figure}

\begin{figure}[t]
 \centerline{\includegraphics[trim=0cm 0cm 0cm
 0cm,width=0.5\textwidth,clip]{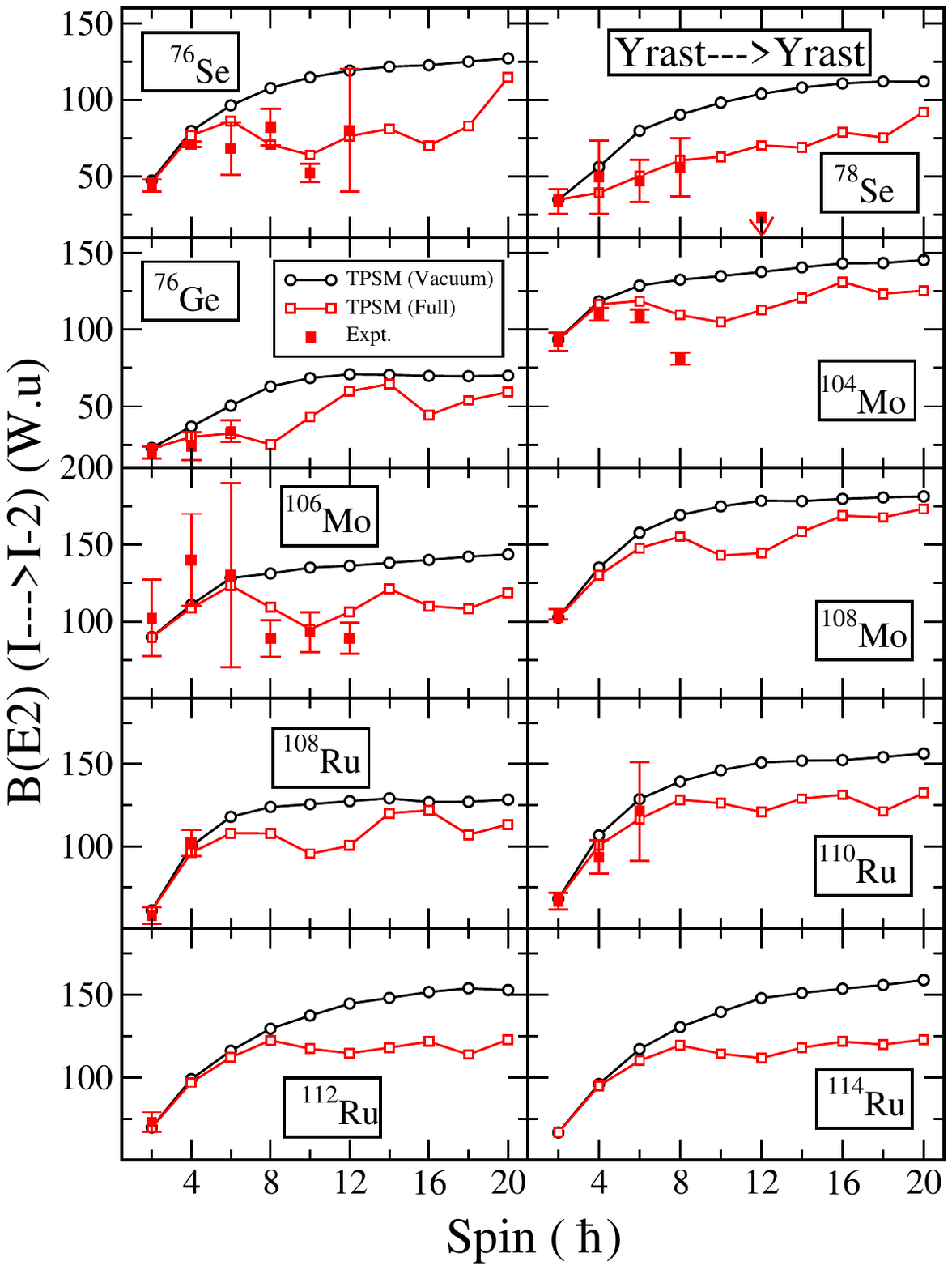}} \caption{(Color
   online) $B(E2)$ transition probabilities (W.u) from yrast-band to the yrast-band for $^{76,78}$Se, $^{76}$Ge, $^{104,106,108}$Mo and $^{108,110,112,114}$Ru isotopes.
   The black  curves show the TPSM values for the vacuum configuration denoted by TPSM (Vacuum).
The red  curves show the TPSM values including the quasiparticle configuration denoted by TPSM (Full).
The bold red squares show the experimental values (Data taken from \cite{Singh1984,Farhan2009,Blach2007,Fren2008,Blach2000,Gurdal2012,Lalkov2015}).
     }
  \label{fig:yyB}
  \end{figure}

\begin{figure}[htb]
 \centerline{\includegraphics[trim=0cm 0cm 0cm
 0cm,width=0.5\textwidth,clip]{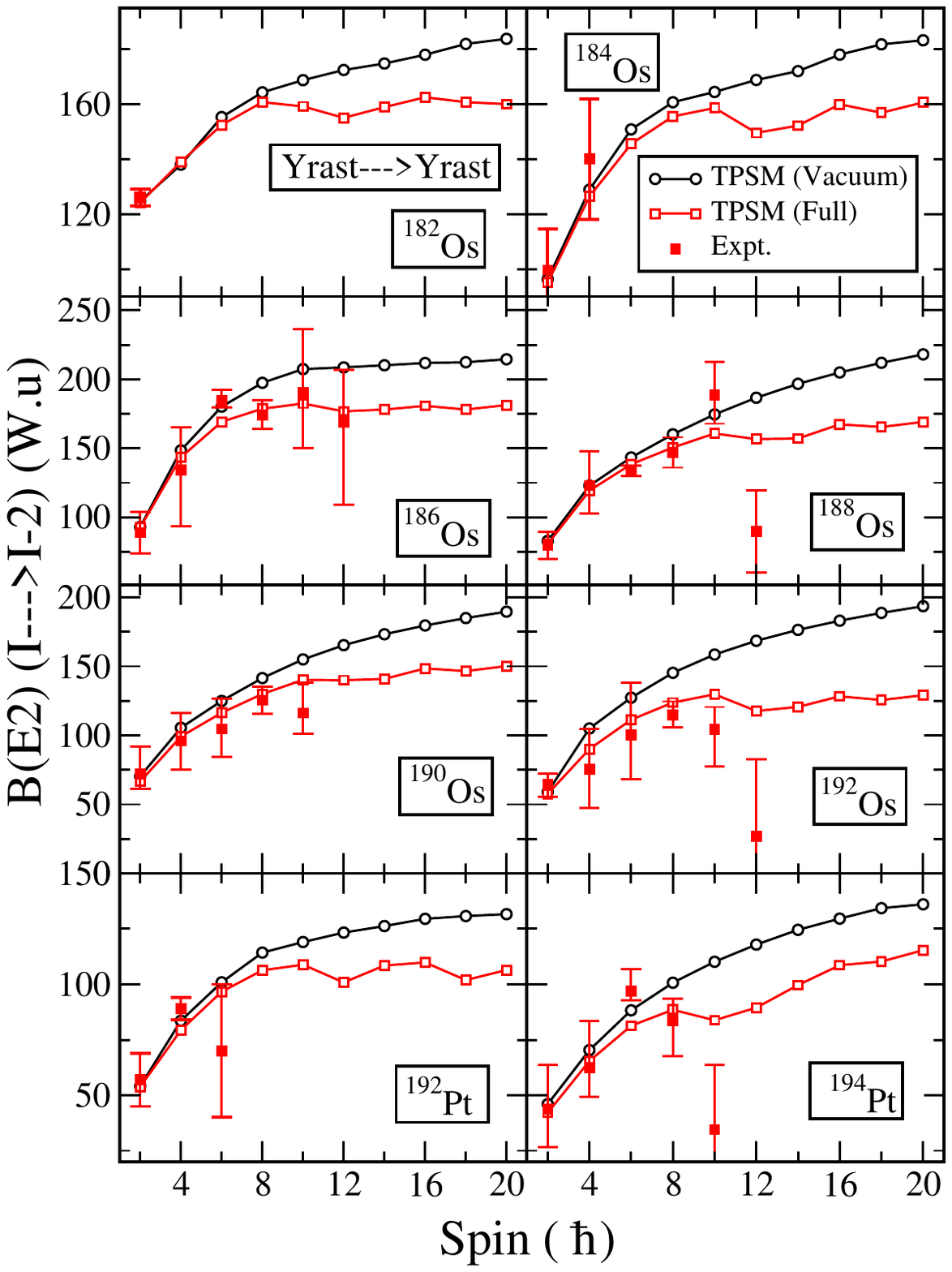}} \caption{(Color
   online) $B(E2)$ transition probabilities (W.u) from yrast-band to the yrast-band for  $^{182-192}$Os and $^{192,194}$Pt isotopes. 
   The black  curves show the TPSM values for the vacuum configuration denoted by TPSM (Vacuum).
The red  curves show the TPSM values including the quasiparticle configuration denoted by TPSM (Full). 
The bold red squares show the experimental values (Data taken from \cite{Singh2015,Baglin2010,Baglin2003,Singh2002,Singh190,Baglin2012,Chen194}).
     }
  \label{fig:yyC}
  \end{figure}

\begin{figure}[htb]
  \centerline{\includegraphics[trim=0cm 0cm 0cm
 0cm,width=0.5\textwidth,clip]{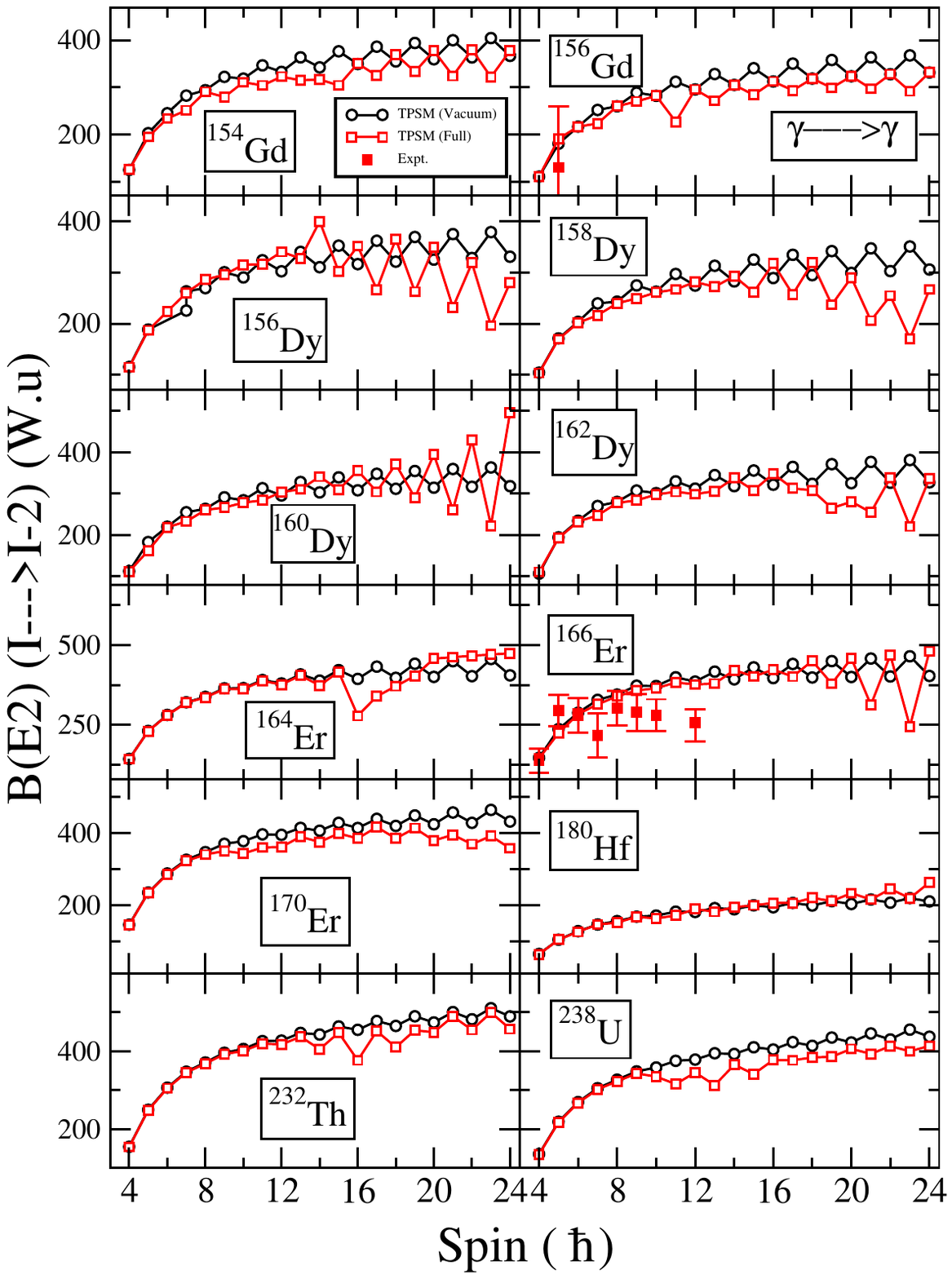}} \caption{(Color
 online) $B(E2)$ transition probabilities (W.u) from $\gamma$-band to the $\gamma$-band for $^{154,156}$Gd, $^{156,158,160,162}$Dy, $^{164,166,170}$Er, $^{180}$Hf, $^{232}$Th and
$^{238}$U isotopes. The black  curves show the TPSM values for the vacuum configuration denoted by TPSM (Vacuum).
The red  curves show the TPSM values including the quasiparticle configuration denoted by TPSM (Full). The bold red squares show the experimental values (Data taken from \cite{Reich2009,Reich2012,Nica2017,Nica2021,Reich2007,Singh2018,Baglin2008,Baglin2018,Wu2003,Browne2006,Martin2002}).
   }
 \label{fig:gg2A}
 \end{figure}

\begin{figure}[htb]
  \centerline{\includegraphics[trim=0cm 0cm 0cm
 0cm,width=0.5\textwidth,clip]{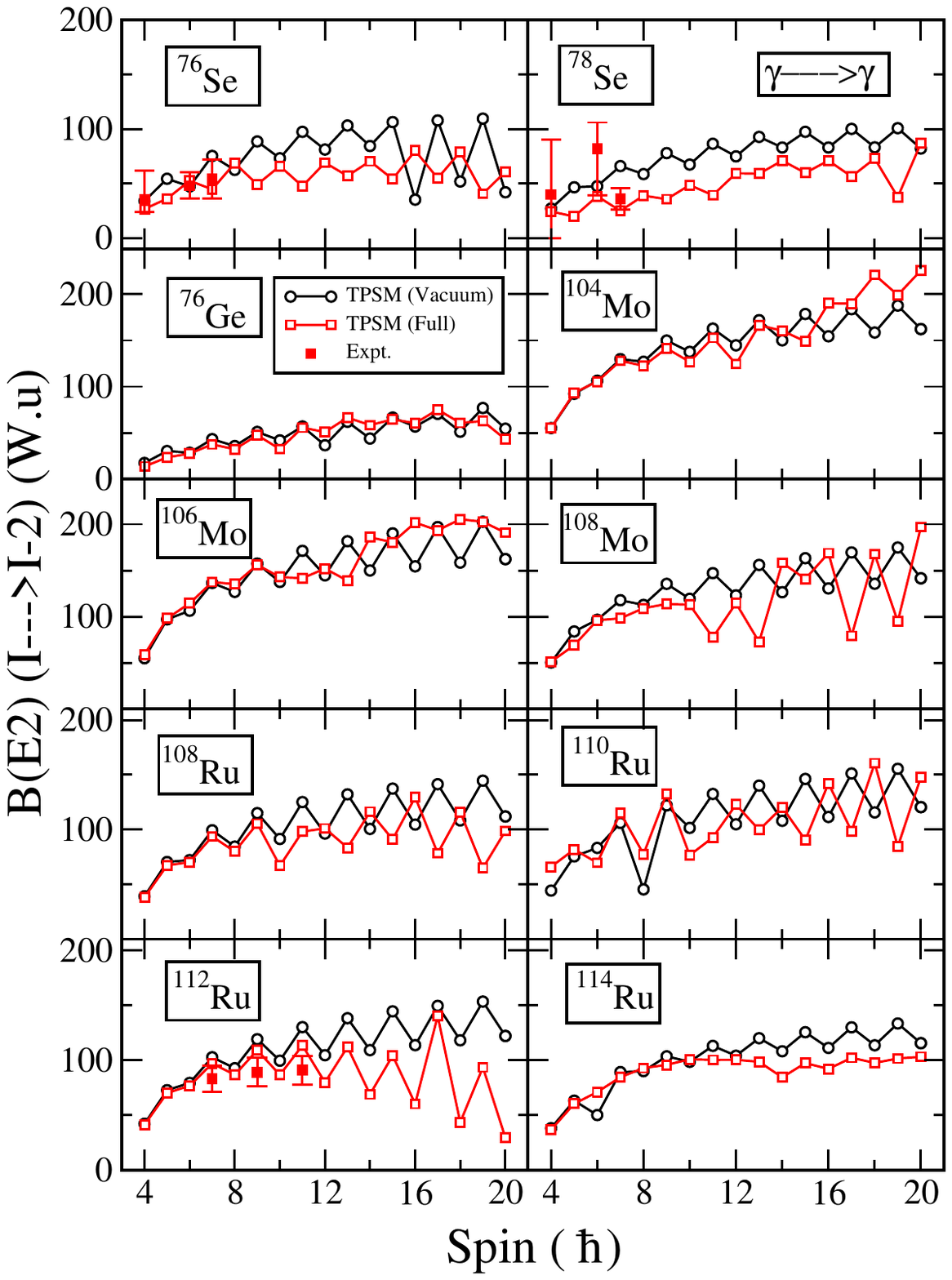}} \caption{(Color
 online) $B(E2)$ transition probabilities (W.u) from $\gamma$-band to the $\gamma$-band for $^{76,78}$Se, $^{76}$Ge, $^{104,106,108}$Mo and $^{108,110,112,114}$Ru isotopes. 
 The black  curves show the TPSM values for the vacuum configuration denoted by TPSM (Vacuum).
The red  curves show the TPSM values including the quasiparticle configuration denoted by TPSM (Full).
The bold red squares show the experimental values (Data taken from \cite{Singh1984,Farhan2009,Blach2007,Fren2008,Blach2000,Gurdal2012,Lalkov2015}).
   }
 \label{fig:gg2B}
 \end{figure}

\begin{figure}[htb]
  \centerline{\includegraphics[trim=0cm 0cm 0cm
 0cm,width=0.5\textwidth,clip]{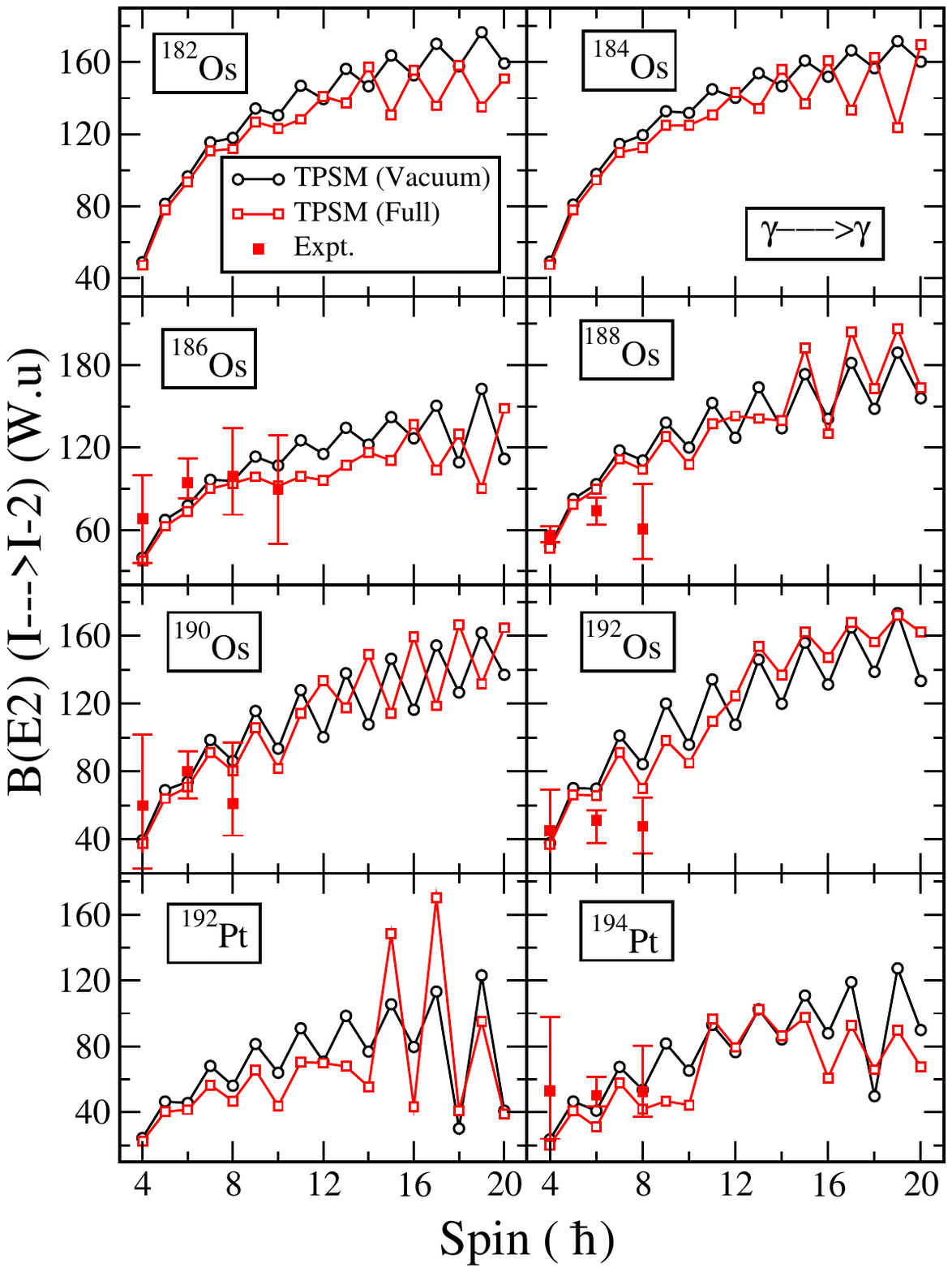}} \caption{(Color
 online) $B(E2)$ transition probabilities (W.u) from $\gamma$-band to the $\gamma$-band for $^{182-192}$Os and $^{192,194}$Pt isotopes. The black  curves show the TPSM values for the vacuum configuration denoted by TPSM (Vacuum).
The red  curves show the TPSM values including the quasiparticle configuration denoted by TPSM (Full).The bold red squares show the experimental values (Data taken from \cite{Singh2015,Baglin2010,Baglin2003,Singh2002,Singh190,Baglin2012,Chen194}).
   }
 \label{fig:gg2C}
 \end{figure}

\begin{figure}[t]
 \centerline{\includegraphics[trim=0cm 0cm 0cm
 0cm,width=0.5\textwidth,clip]{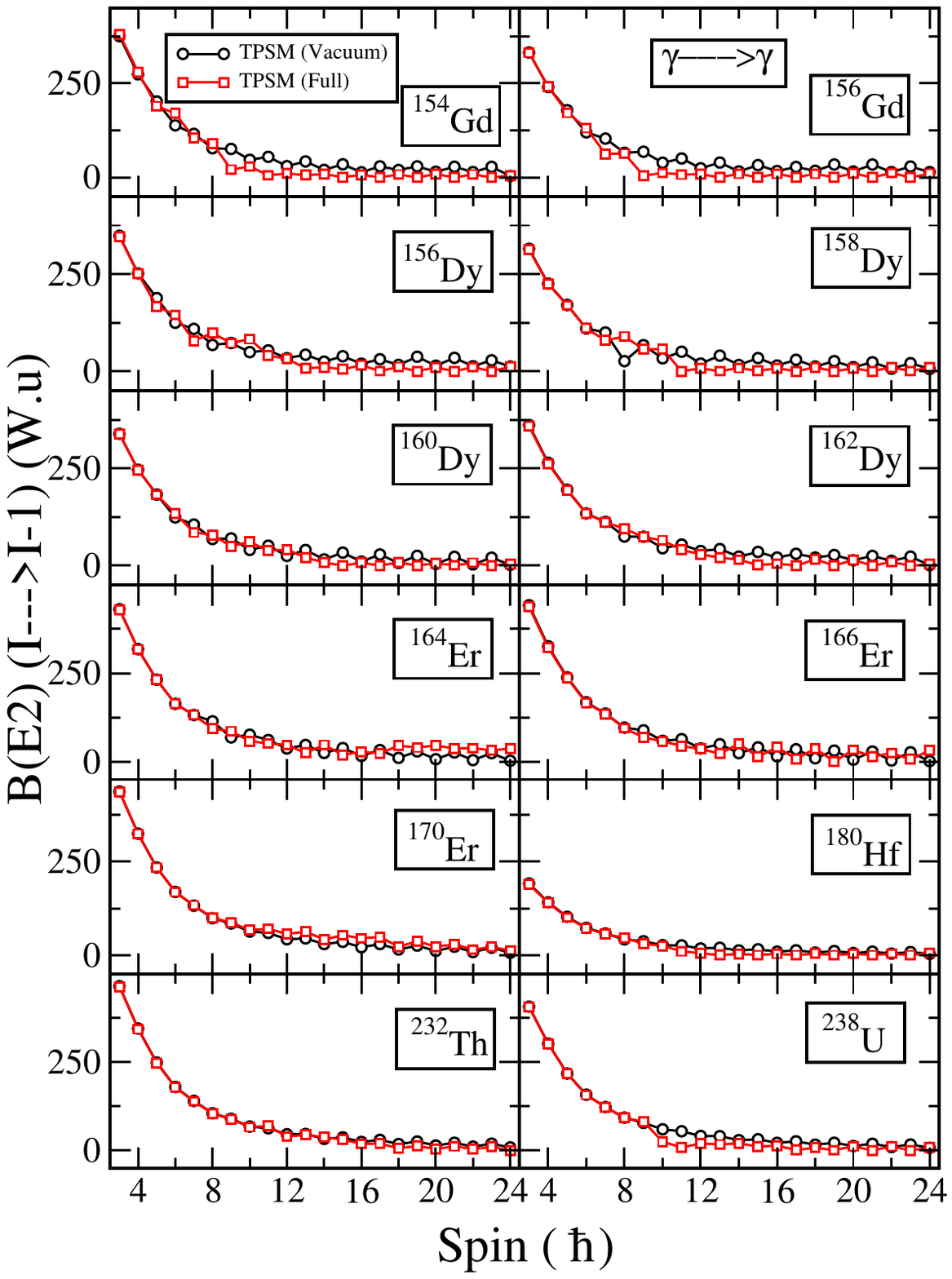}} \caption{(Color
 online) $B(E2)$ transition probabilities (W.u) from $\gamma$-band to the $\gamma$-band for $^{154,156}$Gd, $^{156,158,160,162}$Dy, $^{164,166,170}$Er, $^{180}$Hf, $^{232}$Th and
$^{238}$U isotopes. The black  curves show the TPSM values for the vacuum configuration denoted by TPSM (Vacuum).
The red  curves show the TPSM values including the quasiparticle configuration denoted by TPSM (Full).
   }
 \label{fig:gg1A}
 \end{figure}

\begin{figure}[htb]
 \centerline{\includegraphics[trim=0cm 0cm 0cm
0cm,width=0.5\textwidth,clip]{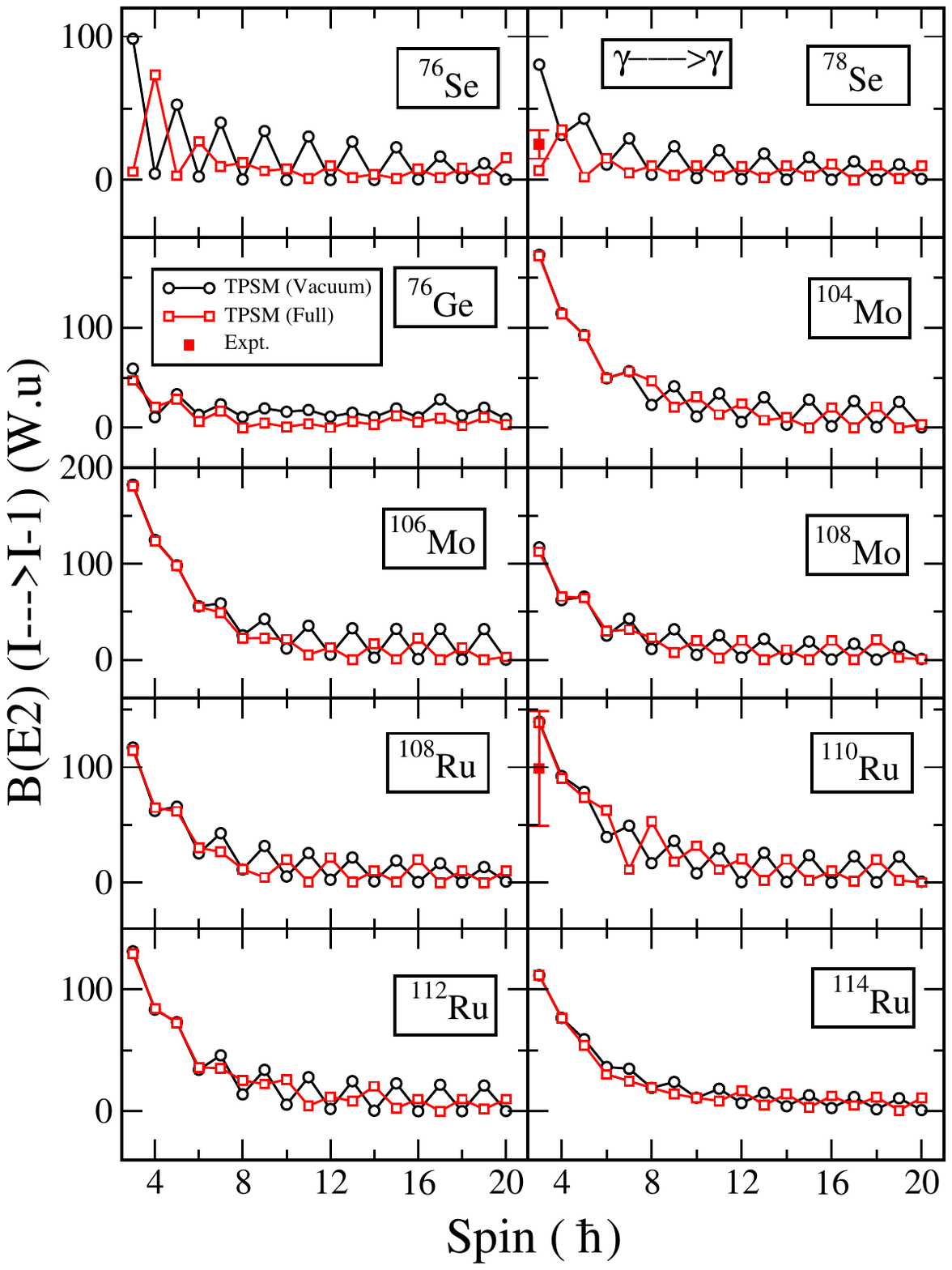}} \caption{(Color
online) $B(E2)$ transition probabilities (W.u) from $\gamma$-band to the $\gamma$-band for $^{76,78}$Se, $^{76}$Ge, $^{104,106,108}$Mo and $^{108,110,112,114}$Ru isotopes. The black  curves show the TPSM values for the vacuum configuration denoted by TPSM (Vacuum).
The red  curves show the TPSM values including the quasiparticle configuration denoted by TPSM (Full). The bold red squares show the experimental values (Data taken from \cite{Singh1984,Farhan2009,Blach2007,Fren2008,Blach2000,Gurdal2012,Lalkov2015}).
  }
\label{fig:gg1B}
\end{figure}

\begin{figure}[htb]
 \centerline{\includegraphics[trim=0cm 0cm 0cm
0cm,width=0.5\textwidth,clip]{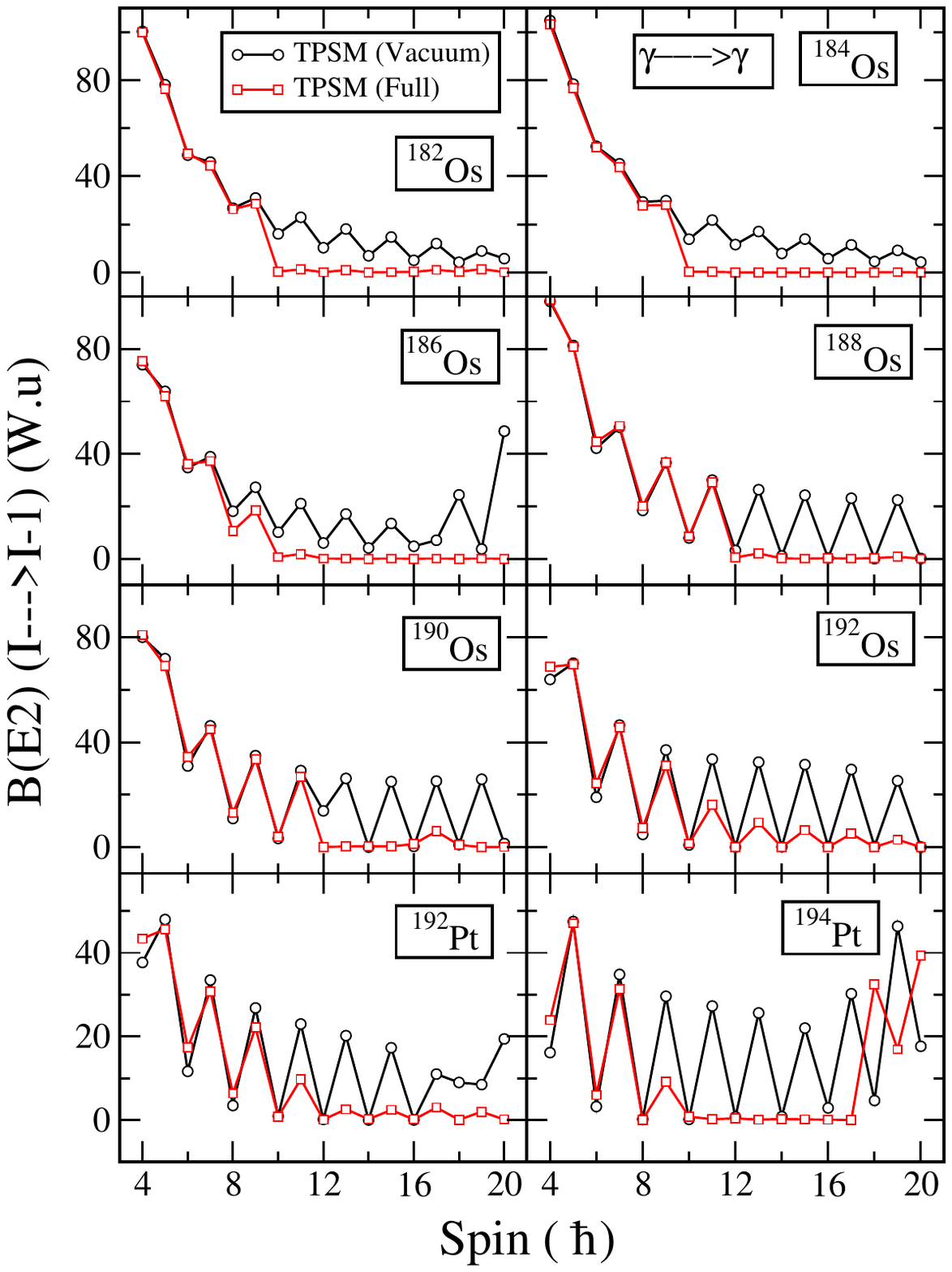}} \caption{(Color
online) $B(E2)$ transition probabilities (W.u) from $\gamma$-band to the $\gamma$-band for $^{182-192}$Os and $^{192,194}$Pt  isotopes. The black  curves show the TPSM values for the vacuum configuration denoted by TPSM (Vacuum).
The red  curves show the TPSM values including the quasiparticle configuration denoted by TPSM (Full). 
  }
\label{fig:gg1C}
\end{figure}

The detailed results of energies and quadrupole transitions of thirty nuclei studied in the present work are presented in Figs.~\ref{fig:energy1}~-~\ref{fig:gy0C}  and Tables \ref{tab:parameters}~-~\ref{tab:02}. 




  
\begin{table*}[htp!]
\LTcapwidth=0.4\textwidth
\caption{\label{tab:parameters}Axial and triaxial quadrupole deformation parameters
  $\epsilon$ and $\epsilon'$  employed in the TPSM calculation. Axial deformations are taken from our earlier
  works \cite{GH14,GH08,bh15,GS12,SJ21} and \cite{raman},
and nonaxial deformations are chosen in such a way that heads
of the $\gamma$-bands are reproduced. In this table, we also provide the energy $E(2^+_1)$, the ratios, ${\frac{E(2^+_2)} {E(2^+_1)}}$ and ${\frac{E(2^+_2)} {E(4^+_1)}}$, $\bar{S}(I=6)=\frac{S(I=6)-S(I=7)}{2}$ and also $B(E2,2_1^+\rightarrow 0_1^+)$ and $B(E2,2_2^+\rightarrow 0_1^+)$  (in W.u.) with both experimental (error bars in curly brackets, data taken from \cite{raman,Singh1984,Farhan2009,Blach2007,Fren2008,Blach2000,Gurdal2012,Lalkov2015,Reich2009,Reich2012,Nica2017,Nica2021,Reich2007,Singh2018,Baglin2008,Baglin2018,Wu2003,Singh2015,Baglin2010,Baglin2003,Singh2002,Singh190,Baglin2012,Chen194,Browne2006,Martin2002}) and TPSM predicted
values.}
\resizebox{1.0\textwidth}{!}
  {
\begin{tabular}{|c|c|c|c|c|c|c|c|c|c|}
  \hline
 Isotope  	&$\epsilon$	&$\epsilon'$	&$\gamma_N$	&$E(2_1^+)$	&${\left[\frac{E(2^+_2)}{E(2^+_1)}\right]}$	&${\left[\frac{E(2^+_2)}{E(4^+_1)}\right]}$	&$\bar{S}(I=6)$	&$B(E2)(2_1^+\rightarrow 0_1^+)$	&$B(E2)(2_2^+\rightarrow 0_1^+)$	\\
 	&	&	&	&	&TPSM (Expt.)	&TPSM (Expt.)	&TPSM (Expt.)	&TPSM (Expt.)	&TPSM (Expt.)	\\
 \hline
$^{76}$Ge	&0.200	&0.160	&38.6	&0.558 (0.563)	&2.1 (1.9)	&0.89 (0.78)	&0.194 (0.124)	&22.42 (29 \{1\})	&4.08 (0.9 \{22\})	\\
$^{76}$Se	&0.260	&0.155	&30.8	&0.551 (0.559)	&2.2 (2.2)	&0.89 (0.91)	&-0.107 (0.487)	&46.08 (44 \{1\})	&0.98 (1.3 \{1\})	\\
 $^{78}$Se	&0.256	&0.150	&30.2	&0.595 (0.613)	&2.2 (2.1)	&0.87 (0.87)	&-0.384 (-0.242)	&34.65 (33.5 \{8\})	&1.85 (0.76 \{6\})	\\
$^{104}$Mo	&0.320	&0.130	&22.1	&0.154 (0.192)	&5.3 (4.2)	&1.64 (1.45)	&-0.328 (-0.184)	&93.61 (92 \{6\})	&19.14	\\
$^{106}$Mo	&0.310	&0.110	&19.5	&0.135 (0.171)	&5.1 (4.1)	&1.55 (1.36)	&-0.285 (-0.091)	&89.79 (102.3 \{25\})	&15.94	\\
 $^{108}$Mo	&0.294	&0.140	&25.4	&0.168 (0.193)	&3.6 (3.0)	&1.16 (1.04)	&-0.306 (-0.031)	&102.48 (104.74 \{10\})
 &16.5	\\
$^{108}$Ru 	&0.280	&0.150	&28.2	&0.202 (0.242)	&3.5 (2.9)	&1.14 (1.07)	&-0.065 (-0.322)	&61.12 (58.0 \{5\})	&21.62	\\
$^{110}$Ru 	&0.290	&0.150	&27.3	&0.188 (0.240)	&3.4 (2.5)	&1.10 (0.92)	&0.189 (-0.011)	&68.11 (66.0 \{5\})	&17.64 	\\
$^{112}$Ru	&0.289 	&0.130	&24.2	&0.161 (0.236)	&3.2 (2.2)	&0.98 (0.81)	&0.854 (0.314)	&69.73 (70.0 \{7\})	&11.89	\\
$^{114}$Ru	&0.250	&0.080	&17.7	&0.222 (0.265)	&2.4 (2.1)	&0.82 (0.79)	&0.909 (0.357)	&66.73	&7.01	\\
$^{154}$Gd	&0.300	&0.100	&18.4	&0.095 (0.123)	&10.6 (8.1)	&3.18 (2.68)	&-0.013 (-0.013)	&165.77 (157.0 \{1\})	&8.4 (5.7 \{5\})	\\
$^{156}$Gd 	&0.341	&0.100	&16.3	&0.092 (0.089)	&12.1 (12.9)	&3.66 (4.00)	&-0.023 (-0.04)	&186.48 (189.0 \{3\})	&9.29 (4.68 \{16\})	\\
$^{156}$Dy 	&0.278	&0.105	&20.7	&0.093 (0.138)	&10.05 (6.5)	&3.06 (2.20)	&0.003 (0.004)	&161.93 (150.0 \{17\})	&9.51	\\
$^{158}$Dy	&0.260	&0.110	&22.9	&0.079 (0.098)	&12.01 (9.6)	&3.63 (2.98)	&-0.012 (0.007)	&190.66 (186.0 \{4\})	&7.58	\\
$^{160}$Dy	&0.270	&0.110	&22.1	&0.074 (0.087)	&12.9 (11.1)	&3.92 (3.40)	&0.003 (-0.008)	&181.55 (195.8 \{25\})	&7.34 (4.46 \{+33-29\})	\\
$^{162}$Dy	&0.280	&0.120	&23.2	&0.077 (0.081)	&11.0 (10.9)	&3.33 (3.34)	&-0.004 (-0.004)	&201.76 (204.0 \{3\})	&7.64 (4.6 \{3\})	\\
$^{164}$Er	&0.317	&0.120	&20.7	&0.087 (0.091)	&9.9 (9.4)	&3.01 (2.87)	&-0.029 (-0.019)	&208.03 (206 \{5\})	&9.51 (5.3 \{6\})	\\
$^{166}$Er	&0.325	&0.126	&21.2	&0.066 (0.081)	&11.1 (9.7)	&3.35 (2.96)	&0.055 (0.007)	&228.13 (217.0 \{5\})	&6.79 (5.17 \{21\})	\\
$^{170}$Er	&0.319	&0.110	&19.0	&0.069 (0.078)	&14.3 (11.8)	&4.31 (3.59)	&0.074 (0.386)	&215.96 (208.0 \{4\})	&8.03 (3.68 \{11\})	\\
$^{180}$Hf	&0.195	&0.090	&24.7	&0.088 (0.093)	&12.6 (12.9)	&3.81 (3.89)	&-0.098 (-0.118)	&149.16 (155.0 \{5\})	&9.49	\\
$^{182}$Os	&0.235	&0.135	&30.0	&0.111 (0.127)	&7.8 (7.0)	&2.45 (2.22)	&-0.142 (-0.373)	&124.09 (126 \{3\})	&16.28	\\
$^{184}$Os	&0.208	&0.108	&27.5	&0.102 (119)	&9.2 (7.7)	&2.81 (2..41)	&-0.074	&95.01 (99.6 \{15\})	&15.17	\\
$^{186}$Os	&0.200	&0.118	&30.7	&0.136 (0.137)	&5.5 (5.6)	&1.81 (1.77)	&-0.258 (-0.101)	&92.67 (92.3 \{23\})	&16.24 (10.1 \{4\})	\\
$^{188}$Os	&0.183	&0.088	&25.8	&0.162 (0.155)	&3.8 (4.1)	&1.40 (1.32)	&-0.064 (0.032)	&80.64 (79.0 \{2\}) 	&14.95 (5.0 \{6\})	\\
$^{190}$Os 	&0.178	&0.092	&27.4	&0.112 (0.186)	&5.3 (3.0)	&1.39 (1.02)	&-0.589	&66.58 (72.9 \{16\})	&13.49 (6.0 \{6\}) 	\\
$^{192}$Os	&0.164	&0.085	&27.4	&0.138 (0.205)	&5.3 (2.4)	&1.68 (0.84)	&0.325 (0.373)	&57.59 (62.1 \{7\})	&11.68 (5.62 \{+21-12\})	\\
$^{192}$Pt	&0.150	&0.087	&30.0	&0.302 (0.316)	&1.7 (1.9)	&0.67 (0.78)	&0.053 (0.393)	&53,82 (57.2 \{12\})	&13.01 (0.55 \{4\})	\\
$^{194}$Pt	&0.125	&0.065	&27.5	&0.321 (0.328)	&1.9 (1.9)	&0.74 (0.77)	&0.046	&42.14 (49.5 \{20\})	&9.05 (0.286 \{+44-35\})	\\
$^{232}$Th 	&0.248	&0.085	&18.9	&0.048 (0.049)	&15.3 (15.9)	&4.61 (4.84)	&0.101 (0.218)	&192.82 (198.0 \{11\})	&7.09	\\
$^{238}$U	&0.210	&0.085	&22.0	&0.048 (0.045)	&20.7 (21.2)	&6.26 (7.16)	&0.010 (0.034)	&274.01 (281.0 \{4\})	&12.1	\\

\hline								

\end{tabular}
}
\end{table*}


\section{Discussion of the Results}\label{sec:discussion}
 
As already stated in the introduction, the main objective of the present work is to evaluate the robustness of the TPSM approach to
account for the features associated  with
triaxiality and  $\gamma$-softness of the collective Bohr Hamiltonian, discussed in the preceding section. As a first step,
some basic properties of the rotational bands are analyzed.
In Table \ref{tab:parameters}, the energies of the 2$^+_1$, $4^+_1$ and $2^+_2$ states along with the transition
probabilities $B(E2, 2^+_1\rightarrow 0^+_1)$ and $B(E2, 2^+_2\rightarrow 0^+_1)$ are listed. 
The model parameters $\epsilon$ for the thirty nuclei are mostly adopted from our earlier works \cite{GH14,GH08,bh15,GS12,SJ21}. These values are
slightly adjusted from the empirical values of Raman et al. \cite{raman} to reproduce the $B(E2, 2^+_1\rightarrow 0^+_1)$ transition
probability in the TPSM approach. In  our earlier works, we focused  on the excitation energies, and no adjustment
to reproduce $B(E2, 2^+_1\rightarrow 0^+_1)$ was performed.
On the other hand, the
$\epsilon'$ parameters,
listed in Table \ref{tab:parameters}, are adjusted to reproduce the band head energy of the $\gamma$-band, $E(2^+_2)$. It is evident from Table \ref{tab:parameters} that the
experimental energies of the $E(2^+_1)$ states, which are
model predictions, are very well reproduced. 
The experimental data on $B(E2, 2^+_2\rightarrow 0^+_1)$ transition probabilities from the head of the $\gamma$- to the
yrast-bands are less complete and accurate. The TPSM results agree within 
the error bars with the experiment values.
The experimental error bars for Os- and Pt- isotopes are too large to make a proper assessment of the predicted values.
In the following subsections, we  turn our attention to various measured properties of near-yrast rotational
bands of the thirty isotopes studied in the present work.
\vspace{-0.5 cm}


\subsection{Relation between Gamma-Rotor and TPSM}\label{sec:GR-TPSM}
In Section \ref{sec:ACM},  we discussed how the staggering pattern emerge
as a consequence of the interaction of the even-$I$-states of harmonic $\gamma$-band with the states of the ground-band and the
states of the  $0^+_2$ band based on the harmonic $\gamma\gamma0$ nature. This excitation has a pulsating nature, that is, it represents a collective 
motion between prolate and oblate via triaxial shapes. Obviously the admixture of such a mode is a measure of $\gamma$-softness.

As the TPSM is based on a mean-field with a fixed triaxial deformation, it does not incorporate such a collective 
excitations in an explicit way. 
It only contains the $0^+_1,~2^+_2,~4^+_3, ~...$ bands, which are generated by projecting the sequences of intrinsic states $K=0$,
$K=2$, $K=4,~...$ from the triaxial quasiparticle vacuum state. In addition projected two- and four-quasiparticle  states
are taken into account. The TPSM staggering pattern is generated by the energy shifts caused by the mixing  of  the $K=2$
and $K=0$  vacuum bands 
and of the quasiparticle bands, where  two-quasiparticle configurations generated from
high-j orbitals  $i_{13/2}$, $h_{11/2}$ and $f_{7/2}$ 
play the decisive role.  As for the phenomenological Gamma-rotor, the competition between the two kinds of admixtures decides
which pattern prevails. We discuss examples of the TPSM band mixing in section \ref{sec:mixing}.  
More examples can be found in Refs. \cite{GH14,SJ21,Na23,SPRTBP}.

The states 
of the TPSM without quasiparticle admixtures always generate the even-$I$-up pattern because of the repulsion between the even-$I$-states   
of the $K=0$ and $K=2$ bands. The  pattern of even-$I$-up can be associated with a narrow Gamma-rotor  potential that 
has a minimum at $\gamma_m>0$ and a $\gamma\gamma 0$-band at  large energy. 
The amplitude of $S(I)$ is substantially smaller than it is for a  rotor with irrotational-flow moments of
inertia that corresponds to the $\gamma$-deformation of the TPSM calculation. For instance, see $S(I)$ 
without quasiparticle admixtures in  Fig.~\ref{fig:gg2E4} for $^{192}$Pt where the TPSM assumes $\gamma_N\approx30^\circ$. 
According to Eq. (\ref{eq:STR}), the triaxial rotor amplitude 
of $S(I)$ is $I - 4.5$ while the TPSM amplitude is approximately $0.1(I - 4.5)$. A similar or even stronger reduction of the amplitude is noted
for the other nuclides in Fig. \ref{fig:gg2E4} and  the Ru isotopes in Fig. \ref{fig:gg2E3} for which  the TPSM assumes $\gamma\approx 30^\circ$.

 Microscopic calculations of the moments of inertia of the three principal axes by means of the cranking model
 give ratios of the moments of inertia  that are not very different from the irrotational-flow ratios \cite{Frauendorf18}, which implies that
 the triaxial rotor model with cranking moments of inertia will give $S(I)$ values that are comparable with 
 the large values obtained from Eq. (\ref{eq:STR}). Thus the reduction must have a different origin. In contrast to the triaxial rotor model, the
 orientation angles of the deformed mean-field of the TPSM are not sharp.
 The overlap between a mean-field state and the same state rotated by an angle $\psi$ has a finite width $\Delta \psi$.
 In Refs. \cite{SF01,SF18}, the author has called 
$\Delta \psi$ the "coherence angle" and discussed its relation with the appearance and termination of rotational bands.
 In the framework of the TPSM, the finite coherence angle
 is reflected by the non-orthogonality of the different $K$ components projected from one and the same mean-field
 configuration and leads to a modified matrix
 in the $K$-space as compared to the one of the triaxial rotor model. The TPSM matrix generates qualitatively
 the same spectrum as the triaxial rotor, 
 consisting of  ground-band ,  $\gamma$-band, $\gamma\gamma4$-band, etc. Quantitatively the energies and transition
 rates differ from the triaxial rotor model one's.  
    

\subsection{Energies}\label{sec:energy}

 The energies of the $I=2$ heads of the $\gamma$-bands are input in the TPSM calculations. All other  energies are calculated.
 The resulting TPSM ratios $E(2^+_2)/E(2^+_1)$ and $E(2^+_2)/E(4^+_1)$ ratios  in Table \ref{tab:parameters} compare  well with the experimental ones.
 The deviations  are caused by the $E(2^+_1)$ and $E(4^+_1)$ energies, which the TPSM tends to slightly overestimate. 

 Figs. \ref{fig:energy1}, \ref{fig:energy2} and \ref{fig:energy3} show the angular momentum as function of the angular frequency, defined as
 \begin{equation}
 \omega(I)=(E(I)-E(I-2))/2~~ \mathrm{associated ~with}~~ I-1/2.
 \end{equation}
 It is evident from the figures that TPSM results are in good agreement with the experimental numbers. The  energies of some of the isotopes 
 have already been reported in our earlier publications \cite{GH14,SJ18,SJ21}. 
 At the frequency of 0.3-0.4 MeV the ground-band interacts with the  s-band, which is a superposition of high-j two-quasiparticle configurations.
 The phenomenon, seen as an up-bend or back-bend,  has been extensively discussed in the literature (see, e.g.,  \cite{SJ18,SF01,SF18}). 
 The crossing frequency is systematically reproduced. The curvature of the crossing depends sensitively on the
 position of the Fermi level, which has the consequence 
 that it is difficult to reproduce. 
 
 The TPSM calculations predict  that $\gamma$-band crosses the two-quasiparticle bands at about the same
 frequency as the g-band.
  The structure of the mixed state has not been studied yet in a systematic way (see Ref.  \cite{SJ18} for $^{156}$Dy).
 In most nuclides, the  $\gamma$-bands are not observed high enough to compare with the TPSM calculations.
 In cases that allow a comparison, the  TPSM describes well the $\gamma$-band crossing  as well as the crossing
 of the ground-band with the s-band.

The calculated energies have been used to evaluate the staggering parameter  of the $\gamma$-bands.
The values of $\bar S(6)$ are listed in Table \ref{tab:parameters}. 
Figs.  \ref{fig:gg2E1}-\ref{fig:gg2E4} display $S(I)$ over an
extended spin range. The table and the figures demonstrate  how well the TPSM calculations account for the
experimental values. The amplitude of $S(I)$ becomes large when the $\gamma$-band enters the region of the 
two-quasiparticle bands which is seen as the up- or back- bends in Figs.   \ref{fig:energy1}, \ref{fig:energy2} and \ref{fig:energy3} (c.f. Sec. \ref{sec:mixing}).

The isotopes of  Gd, Dy, Er, Th and U have  large ratios of $E(2^+_2)/E(4^+_1)$
which are expected for prolate shape and near harmonic $\gamma$-vibrations.  Accordingly, the
staggering pattern in Fig.~\ref{fig:gg2E1},
$S(I)$ is very weak, which is reproduced by the TPSM calculations. The only exceptions are $^{170}$Er and $^{232}$Th,  which show a substantial even-$I$-up pattern associated with triaxiality. 
The TPSM provides  the even-$I$-up staggering for $^{232}$Th though too weak.  
The discrepancies for $^{170}$Er in Table \ref{tab:parameters} and Fig. \ref{fig:gg2E2} reflect the presence of a low-lying $\Delta I=2$ band, which is not 
accounted for by the TPSM. The nature of this band is beyond the scope of the present work.

In Table \ref{tab:parameters}, Se-, Ge-, Mo-, Ru-, and Pt-isotopes have ratios of $E(2^+_2)/E(4^+_1)$ around one or lower, indicative of a 
large triaxiality. 
The TPSM reproduces the positive value of $\bar S(6)$ for $^{76}$Ge, which implies more rigid triaxiality. 
The TPSM does not describe the experimental staggering pattern of  $^{76}$Se, whereas it accounts for the even-$I$-down staggering of $^{78}$Se.
The even-$I$-down pattern of the Mo-isotopes is reproduced by the TPSM, though with a large amplitude. 

The transition from the even-$I$-down to 
the even-$I$-up pattern seen in the Ru-isotopes is present in the TPSM values, although $\bar S(6)$
changes sign two neutrons early ($\bar S(6)=-0.15$ for $^{104}$ Ru), while 	
the other indicators   ${\left[\frac{E(2^+_2)}{E(2^+_1)}\right]}$,	${\left[\frac{E(2^+_2)}{E(4^+_1)}\right]}$ do not change much along the isotope chain
$\biggl({\left[\frac{E(2^+_2)}{E(2^+_1)}\right]}$=2.5,	${\left[\frac{E(2^+_2)}{E(4^+_1)}\right]}$=1.0 for $^{104}$Ru$\biggl)$. 
This indicates that the shell structure of the orbits near the Fermi level plays a significant role in the $N$-dependence of   $\bar S(6)$, because the three characteristics  are 
much stronger correlated in Gamma-rotor model. 

The  $\bar S(6)$ values in the chain of the  Os-isotopes is reasonably well accounted for by the TPSM. 
The changes in other indicators :   ${\left[\frac{E(2^+_2)}{E(2^+_1)}\right]}$,	${\left[\frac{E(2^+_2)}{E(4^+_1)}\right]}$ does not correlate with
  the $N$-dependence of $\bar S(6)$ as expected form the Gamma-rotor model, which points to the influence of the shell structure.
 

 Table \ref{tab:gav} suggests to associate the
Gd-, Dy-, Er-, Th- and U-isotopes with prolate potentials  of the 200-0 class.  However, it seems impossible to find a 
potential that accounts for both the large $E(2^+_2)/E(4^+_1)$ ratios and the substantial even-$I$-up staggering in  $^{170}$Er and $^{232}$Th.
The remaining  nuclides  can  be associated with soft potentials somewhere between 50-30, 50-50,  0-20 and 50-0. It seems inappropriate
to claim "evidence for rigid triaxial deformation at low energy in $^{76}$Ge" \cite{Toh13}. The amplitude of $S(I)$ and the small ratio 
$B(E2, 2^+_2\rightarrow 0^+_1)/B(E2, 2^+_1\rightarrow 0^+_1)$ rather suggest a soft triaxial potential of the type 0-30 between 0-20 and 0-50.

\subsection{In-band $B(E2)$ transitions}\label{sec:intraE2}
 
Figs. \ref{fig:yyA}, \ref{fig:yyB} and \ref{fig:yyC} depict the $B(E2, I\rightarrow I-2)$ values for the transitions 
between the states of the yrast-bands.
 The TPSM calculations without quasiparticle admixtures result in a smooth I dependence,  as expected. The results
 with full the basis give a drop in the $B(E2)$ values at higher spin, which reflects the  crossing between the ground- and the s-band. 
 The latter is composed of two rotational aligned high-j quasiparticles, which reduce the  $B(E2, I\rightarrow I-2)$ value.
 Above the crossing, the yrast line is composed of the s-band, and the full TPSM results are below the vacuum only values.
 The fact that the band crossing  seen  in the $I(\omega)$ plots of
 Figs. \ref{fig:energy1}, \ref{fig:energy2} and \ref{fig:energy3} is accompanied by a drop in the $B(E2, I\rightarrow I-2)$
 values has been discussed in the literature in connection with  the back bending phenomenon
(see, e.g.,  \cite{SJ18,SF01,SF18}). 

For the rare earth nuclei, the first crossing  occurs around $I=12$, and the $B(E2, I\rightarrow I-2)$ values in
Fig. \ref{fig:yyA} stay below the vacuum values above  this spin. 
 For $^{170}$Er, $^{180}$Hf, $^{232}$Th and $^{238}$U the transition probabilities in Fig. \ref{fig:yyA} do not display much of a reduction.

The calculated yrast $B(E2, I\rightarrow I-2)$ values transitions for $^{76,78}$Se, $^{76}$Ge, $^{104,106,108}$Mo and $^{108,110,112,114}$Ru are compared
with the measured values in Fig. \ref{fig:yyB}. The TPSM calculations with full basis show two
drops. In this region the neutron
and proton Fermi surfaces are very close. In most of the cases, two neutrons align first 
quickly followed by the alignment of two protons. It is evident from the figure that TPSM reproduces the measured values
reasonably well. There are two measured transitions, one for $^{76}$Se at $I=8$ and the other for
$^{104}$Mo at  $I=12$ that deviate from the expected trend and from the predicted TPSM values. These
large deviations indicate the presence of shape isomeric states for these nuclei \cite{Moller2009}. 
This is to remind that TPSM approach projects the good angular momentum states
from one and the same mean-field. The truncated multi-quasiparticle basis (\ref{basis})  is too restricted to accommodate 
such a drastic change of the deformation. Similarly, 
the drastic drop of the experimental $B(E2, I\rightarrow I-2)$ values at $I=10$ seen in Fig. \ref{fig:yyC} for the heavy Os- and Pt- isotopes is not reproduced
by the TPSM calculations. The discrepancy seems to indicate that the crossing s-band has a different mean-field deformation, which the
TPSM calculations cannot account for.

\begin{table}[h!]
\LTcapwidth=0.4\textwidth
\caption{\label{tab:mixrat}
Mixing ratios for the inter-band $2_2 \rightarrow 2_1$ and intra-band $3_1 \rightarrow 2_2$  transitions given by expression $\delta=0.835E_\gamma \frac{\langle J_f||\mathcal{M}(E2)||J_i\rangle}{\langle J_f||\mathcal{M}(M1)||J_i\rangle}$. The experimental values are in parenthesis with error bars in curly brackets (Data taken from \cite{Krane1974,Chung1970,Grab1960,Lieder1970,Eldri2018,Uluert2018,Kraci1998,Domi1972}).}
\resizebox{0.49\textwidth}{!}
  {
\begin{tabular}{|c|c|c|c|c|c|c|c|c|c|c|c|}
  \hline
 Isotope  	&$2_2 \rightarrow 2_1$	&$3_1 \rightarrow 2_2$	 &Isotope  	&$2_2 \rightarrow 2_1$	&$3_1 \rightarrow 2_2$	 &Isotope  	&$2_2 \rightarrow 2_1$	&$3_1 \rightarrow 2_2$	\\
	&(Expt.)	&(Expt.)	&	&(Expt.)	&(Expt.)	&	&(Expt.)	&(Expt.)	\\
  \hline									
$^{76}$Ge	&2.099	&0.479	&$^{154}$Gd	&-34.174	&1.587	&$^{182}$Os	&45.917	&1.736	\\
	&(3.5 \{15\})	&	&	&	&	&	&	&	\\
$^{76}$Se	&6.228	&0.176	&$^{156}$Gd 	&-8.27	&2.376	&$^{184}$Os	&22.549	&1.265	\\
	&(5.5 \{5\})	&	&	&(-6.5 \{$^{+2.6}_{-7.9}$\})	&	&	&	&	\\
 $^{78}$Se	&6.299	&0.209	&$^{156}$Dy 	&-9.837	&0.977	&$^{186}$Os	&33.968	&3.627	\\
	&(6 \{$^{+7}_{-2}$\})	&	&	&	&	&	&	&	\\
$^{104}$Mo	&10.383	&0.552	&$^{158}$Dy	&-10.967	&1.174	&$^{188}$Os	&22.549	&4.401	\\
	&(9 \{$^{+4}_{-2}$\})	&	&	&	&	&	&	&	\\
$^{106}$Mo	&7.841	&0.749	&$^{160}$Dy	&-16.634	&1.861	&$^{190}$Os 	&20.513	&7.958	\\
	&(6.2 \{$^{+1}_{-8}$\})	&	&	&(-12.5 \{$^{+2.9}_{-5.0}$\})	&	&	&	&	\\
$^{108}$Mo	&19.745	&0.521	&$^{162}$Dy	&-6.65	&8.449	&$^{192}$Os	&-18.895	&0.398	\\
	&(23 \{$^\infty_{-14}$\})	&	&	&($\ge 29$)	&	&	&	&	\\
$^{108}$Ru 	&14.809	&0.489	&$^{164}$Er	&23.804	&2.369	&$^{192}$Pt	&2.617	&1.147	\\
	&(16 \{$^{+9}_{-4}$\})	&	&	&	&	&	&	&	\\
$^{110}$Ru 	&1.1402	&0.508	&$^{166}$Er	&42.889	&1.03	&$^{194}$Pt	&5.644	&3.022	\\
	&($\infty \{^{>50}_{<-75}\}$)	&	&	&(38 \{$^\infty _{-24}$\})	&	&	&	&	\\
$^{112}$Ru	&-27.363	&0.466	&$^{170}$Er	&56.622	&4.327	&$^{232}$Th 	&56.948	&3.537	\\
	&(-30 \{$^{+10}_{-31}$\})	&	&	&(57 \{$^\infty_{-41}$\})	&	&	&	&	\\
$^{114}$Ru	&-6.225	&0.447	&$^{180}$Hf	&25.568	&0.793	&$^{238}$U	&25.62	&1.771	\\

\hline

\end{tabular}
}
\end{table}

\begin{table*}[h!]
\LTcapwidth=0.4\textwidth
\caption{\label{tab:Q2}  The TPSM and experimental values (experimental values are in parentheses with error bars in curly brackets) of static quadrupole moments $Q=\sqrt{\frac{32\pi}{175}}\langle I=2,M=2|\mathcal{M}(\textrm{E}2)|I=2,M=2\rangle$
  and g factors for thirty nuclei. (Data taken from \cite{NNDC,Ayangeakaa2019,Wu1996,Allmond2008,Corm2019,Smith2004,Menzen1985,Bao2007})}
\resizebox{1\textwidth}{!}
  {
\begin{tabular}{|c|c|c|c|c|c|c|c|c|c|c|c|}
  \hline
  Isotope  	&$Q_{2^+_1}$	&$Q_{2^+_2}$	&g$_{2^+_1}$	 &Isotope  	&$Q_{2^+_1}$	&$Q_{2^+_2}$	&g$_{2^+_1}$	 &Isotope  	&$Q_{2^+_1}$	&$Q_{2^+_2}$	&g$_{2^+_1}$	\\
 	&(Expt.)	&(Expt.)	&(Expt.)	&	&(Expt.)	&(Expt.)	&(Expt.)	&	&(Expt.)	&(Expt.)	&(Expt.)	\\
  \hline												
$^{76}$Ge	&-0.179	&0.188	&0.238	&$^{154}$Gd	&-1.719	&1.719	&0.345	&$^{182}$Os	&-1.382	&1.375	&0.263	\\
	&(-0.181)	&(0.197)	&(0.263 \{21\})	&	&	(-1.82 \{4\})&	&(0.48 \{3\})	&	&	&	&	\\
$^{76}$Se	&-0.084	&0.647	&0.344	&$^{156}$Gd 	&-1.714	&1.717	&0.296	&$^{184}$Os	&-1.415	&1.407	&0.271	\\
	&(-0.34 \{7\})	&	&(0.350 \{27\})	&	&	(-1.93\{4\})&	&(0.41 \{7\})	&	&	(-2.4 \{11\})&	&	\\
 $^{78}$Se	&-0.462	&0.465	&0.351	&$^{156}$Dy 	&-1.970	&1.971	&0.391	&$^{186}$Os	&-1.212	&1.203	&0.261	\\
	&(-0,20 \{7\})	&(0.17 \{9\})	&(0.325 \{24\})	&	&	&	&(0.39 \{4\})	&	&(-1.326)	&(1.203)	&(0.26 \{2\})	\\
$^{104}$Mo	&-1.022	&1.019	&0.220	&$^{158}$Dy	&-1.885	&1.883	&0.347	&$^{188}$Os	&-1.410	&1.409	&0.282	\\
	&	&	&(0.27 \{2\})	&	&	&	&(0.36 \{3\})	&	&(-1.311)	&(1.408)	&(0.29 \{1\})	\\
$^{106}$Mo	&-1.076	&1.075	&0.241	&$^{160}$Dy	&-1.987	&1.984	&0.381	&$^{190}$Os 	&-1.286	&1.286	&0.298	\\
	&	&	&(0.21 \{2\})	&	&	(-1.8 \{4\})&	&(0.35 \{2\})	&	&(-0.947)	&(1.285)	&(0.35 \{1\})	\\
$^{108}$Mo	&-0.852	&0.854	&0.275	&$^{162}$Dy	&-2.063	&1.063	&0.292	&$^{192}$Os	&-1.219	&1.221	&0.276	\\
	&	&	&(0.5 \{3\})	&	&	&	&(0.35 \{2\})	&	&(-0.916)	&(1.221)	&(0.40 \{1\})	\\
$^{108}$Ru 	&-0.811	&0.816	&0.267	&$^{164}$Er	&-2.282	&2.284	&0.299	&$^{192}$Pt	&-0.940	&0.936	&0.195	\\
	&	&	&(0.23 \{4\})	&	&	<0&(2.4 \{3\})	&(0.349 \{8\})	&	&	&	&(0.29 \{2\})	\\
$^{110}$Ru 	&-0.950	&0.954	&0.320	&$^{166}$Er	&-2.321	&2.324	&0.275	&$^{194}$Pt	&-0.635	&0.635	&0.176	\\
	&(-0.74 \{9\})	&	&(0.44 \{7\})	&	&(-1.9)	&(2.2)	&(0.325 \{5\})	&	&(0.409)	&(0.635)	&(0.30 \{2\})	\\
$^{112}$Ru	&-0.766	&0.772	&0.354	&$^{170}$Er	&-2.365	&2.364	&0.287	&$^{232}$Th 	&-2.981	&2.991	&0.287	\\
	&	&	&(0.44 \{9\})	&	&-1.94 \{23\})	&(2.0 \{3\})	&(0.317 \{7\})	&	&	&	&	\\
$^{114}$Ru	&-0.663	&0.669	&0.353	&$^{180}$Hf	&-1.621	&1.616	&0.251	&$^{238}$U	&-2.851	&2.850	&0.281	\\
	&	&	&	&	&	(-2.00 \{2\})&	&(0.31 \{2\})	&	&	&	&	\\ 
\hline

\end{tabular}
}
\end{table*}


\begin{table*}[htp!]
  \renewcommand{\arraystretch}{2.0}
\LTcapwidth=0.6\textwidth
\caption{\label{tab:02} The TPSM energies and B(E2) transition probabilities (W.u) of $0_2^+$ band for thirty nuclei. The experimental values are in parenthesis with error bars are in curly brackets (Data taken from \cite{Farhan2009,Reich2009,Reich2012,Nica2017,Nica2021,Singh2018,Baglin2008,Baglin2018,Singh2002,Singh190,Baglin2012,Chen194,Browne2006,Martin2002}).}
\resizebox{1\textwidth}{!}
  {
\begin{tabular}{|c|c|c|c|c|c|c|c|c|c|c|c|c|c|c}
  \hline
 Isotope  	&Energy	&Energy	&Energy	&B(E2)	&B(E2)	&B(E2)	&B(E2)	&B(E2)	\\
 									
   	&$0^+_2$	&$2^+_3$	&$4^+_3$	&$0_2\rightarrow2_1$	&$0_2\rightarrow2_2$	&$2_3\rightarrow0_1$	&$4_3\rightarrow2_1$	&$4_3\rightarrow2_2$	\\
  \hline									
$^{76}$Ge	&1.902 (1.911)	&2.622 (2.504)	&2.890 (2.733)	&1.639	&2.277	&0.143	&0.569	&12.456	\\
$^{76}$Se	&1.164 (1.122)	&1.612 (1.787)	&1.905	&148.723	&3.957	&0.061	&0.724	&2.402	\\
$^{78}$Se	&1.482 (1.498)	&1.828 (1.995)	&2.266	&2.031 (1.17 \{21\})	&0.152	&0.087 (0.09 \{+3,-6\})	&0.176	&9.320	\\
$^{104}$Mo	&0.901 (0.886)	&1.000	&1.227	&1.914	&0.004	&0.076	&0.002	&15.972	\\
$^{106}$Mo	&0.895 (0.957)	&0.971	&1.259	&0.079	&0.318	&0.058	&0.052	&0.009	\\
$^{108}$Mo	&1.036	&1.170	&1.441	&0.513	&2.307	&0.163	&0.243	& 0.002	\\
$^{108}$Ru 	&0.948 (0.976)	&1.062	&1.303	&29.614	&0.109	&0.074	&0.310	&0.003	\\
$^{110}$Ru 	&1.136 (1.137)	&1.358 (1.396)	&1.571	&249.135	&0.041	& 0.007	&0.003	&0.540	\\
$^{112}$Ru	&1.126	&1.251	&1.483	&67.496	&0.001	&0.001	&0.557	&16.247	\\
$^{114}$Ru	&0.961	&1.087 	&1.348	&115.288	&59.902	&0.0401	&0.363	&18.857	\\
$^{154}$Gd	&0.781 (0.681)	&0.848 (0.815)	&1.197 (1.048)	&18.438 (52 \{8\})	&18.397	&3.046 (6.7 \{6\})	&0.002	&0.003	\\
$^{156}$Gd 	&0.989 (1.049)	&1.15 (1.129)	&1.304 (1.298)	&11.731 (8 \{+4,-7\})	&80.182	&11.268	&3.479 (1.3 \{+5,-7\})	&0.376	\\
$^{156}$Dy 	&0.739 (0.676)	&0.893 (0.829)	&1.017 (1.088)	&9.853	&38.872	&0.032	&0.001	&0.013	\\
$^{158}$Dy	&0.956 (0.991)	&1.008 (1.086)	&1.130 (1.280)	&6.889	&47.854	&0.081 (2.1 \{5\})	&0.074	&0.091	\\
$^{160}$Dy	&1.254 (1.279)	&1.315 (1.349)	&1.576 (1.522)	&1.025	&7.861	&0.086 (0.65 \{8,7\}) 	& 0.160	&0.004	\\
$^{162}$Dy	&1.333 (1.400)	&1.383 (1.453)	&1.488 (1.574)	&1.164	&0.003	&0.01	&0.004	&0.041	\\
$^{164}$Er	&1.214 (1.246)	&1.377 (1.314)	&1.424 (1.469)	&0.354	&5.236	&0.055 (0.23 \{12\})	&0.170	&0.002	\\
$^{166}$Er	&1.486 (1.460)	&1.543 (1.528)	&1.675 (1.679)	& 4.478 (2.7 \{10\})	&1.438	&0.032	&0.091	&8.93 (7.4	\{25\})\\
$^{170}$Er	&0.850 (0.890)	&0.999 (0.960)	&1.102 (1.103)	&0.127	&0.757	&0.080 (0.28 \{3\})	&0.026	&0.133	\\
$^{180}$Hf	&0.803 (1.102)	&1.063 (1.183)	&1.243	&22.641	&0.018	&13.323	&7.149	&0.078	\\
$^{182}$Os	&0.652	&0.878	&1.118	&0.935	&0.042	&15.754	&7.821	&0.031	\\
$^{184}$Os	&0.894 (1.042)	&1.112 (1.204)	&1.189 (1.506)	&0.009	&0.001	&14.477	&7.243	&0.032	\\
$^{186}$Os	&1.206 (1.061)	&1.377 (1.208)	&1.518 (1.461)	&0.010	&0.063	&0.034	&0.155	&0.044	\\
$^{188}$Os	&1.023 (1.086)	&1.150  (1.304)	&1.251	&2.765 (0.95 \{8\})	&2.213 (4.3 \{5\})	&0.040	&0.003	&25.052	\\
$^{190}$Os 	&0.975 (0.912)	&1.099 (1.114)	&1.358	&4.036 (2.4 \{+8,-6\})	&0.012 (24 \{10,-7\})	&0.049	&0.011	&23.242	\\
$^{192}$Os	&0.719 (0.956)	&0.811 (1.127)	&1.039	&0.682 (0.57 \{12\})	&9.740 (30.04 \{+30,-23\})	&0.004	&0.022	&21.411	\\
$^{192}$Pt	&0.907 (1.195)	&1.249 (1.439)	&1.684	&23.653	&20.792	&0.039	&0.222	&7.556	\\
$^{194}$Pt	&1.182 (1.2672)	&1.393	&1.600	&1.967 (0.63 \{+20,-13\})	&3.463 (8.2 [+25, -16])	&0.071	&0.364	&20.337	\\
$^{232}$Th 	&0.727 (0.731)	&0.759 (0.774)	&0.833 (0.873)	&0.433	&8.660	& 1.05 (2.9 \{4\})	&1.01	&7.165	\\
$^{238}$U	&0.914 (0.927)	&0.941 (0.996)	&1.002 (1.056)	&4.632	&2.650	&1.482 (0.38 \{16\})	&6.164	&1.048	\\

  \hline

\end{tabular}
}
\end{table*}


The $B(E2, I\rightarrow I-2)$ values for the transitions within the $\gamma$-bands 
are displayed in Figs. \ref{fig:gg2A}, \ref{fig:gg2B} and \ref{fig:gg2C}. 
As expected, the vacuum-only TPSM values  increase steadily, which is overlaid by an even-$I$-up staggering pattern.
The results  of the full TPSM calculation agree with the vacuum only set for low spin states
 and fall below them for large spin. How staggering pattern changes is discussed in the following.
The TPSM values agree reasonably well with the limited data.

 Figs.  \ref{fig:gg2E1} - \ref{fig:gg2E4} compare the staggering parameter $S(I)$ calculated by
means of Eq. (\ref{eq:staggering1})  
from the energies of the $\gamma$-bands with the staggering parameter calculated by the analog equation  (\ref{eq:SBE22})
from the $B(E2, I\rightarrow I-2)$ values.

The $B(E2)$ staggering calculations without quasiparticle admixtures show even-$I$-down 
pattern, which is opposite to the  even-$I$-up staggering of  the $\gamma$-band energies. 
As we discussed in Sec. \ref{sec:GR-TPSM}, the TPSM results without quasiparticle excitations
show the qualitative features of the triaxial Gamma-rotor.
In Sec. \ref{sec:ACM} we explain that
 the even-$I$-up pattern of energies reflects the mixing of the even-$I$-states of the $\gamma$-band
with the ground-band and that
the same mixing generates the reduction of the $B(E2, I \rightarrow I-2)$ values. Thus the opposite phase of the staggering of the $\gamma$-band energies and intra-band 
$ B(E2,I\rightarrow I-2)$ values appears in a regular way.

 As seen in Figs. \ref{fig:gg2E1}-\ref{fig:gg2E4}, including the quasiparticle admixtures  changes the  $B(E2,I\rightarrow I-2)$  staggering pattern  for a majority
 of nuclides  as follows.
  When the admixtures do not change the energy staggering from even-$I$-up to even-I-down the $B(E2,I\rightarrow I-2)$  staggering 
  remains even-$I$-down as well ( $^{76}$Ge, $^{112}$Ru, $^{170}$Er, $^{182,188}$Os, $^{192}$Pt and $^{232}$Th). 
  When the admixtures reverse the energy staggering to even-$I$- down the $B(E2)$ staggering 
 remains even-I-low for the low I values. Around $I=8$  to 10 the $B(E2,I\rightarrow I-2)$  staggering sets in with the phase opposite to the energy staggering,
 which becomes large. The change of the pattern is discussed  in Sec. \ref{sec:mixing}.

 \begin{figure}[t]
  \centerline{\includegraphics[trim=0cm 0cm 0cm
 0cm,width=0.5\textwidth,clip]{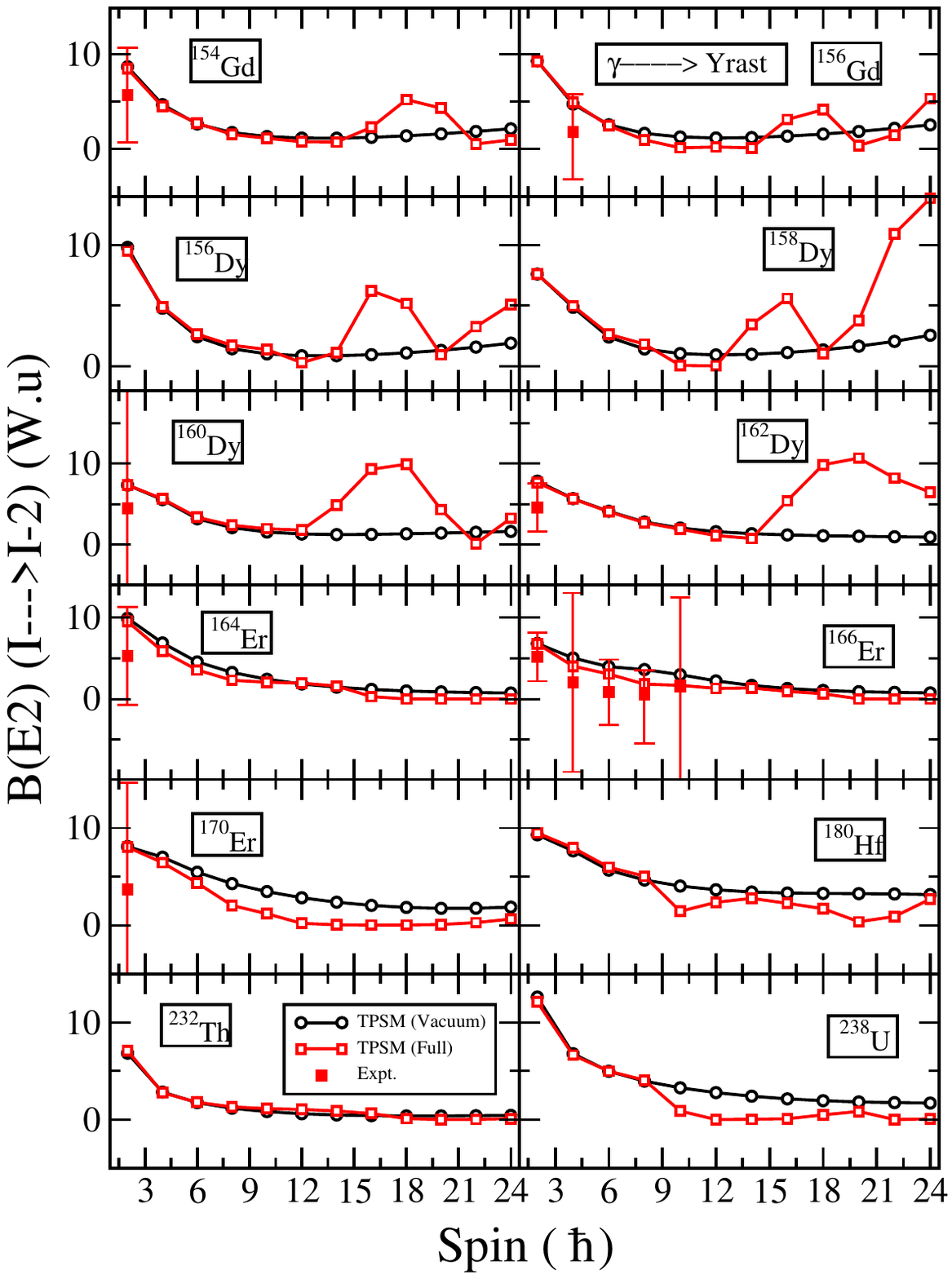}} \caption{(Color
 online) $B(E2)$ transition probabilities (W.u) from $\gamma$-band to the yrast-band for $^{154,156}$Gd, $^{156,158,160,162}$Dy, $^{164,166,170}$Er, $^{180}$Hf, $^{232}$Th and
$^{238}$U isotopes. The black  curves show the TPSM values for the vacuum configuration denoted by TPSM (Vacuum).
The red  curves show the TPSM values including the quasiparticle configuration denoted by TPSM (Full).
The bold red squares show the experimental values (Data taken from \cite{Reich2009,Reich2012,Nica2017,Nica2021,Reich2007,Singh2018,Baglin2008,Baglin2018,Wu2003,Browne2006,Martin2002}).
   }
 \label{fig:gy2A}
 \end{figure}
 \begin{figure}[htb]
  \centerline{\includegraphics[trim=0cm 0cm 0cm
 0cm,width=0.5\textwidth,clip]{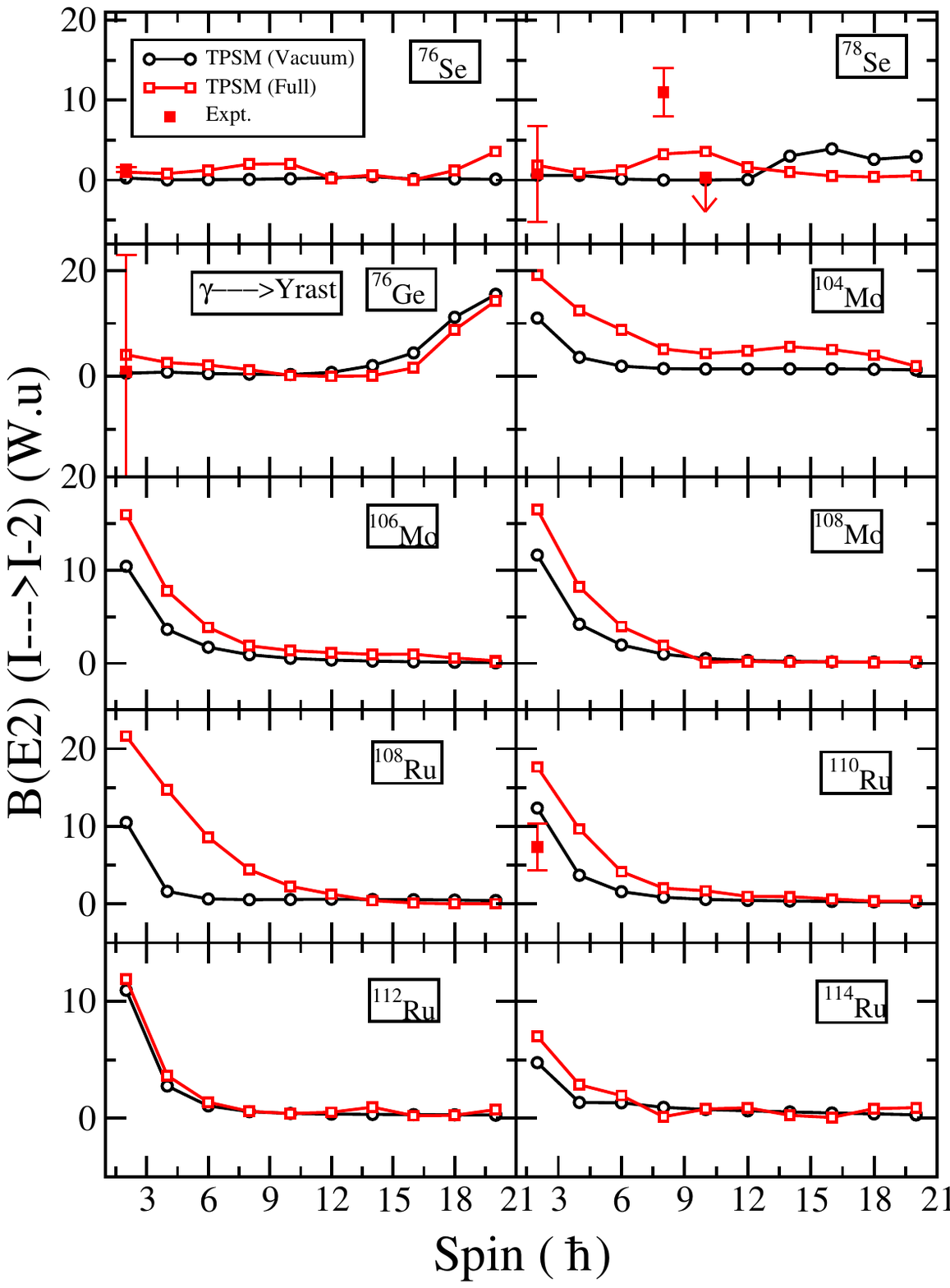}} \caption{(Color
 online) $B(E2)$ transition probabilities (W.u) from $\gamma$-band to the yrast-band for $^{76,78}$Se, $^{76}$Ge, $^{104,106,108}$Mo and $^{108,110,112,114}$Ru isotopes. The black  curves show the TPSM values for the vacuum configuration denoted by TPSM (Vacuum).
The red  curves show the TPSM values including the quasiparticle configuration denoted by TPSM (Full).
The bold red squares show the experimental values (Data taken from \cite{Singh1984,Farhan2009,Blach2007,Fren2008,Blach2000,Gurdal2012,Lalkov2015}).
   }
 \label{fig:gy2B}
 \end{figure}
 \begin{figure}[t]
  \centerline{\includegraphics[trim=0cm 0cm 0cm
 0cm,width=0.5\textwidth,clip]{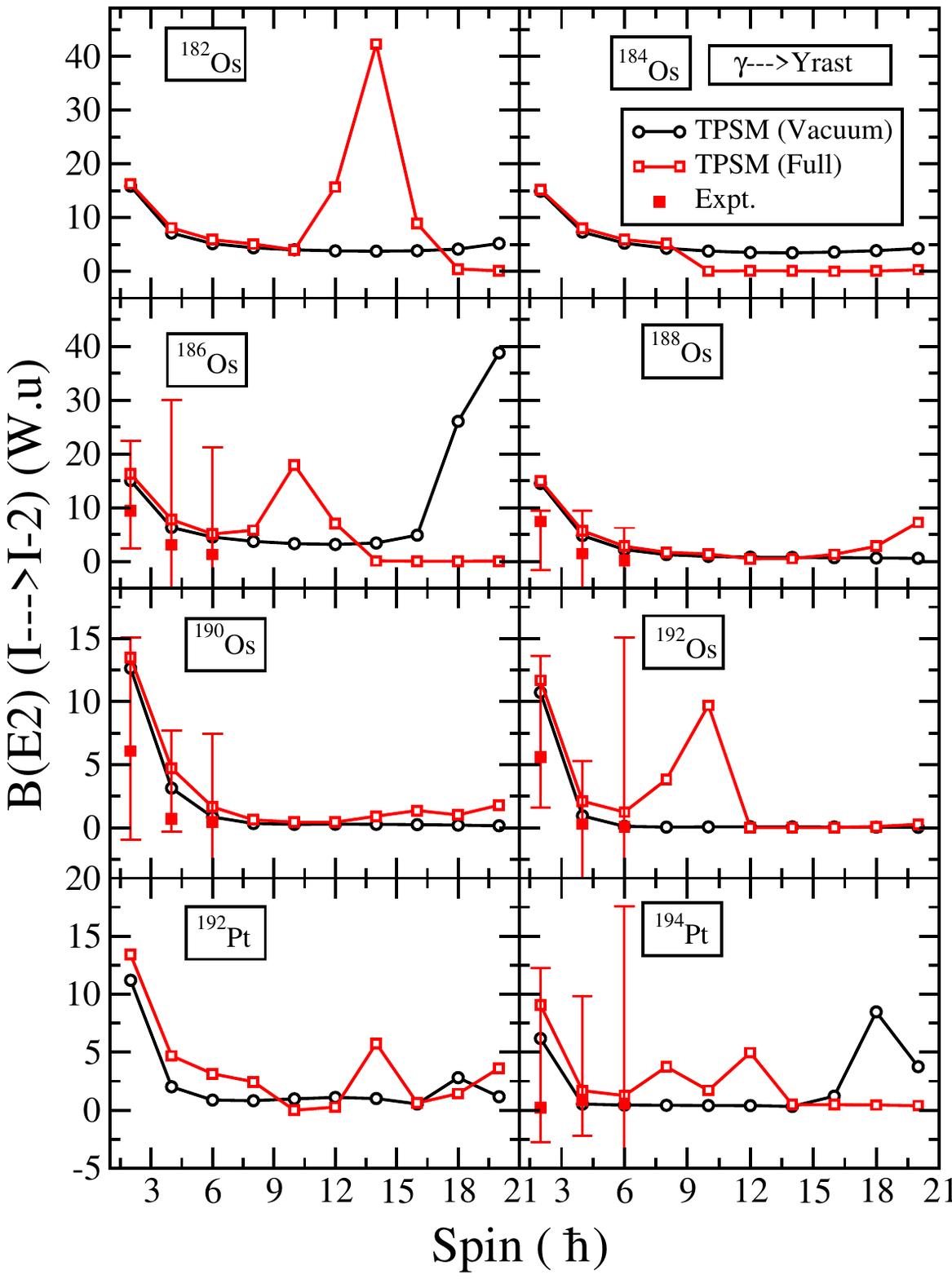}} \caption{(Color
 online) $B(E2)$ transition probabilities (W.u) from $\gamma$-band to the yrast-band for  $^{182-192}$Os and $^{192,194}$Pt  isotopes. The black  curves show the TPSM values for the vacuum configuration denoted by TPSM (Vacuum).
The red  curves show the TPSM values including the quasiparticle configuration denoted by TPSM (Full).
The bold red squares show the experimental values (Data taken from \cite{Singh2015,Baglin2010,Baglin2003,Singh2002,Singh190,Baglin2012,Chen194}).
   }
 \label{fig:gy2C}
 \end{figure}

 \begin{figure}[htb]
  \centerline{\includegraphics[trim=0cm 0cm 0cm
 0cm,width=0.5\textwidth,clip]{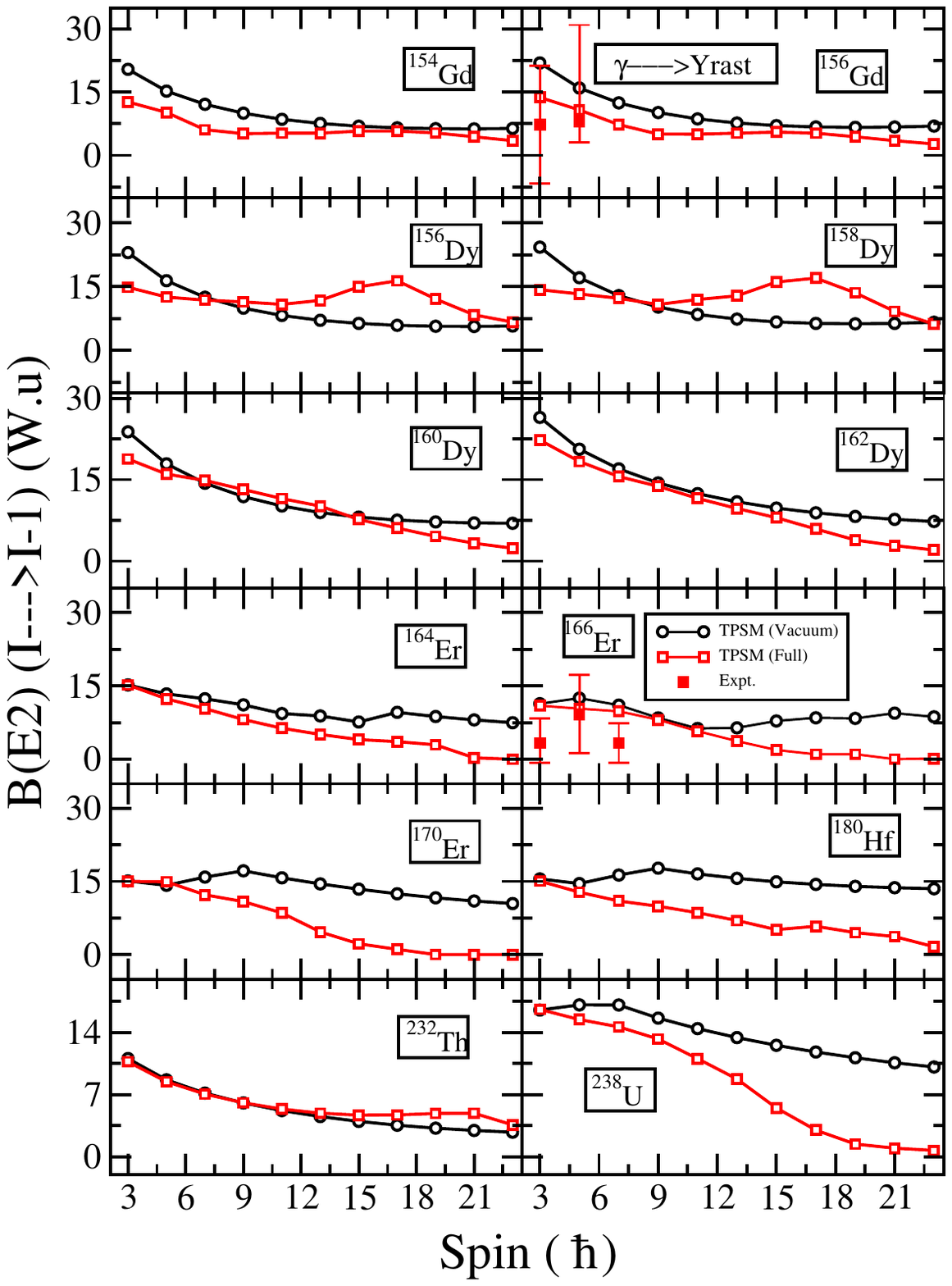}} \caption{(Color
 online) $B(E2)$ transition probabilities (W.u) from $\gamma$-band to the yrast-band for $^{154,156}$Gd, $^{156,158,160,162}$Dy, $^{164,166,170}$Er, $^{180}$Hf, $^{232}$Th and
$^{238}$U isotopes. The black  curves show the TPSM values for the vacuum configuration denoted by TPSM (Vacuum).
The red  curves show the TPSM values including the quasiparticle configuration denoted by TPSM (Full). The bold red squares show the experimental values (Data taken from \cite{Reich2009,Reich2012,Nica2017,Nica2021,Reich2007,Singh2018,Baglin2008,Baglin2018,Wu2003,Browne2006,Martin2002}).
  }
 \label{fig:gy1A}
 \end{figure}

\begin{figure}[htb]
 \centerline{\includegraphics[trim=0cm 0cm 0cm
0cm,width=0.5\textwidth,clip]{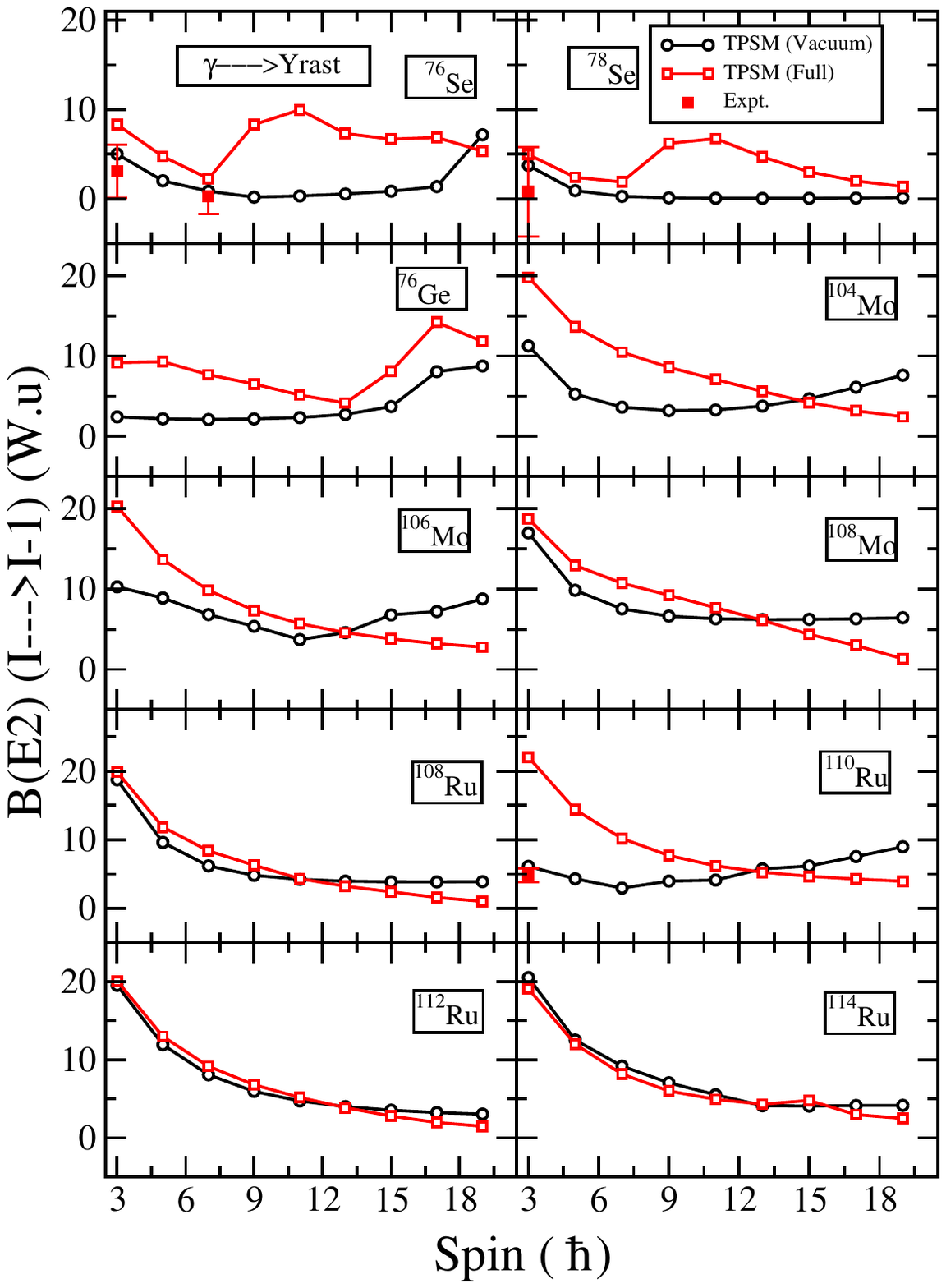}} \caption{(Color
online) $B(E2)$ transition probabilities (W.u) from $\gamma$-band to the yrast-band for $^{76,78}$Se, $^{76}$Ge, $^{104,106,108}$Mo and $^{108,110,112,114}$Ru isotopes. The black  curves show the TPSM values for the vacuum configuration denoted by TPSM (Vacuum).
The red  curves show the TPSM values including the quasiparticle configuration denoted by TPSM (Full). The bold red squares show the experimental values (Data taken from \cite{Singh1984,Farhan2009,Blach2007,Fren2008,Blach2000,Gurdal2012,Lalkov2015}).
  }
\label{fig:gy1B}
\end{figure}
\begin{figure}[htb]
 \centerline{\includegraphics[trim=0cm 0cm 0cm
0cm,width=0.5\textwidth,clip]{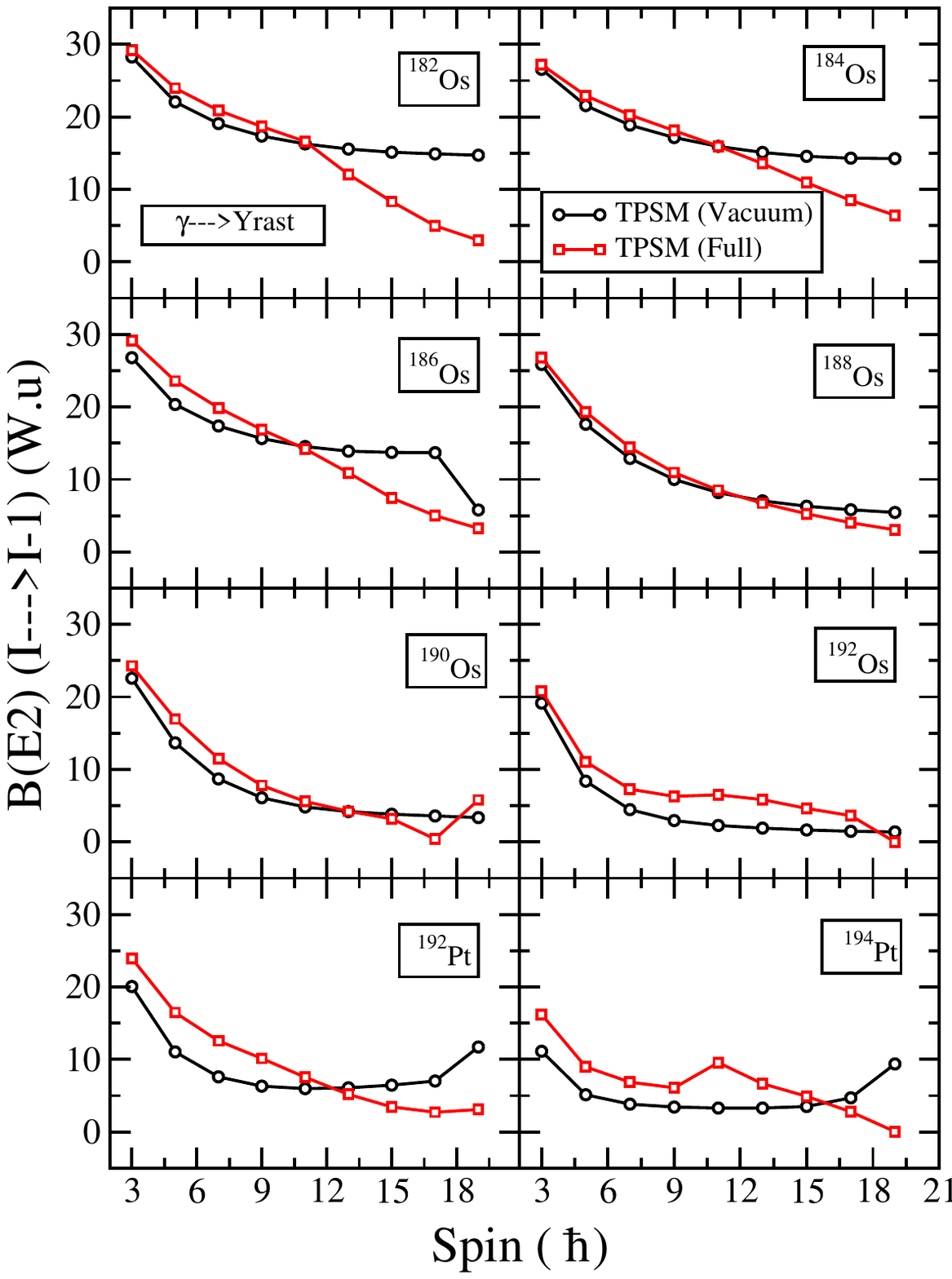}} \caption{(Color
online) $B(E2)$ transition probabilities (W.u) from $\gamma$-band to the yrast-band for $^{182-192}$Os and $^{192,194}$Pt  isotopes. The black  curves show the TPSM values for the vacuum configuration denoted by TPSM (Vacuum).
The red  curves show the TPSM values including the quasiparticle configuration denoted by TPSM (Full). 
  }
\label{fig:gy1C}
\end{figure}

 \begin{figure}[htb]
  \centerline{\includegraphics[trim=0cm 0cm 0cm
 0cm,width=0.5\textwidth,clip]{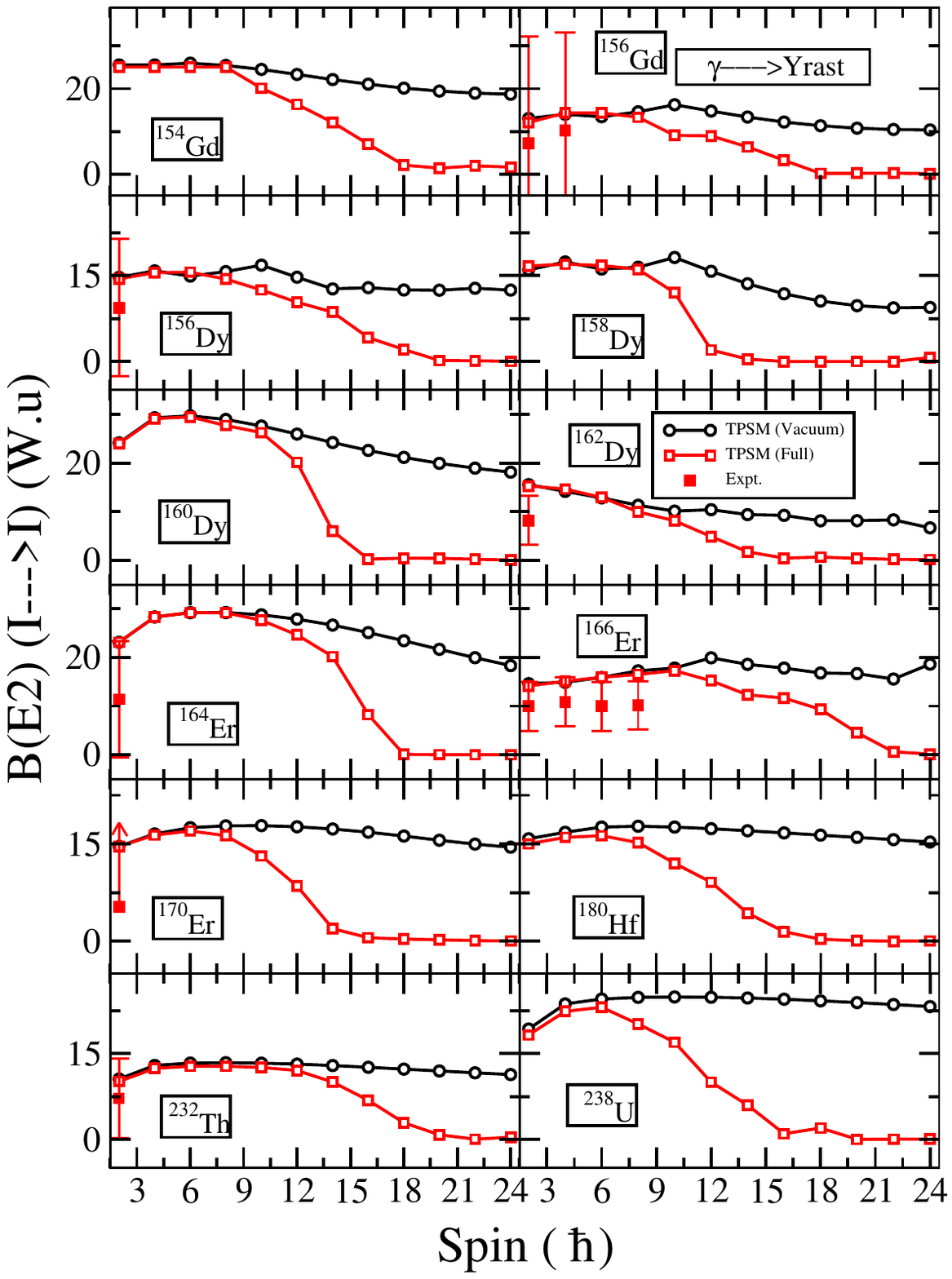}} \caption{(Color
 online) $B(E2)$ transition probabilities (W.u) from $\gamma$-band to the yrast-band for $^{154,156}$Gd, $^{156,158,160,162}$Dy, $^{164,166,170}$Er, $^{180}$Hf, $^{232}$Th and
$^{238}$U isotopes. The black  curves show the TPSM values for the vacuum configuration denoted by TPSM (Vacuum).
The red  curves show the TPSM values including the quasiparticle configuration denoted by TPSM (Full). The bold red squares show the experimental values (Data taken from \cite{Reich2009,Reich2012,Nica2017,Nica2021,Reich2007,Singh2018,Baglin2008,Baglin2018,Wu2003,Browne2006,Martin2002}).
   }
\label{fig:gy0A}
 \end{figure}

\begin{figure}[htb]
 \centerline{\includegraphics[trim=0cm 0cm 0cm
0cm,width=0.5\textwidth,clip]{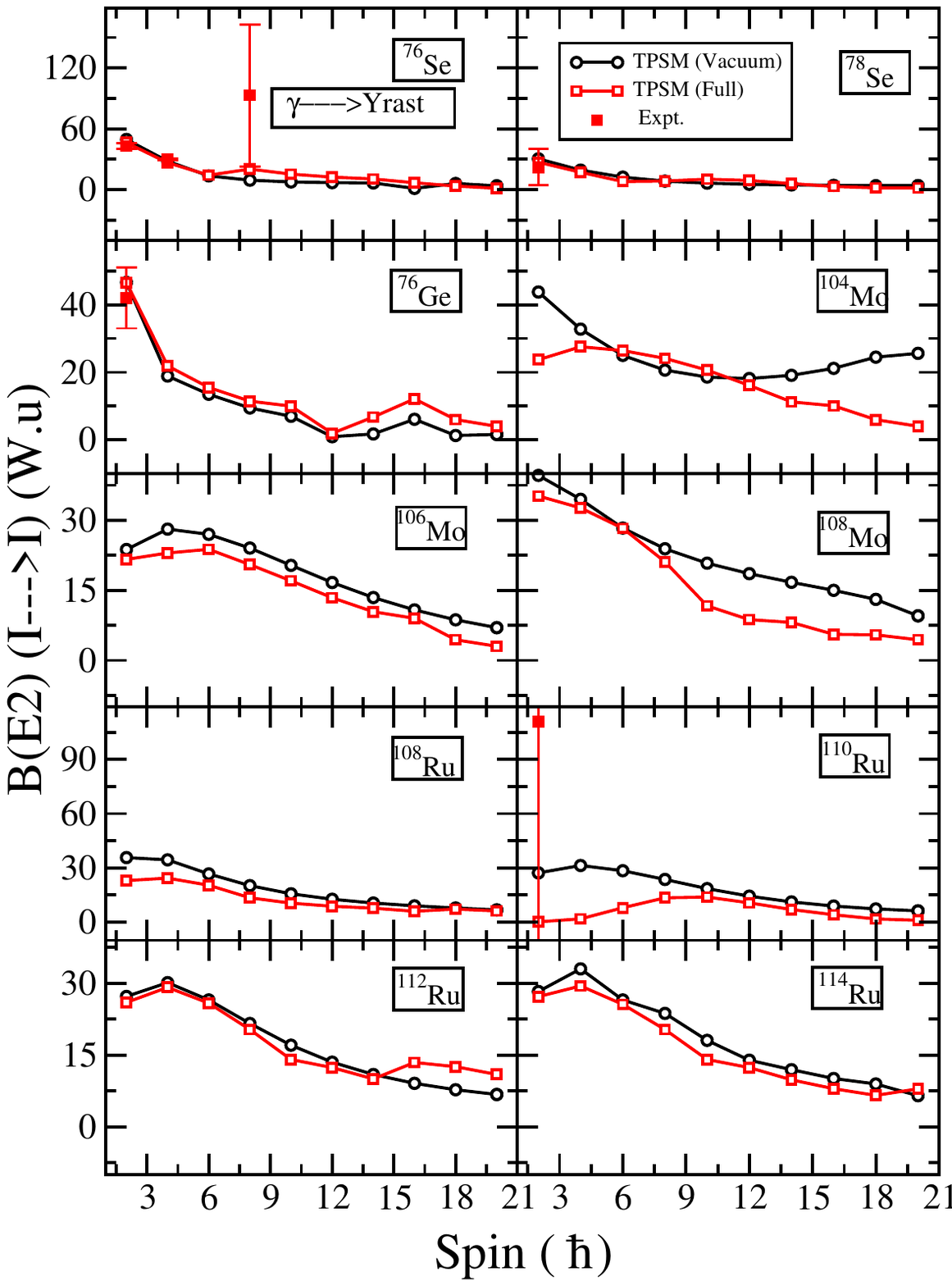}} \caption{(Color
online) $B(E2)$ transition probabilities (W.u) from $\gamma$-band to the yrast-band for $^{76,78}$Se, $^{76}$Ge, $^{104,106,108}$Mo and $^{108,110,112,114}$Ru isotopes.The black  curves show the TPSM values for the vacuum configuration denoted by TPSM (Vacuum).
The red  curves show the TPSM values including the quasiparticle configuration denoted by TPSM (Full). The bold red squares show the experimental values (Data taken from \cite{Singh1984,Farhan2009,Blach2007,Fren2008,Blach2000,Gurdal2012,Lalkov2015}).
  }
\label{fig:gy0B}
\end{figure}

\begin{figure}[htb]
 \centerline{\includegraphics[trim=0cm 0cm 0cm
0cm,width=0.5\textwidth,clip]{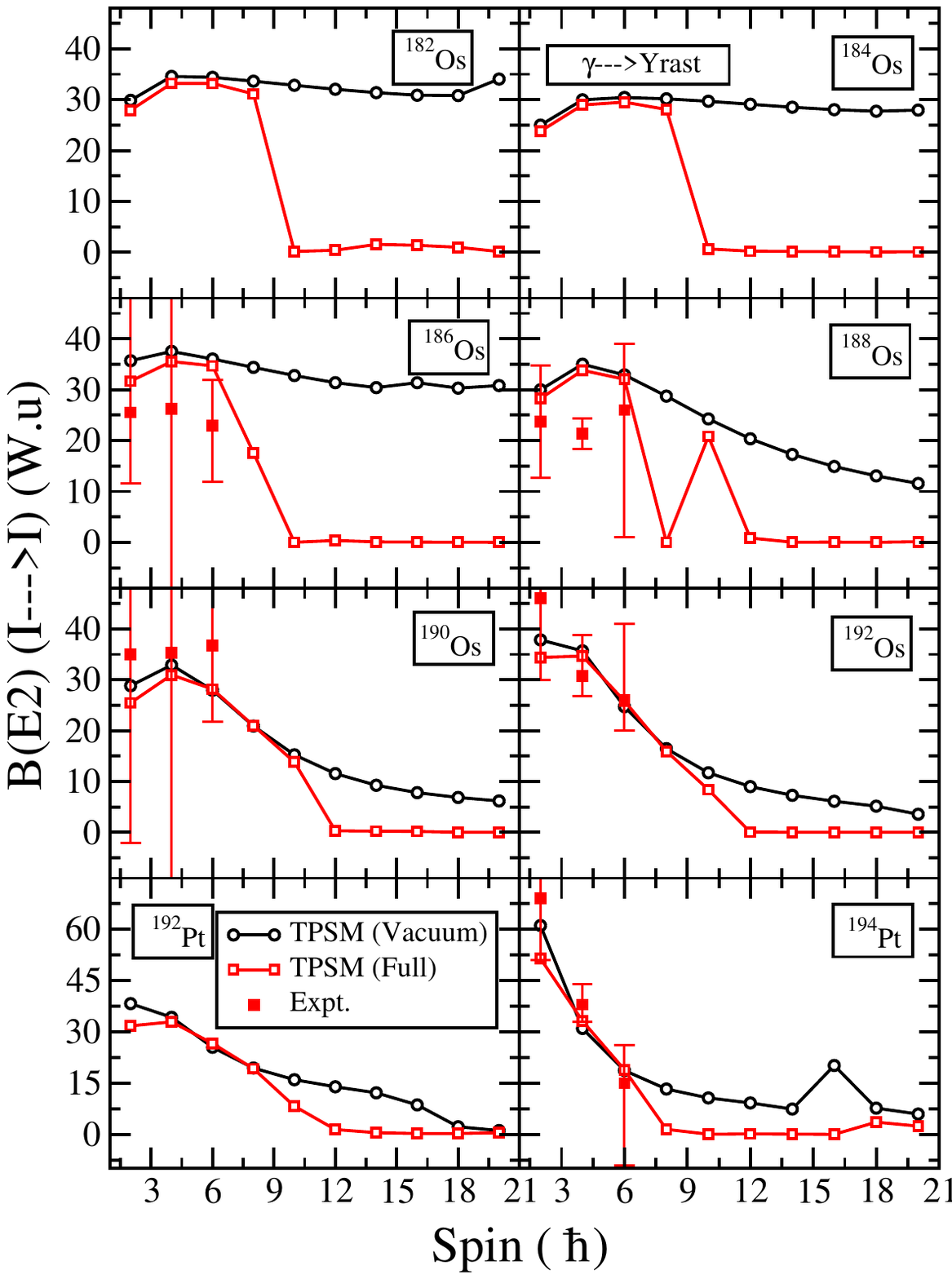}} \caption{(Color
online) $B(E2)$ transition probabilities (W.u) from $\gamma$-band to the yrast-band for $^{182-192}$Os and $^{192,194}$Pt  isotopes.
The black  curves show the TPSM values for the vacuum configuration denoted by TPSM (Vacuum).
The red  curves show the TPSM values including the quasiparticle configuration denoted by TPSM (Full). The bold red squares show the experimental values (Data taken from \cite{Singh2015,Baglin2010,Baglin2003,Singh2002,Singh190,Baglin2012,Chen194}).
  }
\label{fig:gy0C}
\end{figure}

The TPSM values $B(E2, I \rightarrow I-1)$ for the transitions between the $\gamma$-band states are shown in Figs. \ref{fig:gg1A}-\ref{fig:gg1C}.
For the calculations without quasiparticle admixtures, the phase of $S(I)$ is opposite to the phase of $S(I)$ of the energies like for the Gamma-rotor Hamiltonian 
(see Fig. \ref{fig:SBE2}). However, when the quasiparticle admixtures are taken into account there is no clear correlation between the phases of $S(I)$ from the energies and the 
$B(E2, I \rightarrow I-1)$. Comparison with experiment requires to take into account the competing $M1$ component, i. e.,
measuring the mixing ratios. However, to our knowledge there are no such data available.

\subsection{Inter-band $B(E2)$ transition}\label{sec:interE2}

Figs. \ref{fig:gy2A} - \ref{fig:gy2C} show the $B(E2, I\rightarrow I-2$) values for the transitions between the $\gamma$- and yrast-bands. 
The values for the $2^+_2\rightarrow 0^+_1$ transition are also  listed in Table \ref{tab:parameters}. The calculations agree well
with the scarce data,
which are too inaccurate to confirm the predicted decrease with angular momentum. Figs. \ref{fig:gy1A} - \ref{fig:gy1C} 
show the $B(E2, I\rightarrow I-1$) values for the transitions between the $\gamma$- and yrast-bands. The TPSM results again agree
well with the only two data points available. Figs. \ref{fig:gy0A} - \ref{fig:gy0C} show the $B(E2, I\rightarrow I$) values for the 
transitions between the $\gamma$- and yrast-bands, which reproduce the data reasonably well. 
For many nuclei, the TPSM results for low-spin do not change appreciably when the quasiparticles are admixed, and the changes 
are less than the experimental errors. 
For the $\gamma$-soft nuclei of $^{76,78}$Se, $^{76}$Ge,  $^{110}$Ru and $^{104}$Ru \cite{Na23}, the quasiparticle admixtures are essential for reproducing the experimental values.
This demonstrates in a way how the features of $\gamma$-softness come about in the TPSM framework. 
For the $I\rightarrow I-1$ and $I\rightarrow I$ transitions there is a $M1$ component. Table \ref{tab:mixrat} lists the TPSM values 
of the inter-band $2_2 \rightarrow 2_1$ and intra-band $3_1 \rightarrow 2_2$  transitions, which agree with the limited experimental data within the large error bars.

\subsection{Static moments}\label{sec:static}
\vspace{-10pt}
Table \ref{tab:Q2}  lists the static quadrupole moments of the $2^+_1$ and $2^+_2$ states 
and the g factors of the $2^+_1$ states of the considered nuclei. 
The TPSM
reproduces the data reasonably well for all the studied nuclei, except for the $Q$ values of $^{78}$Se and the g-factors in $^{192}$Os and $^{194}$Pt.
The TPSM reproduces the opposite signs of $Q$ found for  the $2^+_1$ and $2^+_2$ states of the Gamma-rotor. It  accounts for the small values   
of $\vert Q \vert$ in $^{76}$Ge which are expected for $\gamma$ close to	30$^\circ$.

\subsection{Higher excitations}\label{sec:high}

Table \ref{tab:02} lists the energies and $B(E2)$ values for the excited states that are central in the discussion of the nature of triaxiality.
The structure of the low-energy spectrum of the collective Gamma-rotor Hamiltonian (\ref{eq:ACM}) is simple. It consists of the bands built on the states $0^+_1$ (ground), $2^+_2~(\gamma),~4^+_3~(\gamma\gamma4)$ and
$0^+_2~(\gamma\gamma0)$. The TPSM spectrum is more complex, because it contains the states originating from the two- and four-quasiparticles, in addition.
  The $\gamma\gamma4$-band, generated by $K=4$ projection from the quasiparticle vacuum, may 
have  larger energy  than a band built on a $K=4$ two-quasiparticle state. The nuclides with very small $B(E2, 4^+_3\rightarrow 2^+_2)$ transition probabilities are examples. 
The soft triaxial nuclei $^{110,112}$Ru, $^{188,190,192}$Os, and $^{192}$Pt have the collective enhancement of ~10-20 W.u., which
 is expected for a $\gamma\gamma4\rightarrow\gamma$ transition. 

Collectivity of the $\gamma\gamma0$ type (pulsating shape)
is not explicitly built into the TPSM. The collective enhancement of the $B(E2, 0^+_2\rightarrow 2^+_2)$ for several nuclides shows that the quasiparticle states admix in a coherent way
to generate a collective state in the shell-model manner. However, the collective state must be more complex than the $\gamma\gamma0$ pulsation because these $0^+_2$ states 
(except $^{192}$Os) have collectively enhanced $B(E2, 0^+_2\rightarrow 2^+_1)$ probabilities, which are quenched for the purely 
collective $\gamma\gamma0$ states of  the Gamma-rotor (see Table \ref{tab:gtrans}).
The diagonalization in the two- and four-quasiparticle space accounts also for the collective $\beta$ mode. The large values of  $B(E2, 0^+_2\rightarrow 2^+_1)$ for the
$N=90$, $92$ nuclei, $^{154,156}$Gd and $^{156,158}$Dy near the transition from small to large $\beta$ values are indications for this type of correlations. For
the Os and Pt isotopes, which are close to the region of prolate-oblate shape coexistence,  the transitions are collectively enhanced as well.  In all cases, the TPSM
generates collective enhancement of the $\beta$ type in a qualitative manner. However, the correlations between the quasiparticle excitations do not  account 
in a quantitative way for the large-scale $\beta$ fluctuations in these nuclei.

For the prolate nuclei of $^{160,162}$Dy, $^{166,168,170}$Er and $^{180}$Hf, the $B(E2, 0^+_2\rightarrow 2^+_1)$ and $B(E2, 0^+_2\rightarrow 2^+_2)$ values fluctuate. This seems to represent a transition from two-quasiparticle states
with a varying amount of quadrupole-correlated admixtures. The interaction of the $\gamma$-band with these bands (and 
likely with higher $0^+_3,~0^+_4,~...$ bands) generates the weak even-$I$-down pattern seen in  the energies for most of the prolate nuclei. It also explains its more erratic
correlation with the staggering pattern of the intra $\gamma$-band $B(E2, I\rightarrow I-2)$. Most of the available data agree
within the error bars with the TPSM results.
However, there are very few measurements to test the TPSM predictions in a consistent manner.

\subsection{Band mixing interpretation}\label{sec:mixing}
In this section, 
we discuss how the triaxiality patterns of the TPSM results emerge using the band mixing argument.
 We address their changes with $N$ by comparing $^{182}$Os with $^{188}$Os.
Additional analysis concerning  $^{156}$Dy, $^{104}$Ru and $^{112}$Ru nuclides are given in Refs. \cite{SJ18,Na23,SPRTBP}.

\begin{figure}[htb]
  \centerline{\includegraphics[trim=0cm 0cm 0cm
 0cm,width=0.5\textwidth,clip]{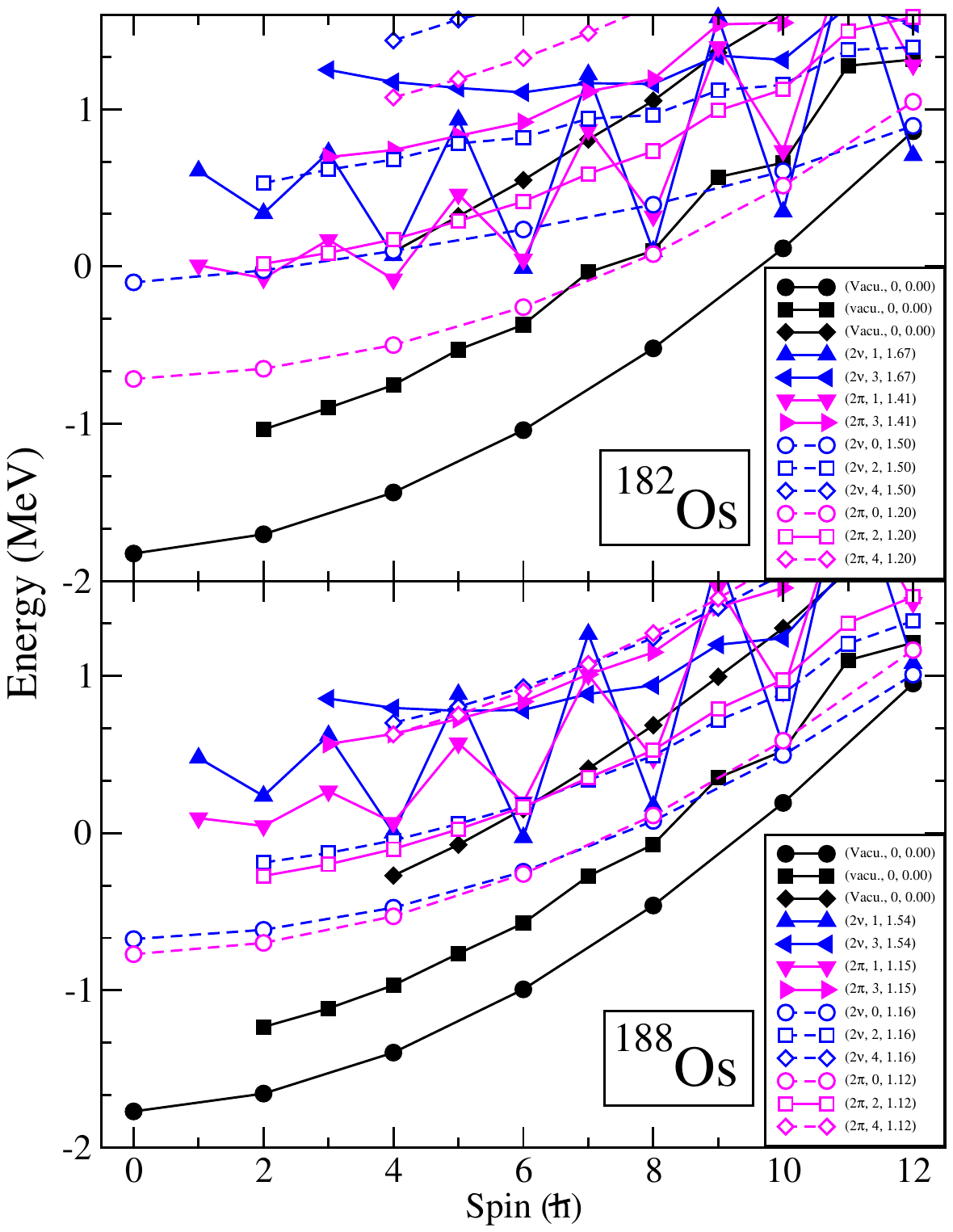}} 
 \caption{(Color online). Band diagrams for $^{182,188}$Os showing the TPSM projected energies before band mixing. 
 The bands are labelled by
three quantities : quasiparticle character, $K$-quantum number and energy of the two-quasiparticle
state. For instance, $(2\pi, ~1, ~1.40)$ in the upper panel designates the $K=1$ state projected from
the $h_{11/2}$ two-quasiproton configuration with the energy of 1.40 MeV.
 The  $K=0, ~2,~4$ states projected from the quasiparticle vacuum are labelled with Vacu.
 The four-quasiparticle states lie above 2 MeV.).
   }
\label{fig:OsBand}
 \end{figure}

\begin{figure}[htb]
  \centerline{\includegraphics[trim=0cm 0cm 0cm
 0cm,width=0.5\textwidth,clip]{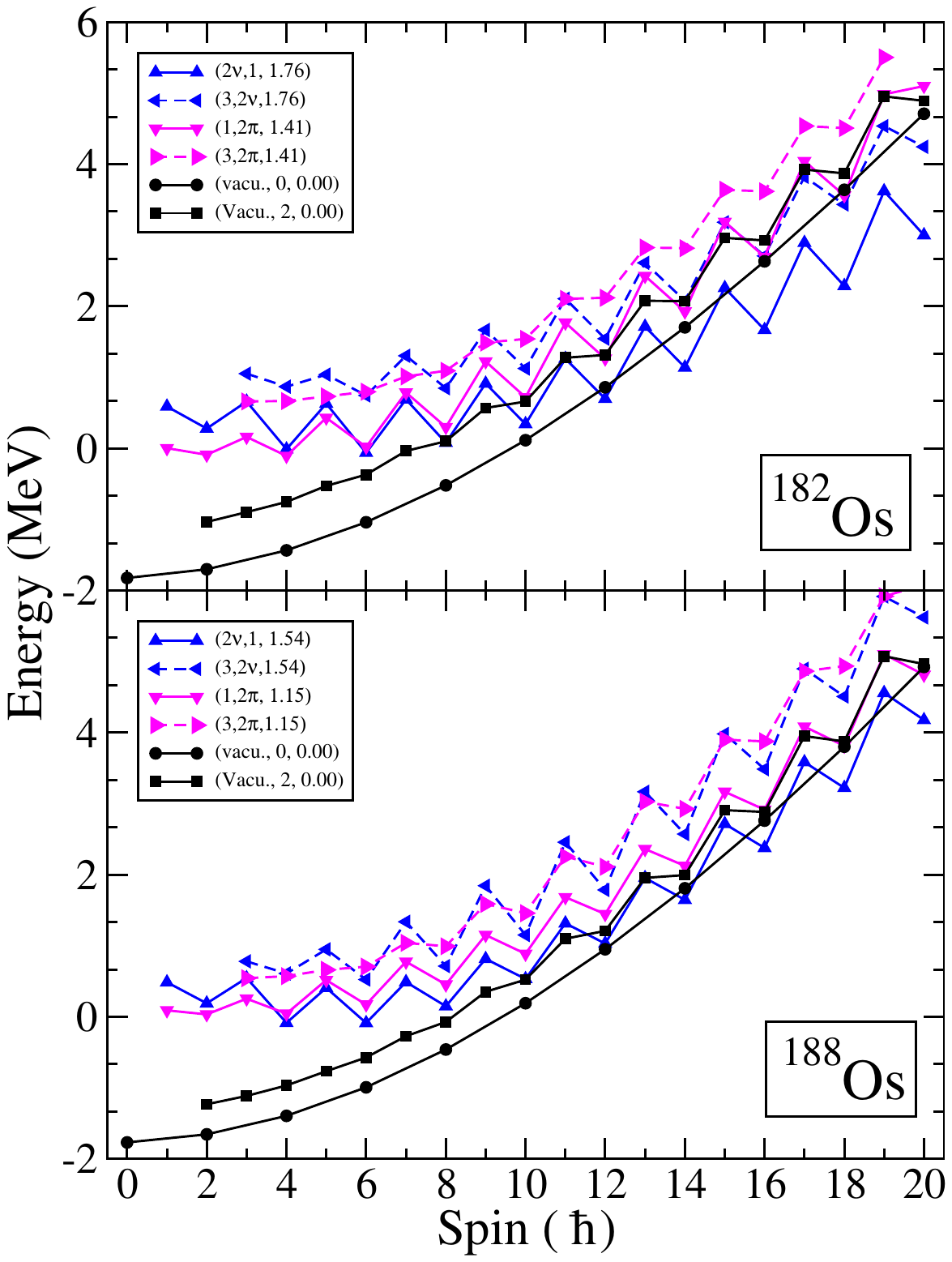}}
 \caption{(Color online). Band diagrams for $^{182,188}$Os after separate pre-diagonalization of the proton $h_{11/2}$ and neutron $i_{13/2}$ two-quasiparticle configurations.
 The states are labelled by their structure at $I=1,~2$, where they are not yet mixed.
  See Fig. \ref{fig:OsBand} for details.
   }
   \label{fig:OsBandDia}
   \end{figure}

The  band diagram in Fig. \ref{fig:OsBand} shows the energies of the projected quasiparticle configurations in $^{182}$Os.
 Around $I=12$, the $K=1$ two-quasineutron configuration crosses the $K=0$ vacuum configuration, which causes the back bending anomaly of
 the yrast states. At such high spin, the $K$ of the two-quasineutron configuration is no longer approximately conserved, which results in a complex
 picture. To present the physics more clearly, Fig. \ref{fig:OsBandDia} presents a modified band diagram, which shows  
  the $h_{11/2}$ two-quasiproton and of the $i_{13/2}$  two-quasineutron bands
 after diagonalizing the TPSM Hamiltonian within their individual subspaces. 
 The lowest two-quasiparticle $(2\nu,~1, ~1.76)$ band with even $I$, which is commonly called the s-band, carries an additional angular moment of 8 compared  the ground-band.
 It can be  seen in Fig. \ref{fig:OsBandDia} by comparing the angular momenta of the ground- and s-band at the same rotational frequency, which is the slope $E(I)$.

\begin{figure}[htb]
  \centerline{\includegraphics[trim=0cm 0cm 0cm
 0cm,width=0.5\textwidth,clip]{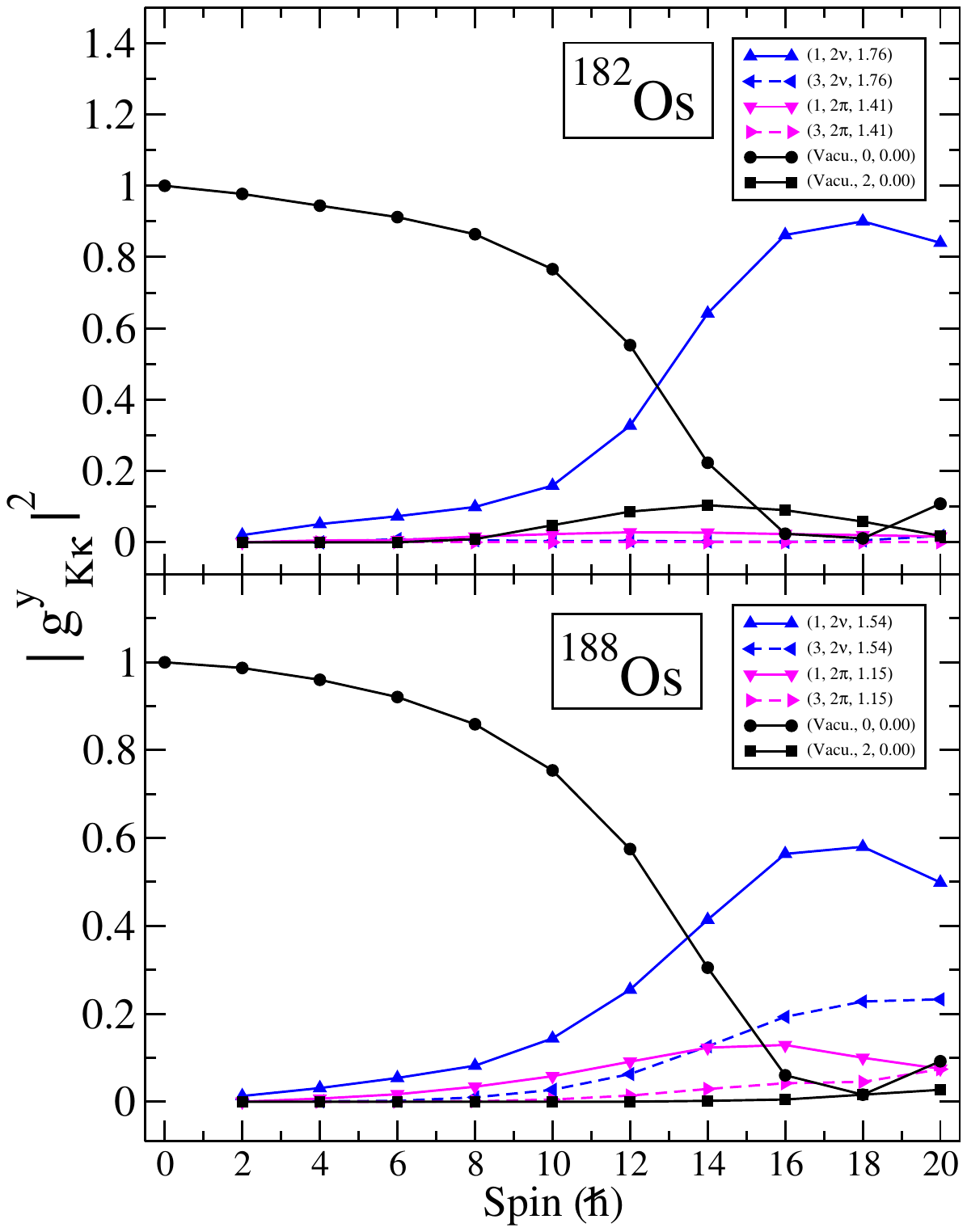}}
 \caption{(Color online). Probabilities (\ref{eq:GKk}) of the pre-diagonalized high-j two-quasiparticle configurations  in Fig. \ref{fig:OsBandDia} for the yrast-bands in $^{182,188}$Os.
 }
\label{fig:OsWfyrast}
\end{figure}

Fig. \ref{fig:OsWfyrast}  shows the probabilities $(g^y)^2$ of the important components of the yrast-bands. The structural change from vacuum to the
neutron s-band, which causes the back-bend  in Fig. \ref{fig:energy3}, is clearly seen.  Compared to
$^{182}$Os, the structural change in
$^{188}$Os is more gradual and involves both the $(2\nu,~1, ~1.54)$  and $(2\nu,~3, ~1.54)$ bands. This is reflected by the later and smoother up-bend in 
Fig. \ref{fig:energy3}.

\begin{figure}[htb]
  \centerline{\includegraphics[trim=0cm 0cm 0cm
 0cm,width=0.5\textwidth,clip]{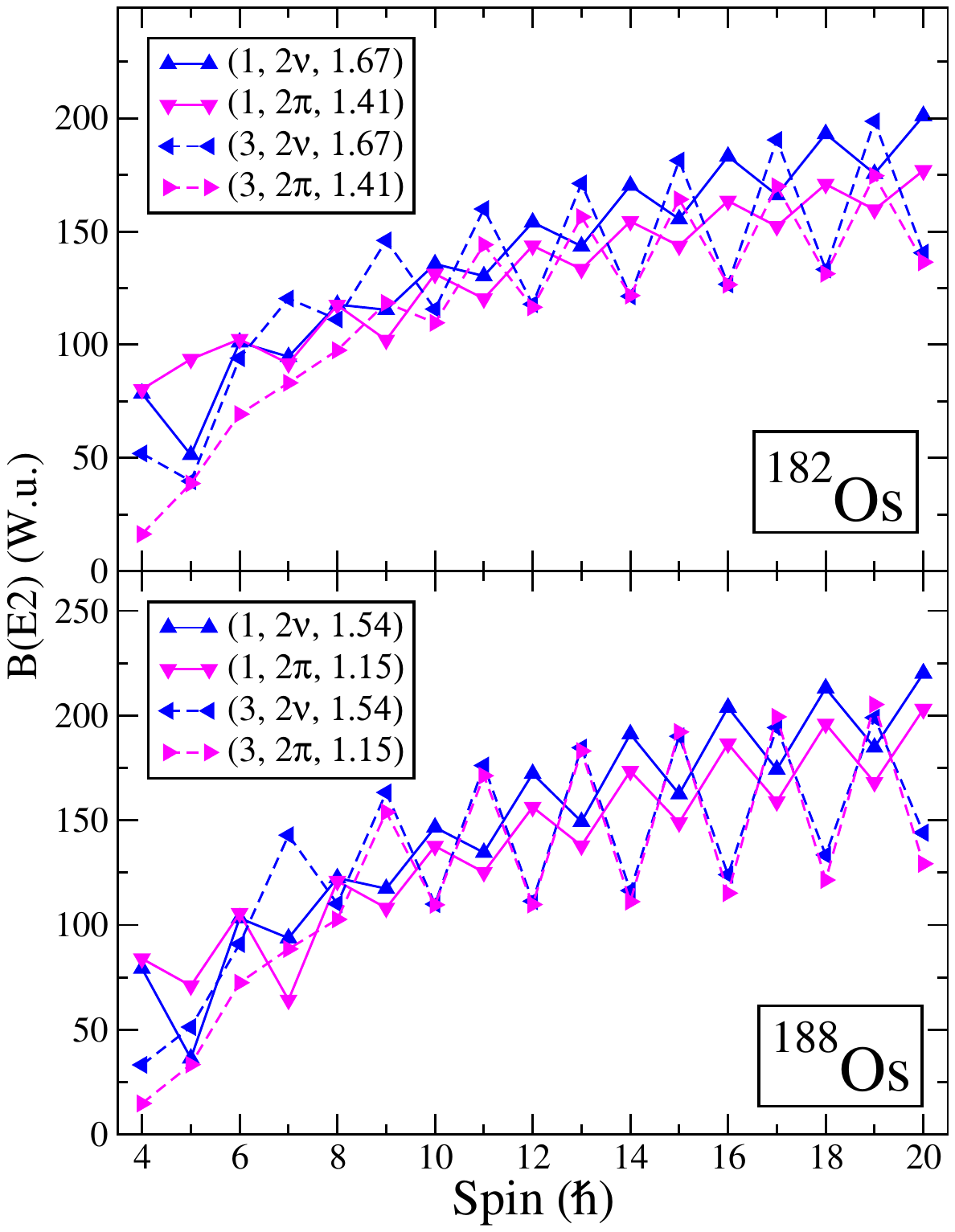}} 
 \caption{(Color online). Intra-band transition probabilities $B(E2, I\rightarrow I-2)$ of the pre-diagonalized high-j two-quasiparticle configurations in Fig. \ref{fig:OsBandDia}. 
   }
\label{fig:OsBE2}
\end{figure}

Fig. \ref{fig:OsBE2} shows the $B(E2, I \rightarrow I-2)$ values between the states  $ I, ~K=1$ projected from the two-quasineutron configuration at 1.67 MeV
  and the two-quasiproton configuration at 1.41 MeV. 
 The intra-band $B(E2, I\rightarrow I-2)$ values for neutron s-band in Fig. \ref{fig:OsBE2} change from 150 W.u. at $I=12$ to 180 W.u. at
 $I=20$. As seen in Fig. \ref{fig:yyC}, this accounts well for  the reduced   $B(E2, I\rightarrow I-2)$ values in the yrast sequence between 150 W.u and 160 W.u. in the same $I$ range.
 The systematic reduction of the $B(E2, I\rightarrow I-2)$ values in the yrast sequence of the other nuclides is explained in the same way by the crossing of the ground-band with the 
 neutron s-band, which has smaller $B(E2, I\rightarrow I-2)$ values.

\begin{figure}[htb]
  \centerline{\includegraphics[trim=0cm 0cm 0cm
 0cm,width=0.5\textwidth,clip]{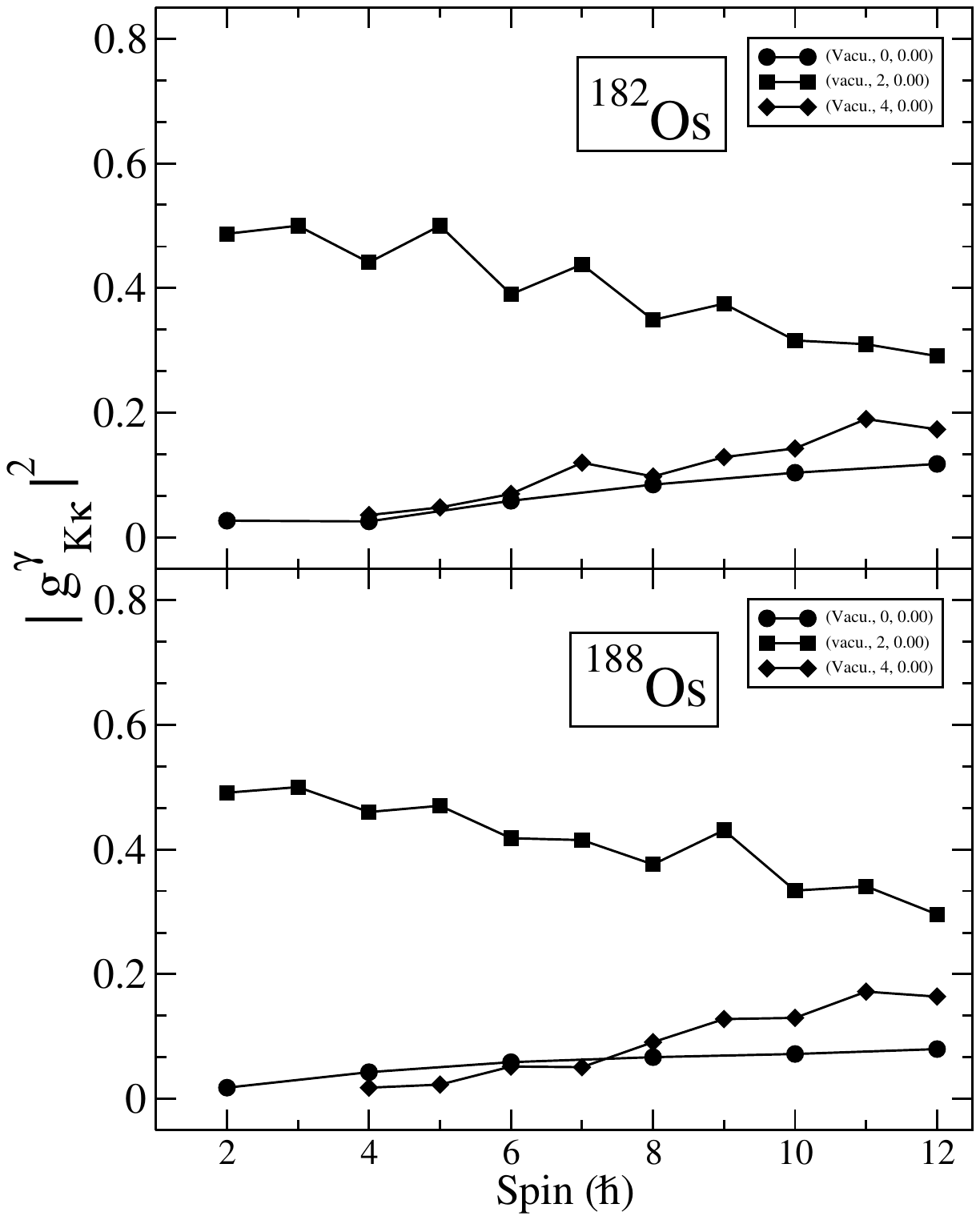}} 
 \caption{(Color online).  Probabilities (\ref{eq:GKk}) of the vacuum components of the $\gamma$-band in $^{182,188}$Os
 after diagonalization within the truncated space of the states projected from the quasi particle vacuum only (the vacuum only cases in figures and tables). 
 }
\label{fig:OsWfvac}
\end{figure}

  Fig. \ref{fig:OsWfvac} illustrates the  band mixing when the basis is truncated to the projected vacuum state only. 
 It displays the probabilities of the   $K=0$ and $K=4$  components admixed to the $K=2$ configuration after diagonaliztion within the truncated space.
  The admixture of the $K=0$ ground-band generates an upward shift of the even-$I$ states of the $\gamma$-band, 
  which is seen in Fig.   \ref{fig:gg2E4} "without quasiparticles". The downward shift 
  by the $K=4$ band, which is about the same for even and odd $I$, does not change the staggering pattern.
  The upward shift  is given by $H_{20}\times g^\gamma_{K=0,vac}$, where $H_{20}$ is the coupling
  matrix element. It is linear in the mixing amplitude $g^\gamma_{K=0,vac}$, which is reflected by the gradual increase of the staggering amplitude with $I$.
  
  The intra-band  $B(E2,I_\gamma\rightarrow I_\gamma-2)$ transitions "without quasiparticles" in Fig. \ref{fig:gg2E4} show
  a staggering which has a phase that is opposite to the energy staggering $S(I)$.
  The reason is the same as for the analog phase flip seen for  the Gamma-rotor (c.f. Fig. \ref{fig:SBE2}) and is explained in the Appendix. 
  It is proportional  to the amplitude of the $K=0$ ground-band admixture and the quadrupole matrix element between $K=0$ and $K=2$, which explains 
  the gradual increase of the staggering amplitude. The opposite
  staggering phase is a consequence of the relative  phases of the amplitudes and the matrix element (see the  discussion at the end of the Appendix).
   
   Fig. \ref{fig:OsWfgamma} shows the composition of the $\gamma$-bands when all quasiparticle configurations are taken into account. 
   The important admixtures are  $(2\nu,~1,~1.76)$ and $(2\nu,~3,~1.76)$.  Above $I$=12 these states dominate the wave function. 
   The consequences of the structural change are complex. In the case of  $^{182}$Os,   Fig. \ref{fig:gg2C} indicate a reduction of the $B(E2,I_\gamma\rightarrow I_\gamma-2)$
   values below the vacuum values, which is expected from Fig. \ref{fig:OsBE2} for the dominating two-quasineutron components in analogy to the yrast-band.
   However for $^{188}$Os there is no such reduction. The reason is a mutual cancelation of the  contributions from the
   $(2\nu,~1, ~1.54)$  and $(2\nu,~3, ~1.54)$ bands.
    
  As seen in Fig. \ref{fig:gg2E4}, below $I$=12 the staggering of the $B(E2)$ values does not change much when the coupling to quasiparticles is taken into account.
  The quasiparticle contributions are proportional to the square of the mixing amplitude because the quadrupole matrix elements between the vacuum and the 
 two-quasineutron configuration are negligibly  small. In contrast, the energy shifts are linear in the mixing amplitudes, and the modification of the energy staggering 
 sets in already below $I$=12. The way the pattern changes is complex because the  $(2\nu,~1,~1.76,~1,54)$,
 $(2\nu,~3,~1.67,~1.76)$ and the $K=0$ and 2 vacuum bands, 
  and their mutual coupling matrix elements are involved.
  
  Comparing the band diagrams in Fig. \ref{fig:OsBandDia},  the following difference is noticed. The neutron s-band
  $(2\nu,~1,~1,54, ~ I ~even)$ in $^{188}$Os 
  encounters the $K=0$ vacuum band later and more gradually than the s-band  $(2\nu,~1,~1,76, ~ I ~even)$ in $^{182}$Os, and the distance between the $K=0$ and 2 
  vacuum bands is  smaller in $^{182}$Os than in $^{182}$Os. One may surmise that this results in a strong mixing of the bands in $^{188}$Os and cancellation 
  of the two-quasineutron components, whereas in $^{182}$Os the mixing is weaker such that downward push by  the s-band $(2\nu,~1,~1,76, ~ I ~even)$ prevails for $I<12$ and changes the
  staggering phase.  Fig. \ref{fig:energy3} reflects the differences 
  by showing a later smooth  up-bend for the former compared to the earlier sharp back-bend for the latter, which supports the interpretation.
  The same scenario is found in the Ru isotopes \cite{SJ21,Na23,SPRTBP}. The band diagrams for $^{104}$Ru and  $^{112}$Ru show the  analog differences as the ones for
  $^{182}$Os and $^{188}$Os. The even-$I$-down of  $\gamma$-softness accompanied by a sharp back-bend is found in $^{104}$Ru \cite{Na23}. 
    The even-$I$-up of  $\gamma$-rigidity accompanied by a smooth up-bend is found in $^{112}$Ru \cite{SJ21}.

\begin{figure}[htb]
 \centerline{\includegraphics[trim=0cm 0cm 0cm
 0cm,width=0.5\textwidth,clip]{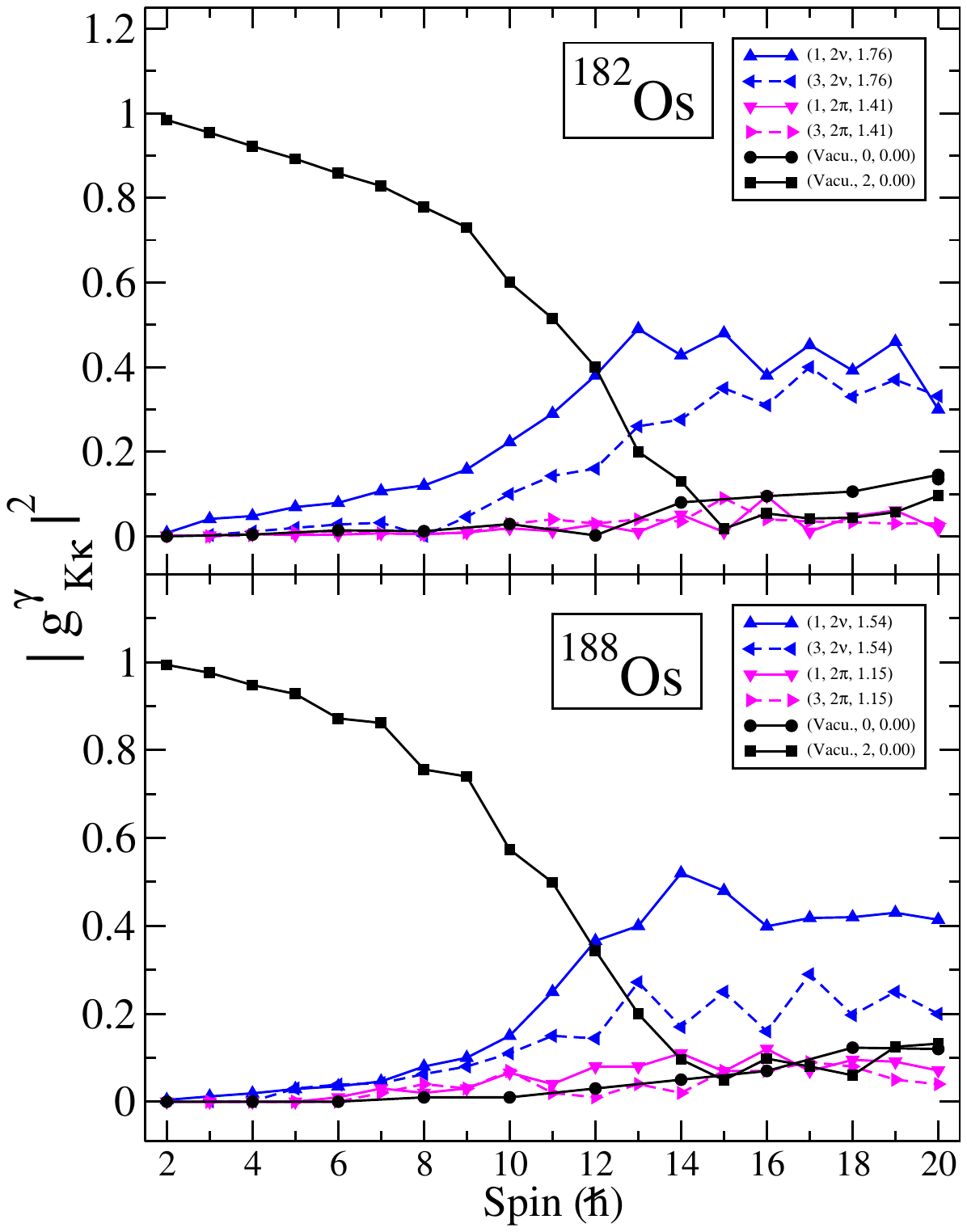}} 
 \caption{(Color online). Probabilities (\ref{eq:GKk}) of the pre-diagonalized high-j two-quasiparticle configurations in  Fig. \ref{fig:OsBandDia} for the $\gamma$-bands in $^{182,188}$Os.
   }
\label{fig:OsWfgamma}
\end{figure}

 

\begin{figure}[htb]
 \centerline{\includegraphics[trim=0cm 0cm 0cm
 0cm,width=1.18\linewidth,clip]{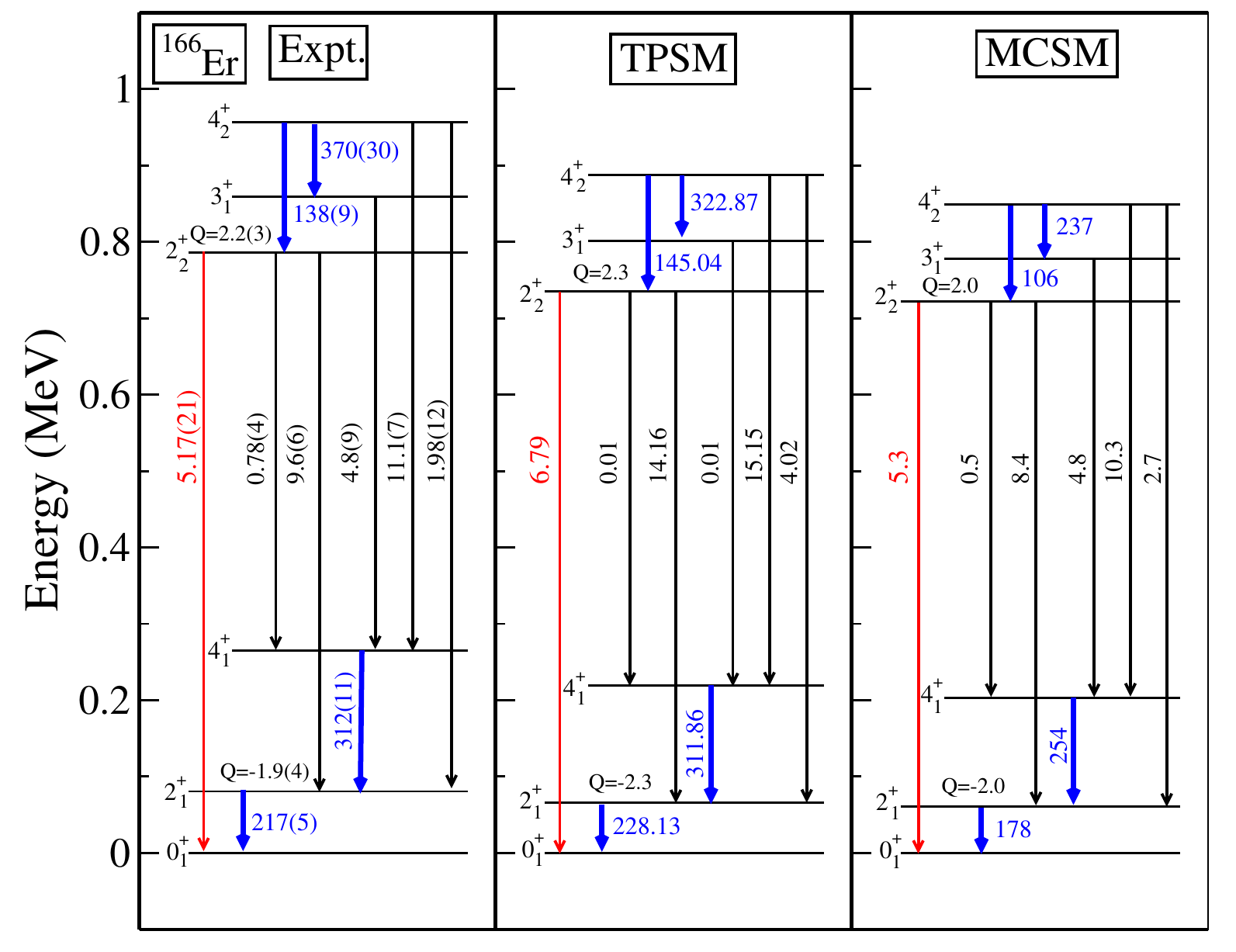}} 
 \caption{(Color online). Experimental energies, reduced transition probabilities $B(E2)$ (in W.u.) and static
 quadrupole moments $Q$ (in W.u.) of $^{166}$Er \cite{NNDC} compared with the MCSM \cite{Otsuka2019} and TPSM calculations.
   }
\label{fig:166Er_MCSM-TPSM}
\end{figure}
\begin{table}[htp!]
\LTcapwidth=0.4\textwidth
\caption{Energy characteristics of several models for $^{166}$Er.
The staggering parameter $S(3)$ is defined by Eq. (\ref{eq:staggering1}). 
G. rot. lists the results for the Gamma-rotor as described in Sec. \ref{sec:ACM} (c.f.  caption of Table \ref{tab:gav}).
T. rot. lists the  results for the triaixial rotor (Davydov) model \cite{AS60} with $\gamma=9^\circ$.
}
\resizebox{0.49\textwidth}{!}
  {
\begin{tabular}{|c|c|c|c|c|}
  \hline
 Model  	 &${\left[\frac{E(2^+_2)}{E(2^+_1)}\right]}$	&${\left[\frac{E(2^+_2)}{E(4^+_1)}\right]}$	&${\left[\frac{E(4^+_3)}{E(2^+_2)}\right]}$	&$ S(3)$ \\
  \hline										
Exp           &9.7          & 3.0      &2.5            &-0.14     \\    %
MCSM       &11.8       & 3.6      &2.8            &-0.11     \\    %
TPSM        &11.1       &3.4         &3.1           &-0.16     \\ %
G. rot.        &11.1        &3.9        & 2.2            &-0.17    \\  %
T. rot          &21.8       &6.5         & 4.0            &-0.16     \\  %
\hline								
\end{tabular}
}
\label{tab:166Er_MCSM-TPSM}
\end{table}

\subsection{Comparison with Monte Carlo Shell Model}
 
The authors of Ref. \cite{Otsuka2019,Tsunoda2021} carried out large-scale Monte Carlo Shell Model (MCSM) calculations for $^{166}$Er, which is considered to be a well deformed axially symmetric nucleus. For the lowest states 
of the ground- and $\gamma$-band, the experimental energies and the $B(E2)$ values for transitions  between them are very well reproduced. 

The MCSM  states are represented by an ensemble of stochastically deformed  Slater determinants, which are projected on to good angular momentum. 
Each of these states has definite intrinsic quadrupole moments  $\langle Q_0\rangle$ and $\langle Q_2\rangle$, where the expectation value is taken 
with the deformed Slater determinant. Multiplying with the probability, each of these states appears in the total wave function, the so called T-plot is generated,
which represents the probability distribution of the intrinsic quadrupole moments. For the $0^+_1$ ground state and the $2^+_2$ single-$\gamma$-band head, the T-plots show distributions
that are centered at a triaxiality parameter of $\gamma=9^\circ$ with an approximate width of $15^\circ$. The same holds for the  $4^+_3$ double $\gamma$ band head \cite{Otsuka2023,Otsuka2023pc}.

The authors of  Refs. \cite{Sun2000,Boutachkov2002} first applied the TPSM approach   to the $\gamma$-bands in well deformed axial nuclei.
 The nucleus $^{166}$Er belongs to this group. Table \ref{tab:166Er_MCSM-TPSM} and Fig. \ref{fig:166Er_MCSM-TPSM} compare our TPSM calculations  with the MCSM results and the experiment.
 It is evident that the TPSM describes the energies and $B(E2)$ values as well as the MCSM. 
 Thus, it  corresponds to the MCSM  picture of a $\gamma$-soft with slight triaxiality and it does not conflict  with the TPSM input.  
   As pointed out in the context of Eqs. (\ref{eq;geo}), the triaxiality parameter $\gamma_N=21^\circ$ of the Nilsson potential corresponds to the
 substantial  smaller  estimate  of $\gamma=13^\circ$ for the triaxiality of the charge and density distributions.
 The value being not far from the MCSM one and is consistent with the similarity of the TPSM and MCSM values for the observables in Fig. \ref{fig:166Er_MCSM-TPSM}.
 In a forthcoming paper we will present the deformation parameters derived from the TPSM results  by means of quadrupole shape invariants
  \cite{KM70} for several nuclides.

The similarity between the TPSM and MCSM does not come as a surprise. The MCSM states are comprised of stochastical 
configurations of the nucleons  in a deformed mean-field,
which are projected on to good angular momentum. The TPSM states are comprised of angular momentum projected   quasi-nucleon configurations  in a deformed mean-field.
In the case of the MCSM, the diagonalization procedure picks out the favorite  configurations that correspond to the appropriate deformation.
In the case of the TPSM, the deformed mean-field is optimized from the outset, and the diagonalization within the space of quasiparticle configurations 
gets the right states.

 The authors of Ref. \cite{Tsunoda2021}  demonstrated  that  the phenomenological Gamma-rotor Hamiltonian discussed in Sec. \ref{sec:ACM}
 with a square well potential and a deep attraction for $4^\circ \leq \gamma \leq 14^\circ$ provides probability distributions $P(\gamma)$ that are
 similar to the ones from the T-Plots, and that the energies and $B(E2)$ values are nearly the same as for the rigid triaxial rotor (Dawydov limit) with $\gamma=9^\circ$
 (see Fig. 1 and 2 of Ref.  \cite{Tsunoda2021}).  The $B(E2$ values correlate well with the  MCSM and the experiment. 
  However, as seen in the line "T. rot." of Table \ref{tab:166Er_MCSM-TPSM}   all three energy ratios  of the rigid triaxial rotor are by factor 1.5-2 larger than 
  the MCSM and experimental ratios. 
 
 The line "G. rot." in Table \ref{tab:166Er_MCSM-TPSM}  lists the energy characteristics of the phenomenological Gamma-rotor of Sec. \ref{sec:ACM}, with the parameters 
 $\chi=100, ~\xi=56$. This potential looks similar to the  $\chi=50, ~\xi=30$ case in Fig. \ref{fig:grot}, except that its 
 minimum is located at $\gamma_m=9^\circ$ and that it is somewhat stiffer. 
 The negative staggering parameter, $S(3)$, which indicate a certain amount of static  triaxiality, is consistent with the MCSM values.
 The three energy ratios are similar to the ones of the MCSM as well.
 However, the ratios 
 $\frac{B(E2,2^+_2\rightarrow 0^+_1)}{B(E2,2^+_1\rightarrow 0^+_1)}$=0.068, 
 $\frac{B(E2,4^+_3\rightarrow 2^+_2)}{B(E2,2^+_2\rightarrow 0^+_1)}$=2.35,
 $\frac{B(E2,2^+_2\rightarrow 2^+_1)}{B(E2,2^+_1\rightarrow 0^+_1)}$=0.150 
 deviate from the respective MCSM ratios of 0.030, 1.89, 0.047. 
 
   For the states $0^+_1,~2^+_2$ and $4^+_3$ of the 100-56 Gamma-rotor, we find that the width of the $\gamma$-probability distribution, defined as
 the distance of the classical turning point $E=V(\Delta \gamma)$ from 0, (see Sec. \ref{sec:ACM} and Appendix) 
 changes as $\Delta \gamma= 24^\circ, ~ 30^\circ, ~36^\circ$, respectively.  The respective density distributions $P(\gamma)$ have well defined 
  maxima at $\gamma= 12^\circ, ~ 18^\circ$ and $24^\circ$.  This is at variance with the T-plots of the TPSM, which look very similar for the three states 
  
Hence, the MCSM quadrupole moments indicate a distribution of triaxial shapes that is about the same for the states $0^+_1,~2^+_2$ and $4^+_3$ . 
 However, the energy ratios of such an effective triaxial rotor deviate from the MCSM ratios.  The MCSM energy ratios are approximately accounted for by an intrinsic  
  shape distributions, the triaxiality of which increases with $I$ for the states $0^+_1,~2^+_2$ and $4^+_3$. The increase  leads to deviations of the $B(E2)$ ratios from the MCSM ones.
 Apparently, the results for both the energies and quadrupole moments from  the microscopic MCSM and TPSM calculations cannot be accounted for by the same
 phenomenological Hamiltonian of the Gamma-rotor type.

\section{Summary and Conclusions}

The present work is a sequel to  our earlier investigation of energy staggering of the $\gamma$-bands 
and its relation to  nuclear triaxiality in the framework of the  triaxial projected shell model (TPSM) approach \cite{GH14,SJ21}.
We have
undertaken a comprehensive investigation  of {\it both} the energies and $B(E2)$ transition probabilities  for
a large set of thirty nuclides.  We  addressed the questions:  (1) which are the observables that characterize the nature of the triaxiality, and (2) how the microscopic TPSM results can be interpreted in
terms of the widely used phenomenology of the collective Bohr Hamiltonian?

The classification of atomic nuclei as ``spherical, axial, triaxial, rigid, soft'' is based on 
the collective Bohr Hamiltonian for the quadrupole degrees of freedom 
of the nuclear shape. To be specific, we discussed its simplified  version,
 the Gamma-rotor, which assumes a fixed deformation $\beta$ and employs two parameters to describe 
 the $\gamma$-dependence of the potential.  In this generic model,
  the location $\gamma_m$ of the potential minimum defines 
 the triaxiality and the distance $\Delta\gamma$ between 
 the semiclassical turning points of the ground-state gives rise to the softness of the mode. 
 We investigated how the energies of the lowest rotational bands and reduced $B(E2)$  probabilities within and  
 among them depend on the triaxiality and softness. 
 
 The staggering phase of the energies of the $\gamma$-band built on the $2^+_2$ state has been widely used in the literature
 to distinguish between "$\gamma$-soft'' and ``$\gamma$-rigid" motion, 
 with even-$I$-states lower for the former case, and even-$I$-states higher for the latter case.
 We demonstrated that this criterion is insufficient. To characterize the mode uniquely, it should be complemented by: the amplitude of the staggering parameter, $S(I)$,  the ratios $E(2^+_2)/E(2^+_1),~E(2^+_2)/E(4^+_1)$,
 the transition probabilities $B(E2, 2^+_2\rightarrow 0^+_1), ~B(E2, 4^+_3\rightarrow 2^+_2), ~B(E2, 2^+_3\rightarrow 2^+_1),~B(E2, 0^+_2\rightarrow 2^+_2)$,
  and the static quadrupole moments $Q(2^+_1),~Q(2^+_2)$.  
  
 The new observation of the present study is that the in-$\gamma$-band transition probabilities
 $B(E2, I\rightarrow I-2)$, stagger with a phase that is opposite to the phase  of
 the energies.  The appearance of these staggering patterns
  could be explained  in terms of the interaction of the harmonic  single $\gamma$-band with the harmonic ground-band and 
  the harmonic  double $\gamma\gamma 0$-band, which is the band based on the $0^+_2$ state.  It represents two-phonons 
  with opposite projection on the symmetry axis and, classically,  a pulsation between prolate and oblate shape. The deviations
  of the potential from the quadratic form cause couplings between the bands. This band mixing interpretation connects  the phenomenology and the microscopic TPSM results.

 We have calculated  the staggering of the $\gamma$-bands in the thirty nuclei and
 disregarding the  quasiparticle admixtures, the  staggering pattern of the $\gamma$-band is the same as for the 
 triaxial nuclei according to phenomenology. The energies are always odd-$I$-down and for the intra-band, $B(E2, I\rightarrow I-2)$ values are even-$I$-down. This is expected because the TPSM uses a
 static triaxial deformation. 
 The coupling between the $K=0$ and $K=2$ bands projected from the quasiparticle vacuum causes an upward shift 
 of the even-$I$ members of the $\gamma$-band and is in complete analogy to the Bohr Hamiltonian.
 However, the staggering amplitude is much smaller than for the collective  Hamiltonian with the same
 position of the $\gamma$-band relative to the ground-band.
 
 In the context of the collective Bohr Hamiltonian, the reversal of the energy  staggering to the 
 even-$I$-down pattern of $\gamma$-softness  reflects the coupling of the $\gamma$-band with
the $\gamma\gamma0$ pulsating mode, which incorporates the fluctuations of the $\gamma$-degree of freedom.
In the TPSM context, it is the coupling to the set of  two- and four- quasiparticle states, which modifies the vacuum pattern.
The inclusion of the quasiparticle  states into the TPSM vacuum configuration space reverses the phase of $S(I)$  
  for all selected nuclei, except for the six
nuclei of $^{76}$Ge, $^{112}$Ru, $^{188,192}$Os, $^{192}$Pt and $^{232}$Th.
 It is remarkable that TPSM  reproduces the $(N-Z)$ dependence of
the soft-rigid characteristic of $S(I)$ in all cases.

In the cases of  $^{76}$Ge \cite{SJ21}, $^{104,112}$Ru \cite{Na23,SPRTBP} and $^{188,192}$Os, for which we studied the
micro composition,
it turned out that the coupling to the  bands projected from the two-quasiparticle configurations of the high-j orbitals $f_{7/2}$, $g_{9/2}$,  $h_{11/2}$ and $i_{13/2}$ 
dictates the staggering patterns of the $\gamma$-band energies and intra-band $B(E2, I\rightarrow I-2)$ values. We could not identify a simple 
intuitive mechanism yet,  rather the interference between several terms seems to determine the final result. 
Nevertheless, we found a correlation for the Ru- and Os- isotopes. For nuclei with a sharp back-bend in the yrast
sequence the energy staggering of the $\gamma$-band
shows the even-$I$-down pattern of $\gamma$-softness and for nuclei with a smooth up-bend in the yrast sequence, the energy staggering of the $\gamma$-band
shows the odd-$I$-down pattern of $\gamma$-rigidness.


Combining the present study with our  previous work \cite{SJ21,Na23,SPRTBP}, it can be concluded that TPSM approach
describes the experimental energies and the available reduced $B(E2)$ transition probabilities of the yrast-
and $\gamma$-bands in
   the selected thirty  nuclei quite well. The same holds for the static quadrupole moments   and g-factors of the $2^+_1$ states. The inter- and intra-band $B(E2)$
   values were calculated in a systematic way  and have been listed for future experimental comparisons.
   In order to elucidate the nature of the collective $\gamma$-mode in a more complete manner, the
   experimental transition probabilities connecting the states of the $\gamma$-band with bands built on excited
   $0^+_n$ and $4^+_n$ states are essential. 
   This was demonstrated in our TPSM analysis of the rich COULEX data for the $\gamma$-soft triaxial nucleus $^{104}$Ru \cite{Na23}.

It may be argued that the coupling to two- and four- quasiparticle excitations 
is a  shell model  representation of the  fragmented   $\gamma\gamma0$-mode of the collective model. 
 However, this interpretation needs to be substantiated by further analysis. The present work found that
 for the transitional nuclei $^{154,156}$Gd , $^{188,190,192}$Os and $^{194}$Pt, the experimental $B(E2)$ values for the transitions from
the rotational band on the $0^+_2$ state to the yrast and $\gamma$-bands are collectively enhanced.   TPSM values are enhanced as well, however, 
the values noticeably deviate from experiment.  The $0^+_2$ states in these nuclei are associated with the
substantial changes in the $\beta$ deformation.
It seems that the TPSM quasiparticle configuration space is too restrictive  to quantitatively accommodate such changes. 
Improvement of the results can be achieved with the development of generator
coordinate method (GCM) by considering TPSM wavefunctions as the generating states and $\beta$ and $
\gamma$ as  generator coordinates. This extension
is presently being pursued along the lines of Refs. \cite{Chen2016,Chen2017}.

The present work can be considered as a part of the more general challenge in nuclear physics on how to associate the
 results from a large-scale matrix diagonalization with the intuitive collective model. 
 The authors of Refs.  \cite{Otsuka2019,Tsunoda2021} introduced the T-plots for relating their MCSM calculations
 to the  collective shape dynamics. It was noted that $^{166}$Er, traditionally  classified
 as well deformed axial, has a T-plot
 that indicates a triaxial distribution around $\gamma=9^\circ$. We compared the MCSM energies and the
 $B(E2)$ values for the lowest 
 states of the ground- and $\gamma$- bands with the one's from the TPSM, and found that they agree rather well. This
 seems to indicate that 
 the TPSM  and MCSM results corresponds to a similar shape distribution. 
 
  In the present work, we followed the standard route  
  of calculating  energies and transition probabilities for the individual states and comparing them with the corresponding
  experimental values. Other approaches, for instance,
  the quadrupole shape invariant analysis will provide a complementry perspective on the relation between the TPSM and
  the collective model  as we demonstrated for 
$^{104}$Ru  \cite{Na23}. We are in the process of performing a systematic shape invariant analysis for a set of nuclei
discussed in this manuscript for which the Coulomb excitation data is available, and the results will be
presented in a forthcoming publication.

\section*{Appendix: Detailed discussion of the Gamma-rotor model}\label{sec:app1}

Section \ref{sec:ACM} presented the results of the Gamma-rotor model for a selection collective potentials,  which were used to classify the 
nature of the triaxiality  of the nuclear shape. A summary of characteristic relations between energies and transition probabilities between 
of the lowest  collective excitations of the quadrupole type was given. Here we provide the details of how these characteristic come about.

The potential $\chi-\kappa=200-0$ represents a rigid prolate nucleus with a harmonic $\gamma$-vibration  far above the $4^+_1$ level
of the ground-band and at about twice the energy are the two-phonon bands. The $K=4$ band labeled by $4^+_3$ in Fig. \ref{fig:grot}
represents a traveling wave generated by adding a second $K=2$ on top of the first with the same angular momentum along the symmetry axis.
Transforming the ratio $B(E2,4^+_3\rightarrow 2^+_2)/B(E2,2^+_2\rightarrow 0^+_1)=0.094/0.033$ in Table \ref{tab:gtrans} into the ratio 
 of the matrix elements for transitions between states in an axial symmetric potential (see \cite{BM75,RW10}), 
 one obtains 1.42, which is approximately the ratio 
 expected  for the two- and one-phonon states of a harmonic vibration. 
 The $K=0$ two-phonon state is generated by putting the second phonon with opposite angular momentum 
 on top of the first, which represents  a pulsating wave.   
 Transforming the ratio $B(E2,0^+_2\rightarrow 2^+_2)/B(E2,2^+_2\rightarrow 0^+_1)=0.195/0.033$ in Table \ref{tab:gtrans}
   into the ratio of the matrix elements for
 transitions between the states, one obtains 1.08, which is close to 1 expected for a harmonic vibration.
 The potential $\chi (1-\cos{3\gamma})=3/2\gamma^2+...$ is not exactly harmonic. The distances  between the
 bands are large, such that the couplings of the harmonic bands with higher terms is small, which is reflected by the small value of $\bar S(6)=-0.03$ in Table \ref{tab:gav}.
 
  The difference of the harmonic ratios can be understood by the following
 classical consideration.  The $K=2$ one-phonon state correspond to  a traveling wave in the plane perpendicular to the symmetry axis with $x=A\cos{\omega t}$ and $y=A\sin{\omega t}$.
 The average radiation power is $\propto\overline{ x^2} +\overline{ y^2}+2\overline{xy}= A^2$. The $K=4$ two-phonon state correspond to a traveling wave with the amplitude $\sqrt{2}A$. The radiation power of the $K=4$ two-phonon state is 2$A^2$. The $ K=0$ two-phonon state correspond to a pulsating wave $x=\sqrt{2} A\cos{\omega t}$, which gives a radiation power
 $\propto \overline{x^2}=A^2$,  corresponding to a ratio of 1. 
 
 The second limiting case is the $\gamma$-independent potential (0-0) of the Wilet-Jean model \cite{Wilets}, which is discussed in detail in the textbook by Rowe and Wood 
\cite{RW10} (p. 117, 222 ff.) The states carry the seniority quantum number $v$. The quadrupole operator changes the seniority by 1, 
which is reflected by the $B(E2)$ values 
in Table \ref{tab:gtrans}. The static quadrupole moments vanish because they are matrix elements between states with  the same $v$, and the 
$B(E2,2^+_2\rightarrow0^+_1)=B(E2,4^+_3\rightarrow2^+_2)=0$ because $v$ changes by 2.
The states organize into $\Delta I=2$ bands of good signature, which are connected by 
strong $B(E2,I\rightarrow I-2)$ values   (cf. also Fig. 2.3 of Ref. \cite{RW10}). The band based on the $2^+_2$ state has $v=2+I/2$ and the one on $3^+_1$ has
$v=3+I/2$. As the energy of the states is $E(v)=v(v+3)$, the two $\gamma$-band  branches have a pronounced even-$I$-down staggering
\begin{equation}\label{eq:SWJ}
S(I)=\frac{1}{8}\mp\left(\frac{I}{4}+\frac{9}{8}\right),~~I=\begin{array}{c} even\\ odd\end{array}.
\end{equation}
The $\gamma$-band 
head has the same energy as the $4^+_1$ state of the ground-band,  $E(2^+_2)=E(4^+_1)$, which is another signature of the $\gamma$-independent potential. 

The shallow potential (10-0) and the soft potentials 20-0 and 50-0 illustrate the transition to rigid prolate limit (for more details see Ref. \cite{caprio11}).  The  static quadrupole  moments quickly approach the limit for prolate shape as a consequence of the seniority mixing.
 The $B(E2,2^+_2\rightarrow0^+_1)$ and $B(E2,4^+_3\rightarrow2^+_2)$ first increase because of 
the seniority mixing,  and then they decrease because the amplitude of the $\gamma$-vibration decreases with $\chi$. 
The $B(E2, 2^+_2\rightarrow 2^+_1)$ values decrease  due to the seniority mixing.
 The potential term $\cos 3\gamma$ couples states of $v=\pm1$ \cite{RW10}. The coupling of the even-$I$ branch  of the $\gamma$-band with the ground-band pushes them up.
The even-$I$ band on top of the $0^+_2$ band does not couple because $v$ differs by 2,
which reduces the staggering.  For large values of the $\chi$ the higher order $\Delta v=\pm 1$ couplings quench the staggering.  
The repulsion caused by the coupling pushes the $2^+_2$ band head above the $4^+_1$ state of the ground-band. 

The prolate harmonic limit provides another perspective that is instructive for interpreting the TPSM results. First 
inspect the probability density $P(\gamma)$ of the wave functions of the lowest bands obtained by integration over the angle degrees of freedom.
Fig. 9 of Ref. \cite{caprio11} shows the case of potential 50-0. The density  $P(\gamma)$ of the ground band is centered at $\gamma=0^\circ$ 
(after dividing out the volume element  $\sin 3\gamma$).
For the $\gamma$-band built on the $2^+_2$, the density has a maximum around 20$^\circ$, which indicates that 
the state represents a wave that  travels around the symmetry axis. The double $\gamma$-band ($\gamma\gamma 4$) built on the $4^+_3$ state has a maximum
at $27^\circ$, which correspond to a traveling wave with a larger amplitude. (The  concept of traveling  (tidal) waves 
 has been developed to microscopically  calculate  unharmonic many  phonon excitation (see e.g. Ref. \cite{Frauendorf15}).)
For the  double $\gamma$-band ($\gamma\gamma0$) built on the 0$^+_2$ state the density $P(\gamma)$ has two maxima with a zero at $15^\circ$ in between.
The $0^+_2$ state represents a vibration of the triaxial shape between the prolate and oblate turning points (see \cite{BM75}). The structure of the wave functions for other $\chi$ values is qualitatively the same.
For larger $\chi$ the distributions $P(\gamma)$ are squeezed such that they fit into the potential (see 200-0 in Fig. 9 of \cite{caprio11}).
 For smaller $\chi$ the distributions are shifted to larger $\gamma$ until they become symmetric about $30^\circ$ for $\chi=0$.
 
The deviation of ``$1-\cos 3\gamma$'' from the leading order term $3\gamma^2/2$ couples the even-$I$-states of the
$\gamma$-band with the states of the ground-band and of the $0^+_2$ band, which, respectively, shifts them up or down. 
The down shift by the repulsion from the upper level prevails because it has a larger coupling matrix element. The reason is that
the difference  $\vert1-\cos 3\gamma-3\gamma^2/2\vert$
becomes larger with increasing $\gamma$, and the probability density $P(\gamma)$ of the ground-band is closer to zero localized than the one of the $0^+_2$ band, 
which has a zero and reaches further out 
(compare the cases g, $\gamma$ and $\gamma\gamma 0$) for the potential  200-0 in Fig. 9 of \cite{caprio11}). With decreasing $\chi$, the potentials become shallower, 
which reduces distance between the bands. The level shift become larger because the  energy differences decrease and the coupling matrix elements increase. 
The even-$I$-down staggering pattern of the $\gamma$-band evolves.

The potentials with $\chi$=0 represent the cases of maximal triaxiality.  
The term $\cos^23\gamma$ of the potential is symmetric with respect to $\gamma\rightarrow60^\circ \gamma$. It changes the seniority by $\Delta v=0, \pm2$.
The states have good $\gamma$-parity, which is reflected by $Q(2^+_1)=Q(2^+_2)=0$, $B(E2, 2^+_2\rightarrow 0^+_1 )=0$ and $B(E2, 4^+_3\rightarrow 2^+_2 )=0$
in Table \ref{tab:gtrans}. For small $\kappa$ the diagonal term $\langle Iv\vert \cos ^2 3\gamma\vert Iv\rangle$ dominates, which is larger for even $I$ ($\gamma$ parity even)
than for odd $I$ ($\gamma$-parity odd). With increasing $\kappa$ level shifts decrease the even-$I$-down staggering of the $\gamma$-band until it disappears and changes into
the even-$I$-up pattern (see 0-20 and 0-100 in Fig. \ref{fig:gband} and Table \ref{tab:gtrans}).  In the case of $\kappa<0$ with a barrier at $\gamma=30^\circ$ the shifts increase
the staggering (see 0- (-20) in Fig. \ref{fig:gband} and Table \ref{tab:gtrans}). With increasing $\kappa$ the $\Delta v=\pm2$ terms become important and the structure of 
the bands based on the $0^+_1,~2^+_2,~4^+_3$ state quickly approach the ones of the triaxial rotor with $\gamma=30^\circ$. 

 The case is discussed as the Meyer-ter-Vehn limit in Ref. \cite{RW10} and is known as the symmetric top  in molecular physics.
 The ratios of the moments of inertia are 4:1:1 for the medium, short, long axes, respectively, which means the eigenfunctions are the ones of 
a symmetric rotor, where $K$ is the angular momentum projection on the medium axis with the largest moment of inertia. The energy ratios are
\begin{equation}\label{eq:ETR}
\frac{E(I,K)}{E(2^+_1)}=\frac{4I(I+1)-3K^2}{12}.
\end{equation}
The corresponding ratio $E(2^+_2)/E(4^+_1)=0.75$ represent the limit of maximal triaxiality. 
For the potential 0-200 the ratio is 0.81 and for the potential 0-20 it is 0.97.
The Meyer-ter-Vehn limit of the staggering parameter is
\begin{equation}\label{eq:STR}
S(I)=\frac{1}{6}\pm\left(I-\frac{5}{2}\right),~~I=\begin{array}{c} even\\ odd\end{array}.
\end{equation}
The value $\bar S(6)=4$ is to be compared with 3.87 for the  potential 0-200  in Table \ref{tab:gtrans}. The quadruple operator is $(Q_2+Q_{-2})/\sqrt{2}$ with 
respect to the quantization axis ``m''. The ratios of reduced transition probabilities are given by the corresponding ratios of the squares of the Clebsch-Gordan coefficients,
(quoted e.g., Fig. 2.6 of Ref. \cite{RW10}). For the lowest transitions, the $B(E2)$ values divided by the \mbox{$B(E2, 2^+_1\rightarrow 0^+_1)$} are 
  $B(E2, 2^+_2\rightarrow 2^+_1)$=1.43 (1.42, 1.42), $B(E2, 4^+_2\rightarrow 2^+_2)$=0.60 (0.58, 0.73), 
  $B(E2, 4^+_1\rightarrow 2^+_1)$=1.39 (1.43, 1.42), $B(E2, 3^+_1\rightarrow 2^+_2)$=1.79 (1.71, 1.45), $B(E2, 4^+_3\rightarrow 3^+_1)$=0.56 (0.60, 0.86), where
 the ratios for the potentials 0-200 and 0-20 are quoted in parenthesis.  The $B(E2)$ ratios for the Meyer-ter-Vehn  limit are 
 more rapidly approached than for the energies. 
 
The transition from the symmetric triaxial to the prolate potentials  involves the competition between 
 the two potential terms discussed. Consider the soft potentials 0-50, 50-50 and 50-0.  The symmetry with respect to $\gamma\rightarrow60^\circ-\gamma$  of 0-50
 is progressively broken in 50-50 and 50-0. 
  As a consequence, the  $B(E2, 2^+_2\rightarrow 0^+_1 )$ and $B(E2, 4^+_3\rightarrow 2^+_2 )$ 
become quickly large and the $B(E2, 2^+_2\rightarrow 2^+_1 )$ decrease.
 The even-$I$-up staggering of the $\gamma$-band is reduced, and it reverses to the even-$I$-down pattern for the potentials. Along the sequence the state $2^+_2$ moves away from the state $4^+_1$ to larger energies. 
Both trends can be taken as signatures of decreasing triaxiality. The other cases between the prolate and symmetric triaxial limits exemplify the changes.
Hence, a small staggering parameter indicates either prolate potential or a transitional one, where the former has a large and the latter a small ratio $E(2^+_2)/E(4^+_1)$.
For all the cases, the $0^+_2$ states represent a pulsating $\gamma$ vibration with a zero in the $P(\gamma)$ probability density, the position of which depends on the 
potential parameters (see Fig. 9 of \cite{caprio11}). The $B(E2, 0^+_2\rightarrow 2^+_2 )$ values are always substantial, being larger for more flatter potentials. 
 
The reversal of the phase of $S(I)$ with increasing $\kappa$ is understood as follows: The term symmetric $(\cos3\gamma)^2$ couples the $\gamma$-band strongly to the ground-band
because both the states do not have a zero in the $\gamma$ degree of freedom. The coupling to   the $0^+_2$ band is weaker because its states have a zero in the 
$\gamma$ degree of freedom. Hence the repulsion between the even-$I$-states of the ground- and $\gamma$-band prevails which generates even-$I$-up staggering pattern. 
As discussed above,  the ``$\cos3\gamma$'' term generates the even-$I$-down staggering pattern, and the competition of both terms determine the phase of $S(I)$.

 

The potential 20-50 represents a case incipient of prolate-oblate shape coexistence. There is a regular rotational ground-band with $P(\gamma)$ localized on the prolate side. 
The $0^+_2$ state  comes low in energy. The second bump of $P(\gamma)$ on the oblate side is larger than the first on the prolate side. The  $\Delta I=2$ sequence 
built on the $0^+_2$ state can be interpreted as the "oblate" rotational band.  However, one can alternatively interpret the sequence as a band built on a large-amplitude 
$\gamma$-vibration, which applies to all potentials (see discussion of the WJ limit related to Fig. 4.15 of Ref. \cite{RW10}).

Now the mixing of the even-$I$ members of the harmonic  ground-band (g), $\gamma$-band ($\gamma$)  
and double-$\gamma$-band ($\gamma\gamma0$) is considered in detail, 
\begin{eqnarray}\label{eq:mixing}
\psi(I,M,\Omega,\gamma)=c_g\psi_g(I,M,\Omega,\gamma)\nonumber\\
+c_{\gamma\gamma0}\psi_{\gamma\gamma0}(I,M,\Omega,\gamma)\nonumber\\
+\sqrt{1-c_g^2-c_{\gamma\gamma0}^2}\psi_\gamma(I,M,\Omega,\gamma),
\end{eqnarray}
where $c_g$ and $c_{\gamma\gamma0}$ are the mixing amplitudes.
The basis states are eigenfunctions of the Gamma-rotor Hamiltonian (\ref{eq:ACM}) with the harmonic prolate potential $\chi 3/2\gamma^2$,
\begin{eqnarray}
\psi_g(I,M,\Omega,\gamma)=N(I)D^I_{M0}(\Omega)\phi_g(\gamma),\\
\psi_\gamma(I,M,\Omega,\gamma)=N(I)D^I_{M2}(\Omega)\phi_{\gamma}(\gamma),\\
\psi_{\gamma\gamma0}(I,M,\Omega,\gamma)=N(I)D^I_{M0}(\Omega)\phi_{\gamma\gamma0}(\gamma),
\end{eqnarray}
where $N(I)$ are the normalization factors of the $D$ functions. The $\psi(\gamma)$ are  the normalized wave functions in the $\gamma$-degree of freedom,
the probability distributions are shown in Fig. of Ref. \cite{caprio11}. Their phases are chosen such that they are positive for small $\gamma$. 
The reduced transition matrix elements for the $I\rightarrow I-2$ transitions 
in units of $3ZR_0^2\beta/4\pi$ are the following
(cf. Ref. \cite{BM75} Eq. (4-91)).
\begin{eqnarray}
\langle I-2,\gamma\vert\vert {\cal M}(E2)\vert\vert I, \gamma\rangle=\nonumber\\
\sqrt{2I+1}\langle I~2~2~0\vert I-2~2\rangle\langle\psi_\gamma\vert\cos\gamma\vert\psi_\gamma\rangle~>0.
\end{eqnarray}
\begin{eqnarray}
\langle I-2,g\vert\vert {\cal M}(E2)\vert\vert I ,\gamma\rangle=\nonumber\\
\sqrt{2I+1}\langle I~2~2~-2\vert I-2~0\rangle\langle\psi_g\vert\sin\gamma\vert\psi_\gamma\rangle/\sqrt{2}~>0.
\end{eqnarray}
The matrix element is positive because both $\psi_g(\gamma)$ and $\psi_\gamma(\gamma)$ do have a zero and are positive
over the whole $\gamma$ range.
 \begin{eqnarray}
\langle I-2,\gamma\gamma0\vert\vert {\cal M}(E2)\vert\vert I ,\gamma\rangle=\nonumber\\
\sqrt{2I+1}\langle I~2~2~-2\vert I-2~0\rangle\langle\psi_{\gamma\gamma0}\vert\sin\gamma\vert\psi_\gamma\rangle/\sqrt{2}~<0.
\end{eqnarray}
The function  $\psi_{\gamma\gamma0}(\gamma)$ is positive for small $\gamma$ and  negative for large $\gamma$,
while  $\psi_g(\gamma)$ is positive over the whole $\gamma$ range.
Integrating the $\sin\gamma$ function results in a negative value because it weights more the large $\gamma$ values.   
As discussed for the prolate regime, 
$\vert \langle I-2,\gamma\gamma0\vert\vert {\cal M}(E2)\vert\vert I ,\gamma\rangle\vert =\langle I-2,g\vert\vert {\cal M}(E2)\vert\vert I ,\gamma\rangle$
in the harmonic limit.

The coupling is generated by the difference of the potential from its prolate
quadratic approximation
\begin{equation}
V_c(\gamma)=\chi(1-\cos3\gamma)+\kappa(\cos^23\gamma-1)-\chi\frac{3\gamma^2}{2}.
\end{equation}
For a prolate potential ($\kappa=0$)
$V_c(\gamma)$ is weakly negative for small $\gamma$ and 
strongly negative for large $\gamma$. The coupling matrix element 
$V_{g\gamma}=\langle\psi_g\vert V_c(\gamma)\vert\psi_\gamma\rangle<0$.
The matrix element $V_{\gamma\gamma0\gamma}=\langle\psi_{\gamma\gamma0}\vert V_c(\gamma)\vert\psi_\gamma\rangle>0$ 
because $\psi_{\gamma\gamma0}(\gamma)<0$ for large $\gamma$ where 
$\vert V_c\vert$ is large. For $\psi_{g}(\gamma)$ is predominantly localized in the 
small-$\gamma$ region and $\psi_{\gamma\gamma0}(\gamma)$ is predominantly localized in the 
large-$\gamma$ region $\vert V_{g\gamma}\vert<V_{\gamma\gamma0\gamma}$. 
With $E_\gamma(I)-E_g(I)= E_{\gamma\gamma0}-E_\gamma(I)$ the diagonalization
of the 3 x 3 matrix gives a down-shift of the middle eigenvalue, which represents the 
$\gamma$-band with admixtures. As expected, the repulsion from the $\gamma\gamma0$
state prevails over the one from the g-state because of the larger coupling matrix element 
of the former. The   mixing amplitudes are both negative, 
where $\vert c_{\gamma\gamma0}\vert>\vert c_g\vert$, which {\it increases} the reduced matrix element by  
$\vert c_{\gamma\gamma0}-c_g\vert\langle I-2,g\vert\vert {\cal M}(E2)\vert\vert I, \gamma\rangle$.
 
Now consider a triaxial potential by increasing $\kappa$ while keeping $\chi$ constant (see 50-0, 50-50, 50-100 in Fig. \ref{fig:grot}).
The term  $\kappa(\cos^23\gamma-1)$, which  decreases $V_c$, is symmetric about $30^\circ$. As a cosequence,the coupling matrix element 
$V_{g\gamma}=\langle\psi_g\vert V_c(\gamma)\vert\psi_\gamma\rangle$ becomes more and more negative. 
The matrix element $V_{\gamma\gamma0\gamma}=\langle\psi_{\gamma\gamma0}\vert V_c(\gamma)\vert\psi_\gamma\rangle$ does not
change much because $\psi_{\gamma\gamma0}(\gamma)$ is antisymmetric about $30^\circ$. Hence for a certain value of $\kappa$ 
coupling matrix element reverses order to $\vert V_{g\gamma}\vert>V_{\gamma\gamma0\gamma}$ and the staggering  
of the energies and of the $B(E2)$ values change to the even-$I$-up pattern of triaxial potentials. 

 For the limiting case of the WJ potential the staggering of the inband $B(E2)$ values 
reflects the seniority conservation. The $B(E2)$ transition operator (\ref{eq:E2}) has the selection rule $\Delta v=\pm1$. As seen
in Fig. \ref{fig:gband},  the transitions $I_1\rightarrow I_1-2, ~I$-odd are allowed while the transitions   $I_2\rightarrow I_2-2, ~I$-even
are forbidden. This results  in a staggering pattern of $SE22(I)$ with the phase opposite to the one of $S(I)$. In analogous conjecture holds for $SBE21(I)$. 
For the symmetric top limit the reversed staggering pattern  appears as a consequence of the fact that the ratios of  $B(E2)$ values follow the Alaga rules.

\section*{Acknowledgments}

The authors are thankful to Science and
Engineering Research Board (SERB), Department of Science and
Technology (Govt. of India) for providing financial assistance under the 
Project No.CRG/2019/004960, and for the INSPIRE fellowship to one of the author (NN). 
SF thanks Prof. M. Caprio for providing the ACM code and the help for applying it, and Prof. T Otsuka for private communications.

\bibliographystyle{epj}
\bibliography{BE2_transitions_epja}

\end{document}